\documentclass[aps,prb,showpacs,amsmath,amssymb,notitlepage,superscriptaddress]{revtex4-1}

\usepackage{graphicx}
\usepackage{bm}
\usepackage{epsfig}
\usepackage{color}

\usepackage{comment}
\usepackage{mathtools,pbox}
\usepackage{tabularx}
\usepackage{ragged2e}
\newcolumntype{Y}{>{\RaggedRight\arraybackslash}X}
\usepackage{booktabs}

\newcommand{\vR}{{\mbox{\boldmath$R$}}}

\newcommand{\vg}{\mbox{\boldmath$\gamma$}}
\newcommand{\hvg}{\hat{\mbox{\boldmath$\gamma$}}}

\newcommand{\vH}{\mbox{\boldmath$H$}}

\newcommand{\vk}{{\mbox{\boldmath$k$}}}
\newcommand{\vsig}{\mbox{\boldmath$\sigma$}}
\newcommand{\vv}{{\mbox{\boldmath$v$}}}

\newcommand{\vsk}{{\mbox{\boldmath$k$}}}

\newcommand{\vq}{\mbox{\boldmath$q$}}
\newcommand{\vd}{\mbox{\boldmath$d$}}
\newcommand{\vh}{\mbox{\boldmath$h$}}

\def\bk{ \bm{k} }

\begin{document}

\title{Superconductivity and spin-orbit coupling in non-centrosymmetric materials: a review}

\author{M. Smidman}
\affiliation{Center for Correlated Matter and Department of Physics, Zhejiang University, Hangzhou 310058, China}
\author{M. B. Salamon}
\affiliation{UTD-NanoTech Institute, The University of Texas at Dallas, Richardson, Texas 75083–0688, USA}
\author{H. Q. Yuan}
\email{hqyuan@zju.edu.cn}
\affiliation{Center for Correlated Matter and Department of Physics, Zhejiang University, Hangzhou 310058, China}
\affiliation{Collaborative Innovation Center of Advanced Microstructures, Nanjing 210093, China}
\author{D. F. Agterberg}
\email{agterber@uwm.edu}
\affiliation{Department of Physics, University of Wisconsin, Milwaukee, WI 53201, USA}


\begin{abstract}
In non-centrosymmetric superconductors, where the crystal structure lacks a centre of inversion, parity is no longer a good quantum number and an electronic antisymmetric spin-orbit coupling (ASOC) is allowed to exist by symmetry.  If this ASOC is sufficiently large, it has profound consequences on the superconducting state. For example, it generally leads to a superconducting pairing state which is a mixture of spin-singlet and spin-triplet components. The possibility of such novel pairing states, as well as the potential for observing a variety of unusual behaviours, led to intensive theoretical and experimental investigations. Here we review the experimental and theoretical results for superconducting systems lacking inversion symmetry. Firstly we give a conceptual overview of the key theoretical results. We then review the experimental properties of both strongly and weakly correlated bulk materials, as well as two dimensional systems. Here the focus is on evaluating the effect of ASOC on the superconducting properties and the extent to which there is evidence for singlet-triplet mixing. This is followed by a more detailed overview of theoretical aspects of non-centrosymmetric superconductivity. This includes the effects of the ASOC on the pairing symmetry and the superconducting magnetic response, magneto-electric effects, superconducting finite momentum pairing states,  and the potential of non-centrosymmetric superconductors to display topological superconductivity.
\end{abstract}


\maketitle
\section{Introduction} \label{Introduction}

Research on superconductors without inversion symmetry significantly intensified with the discovery of superconducting CePt$_3$Si. This was despite the vast number of known superconductors that also do not have an inversion center (see Table I for a list). The interest in CePt$_3$Si stemmed from its unusual superconducting behaviour (which is described in more detail below). With the flurry of theoretical activity that followed, it became clear that superconductors without parity symmetry can exhibit properties not anticipated or even permitted in inversion symmetric materials. The field grew from the experimental perspective as well, with many new non-centrosymmetric superconducting materials being discovered, some of which demonstrating new and novel physics.  In this review, we provide an overview of these developments.

In particular, after a brief summary of theoretical implications, we begin with an overview of relevant materials, highlighting materials that display unusual physics. These materials include weakly correlated materials (for example d-electron materials), strongly correlated materials (such as heavy electron materials), two-dimensional materials (MoS$_2$), and topological superconductors. This will be followed by a detailed overview of the role of spin-orbit coupling on the single electron physics, which is an essential ingredient in microscopic descriptions. Following this overview, the theory of the superconducting state will be developed, including discussion on the symmetry classification of the gap functions, the role of magnetic fields (spin susceptibility, magneto-electric effects, critical fields, and finite momentum pairing phases), and an overview of topological superconductivity will be given.

\subsection{Conceptual description of key results in non-centrosymmetric superconductivity} \label{conceptual}

\begin{figure}[t]
\begin{center}
  \includegraphics[width=0.5\columnwidth]{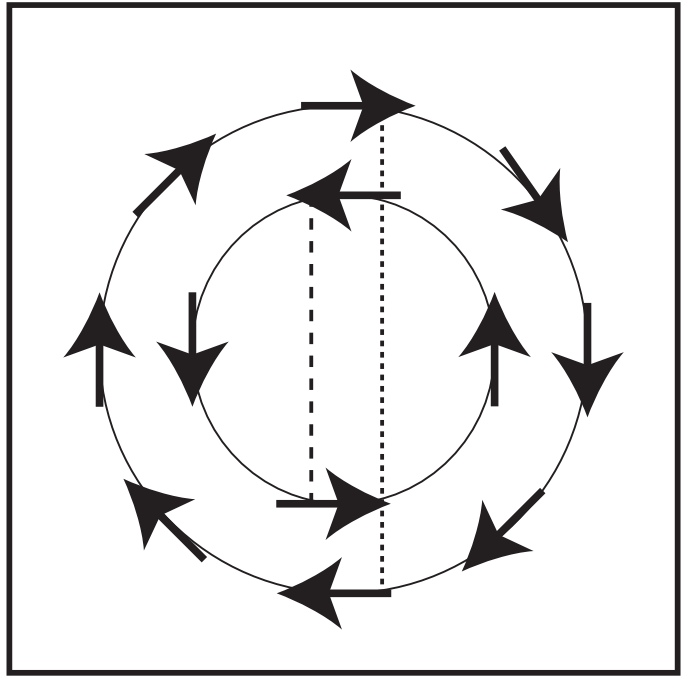}	
\end{center}
	\caption{Helicity bands with a Rasbha spin-orbit interaction. The arrows indicate the directions of the spin eigenstates. The dashed lines indicate Cooper pairs made from these states as discussed in the text.}
  \label{fig1}	
\end{figure}

As a conceptual introduction to what follows, it is useful to consider a Rashba spin-orbit coupling \cite{ras60} and the consequences of this coupling on superconductivity in two-dimensions \cite{NCSGorkov}. Rashba spin-orbit coupling is known to occur when a mirror symmetry is broken, for example in 2D electron gases. It is an example of the antisymmetric spin-orbit coupling (ASOC) that plays an important role later. We will use $\vg({\bf k})$ to represent the ASOC throughout this article (we also often use $\hvg({\bf k})$ to represent the normal vector along the direction $\vg(\vk)$). The Rashba spin-orbit coupling has $\vg(\vk)=\alpha(k_y,-k_x,0)$ , leading to the interaction  $\vg(\vk)\cdot \vsig=\alpha (k_y \sigma_x-k_x\sigma_y)$ where $\sigma_i$ are Pauli matrices corresponding to the usual spin-1/2 operators for a fermion quasi-particle. When added to a usual quadratic dispersion, $\hbar k^2/(2m)$, this yields single-particle states (called helicity states) with energies $\epsilon_{\pm}=\hbar k^2/2m \pm \alpha|k|$, leading to circular bands in momentum space that have the fermion spins polarized tangential to the momentum as shown in Fig.~\ref{fig1}. The relationship between these helicity bands (labeled by $|\vsk,\pm>$ where $\vsk$ is the momentum of the state) and the original spin 1/2 basis (labelled by $|\vsk,\uparrow>,|\vsk,\downarrow>$) can be expressed as
\begin{eqnarray}
|\vsk,\uparrow>= &&|\vsk,+>+ie^{i\phi_k}|\vsk,-> \nonumber \\
|\vsk,\downarrow>= &&ie^{-i\phi_k}|\vsk,+>+|\vsk,->
\end{eqnarray}
where $\phi_k$ is the polar angle in momentum space (note that $e^{i\phi_{-k}}=-e^{i\phi_k}$). With the spin-orbit interaction present, it is instructive to examine the formation of Cooper pairs, specifically, the two Cooper pairs shown in Fig.~\ref{fig1}. If the Rashba spin-orbit coupling is much larger than the pairing gap, this figure reveals that the state $|\vsk_1,\rightarrow>$ (here $\rightarrow$ refers to the spin direction of the helicity eigenstate for momenta $\vsk_1$ and $\vsk_2$) can be paired with $|-\vsk_1,\leftarrow>$ but not with $|-\vsk_1,\rightarrow>$ (since this state is not degenerate with $|\vsk_1,\rightarrow>$, so this pairing would have a large kinetic energy cost), while the state $|\vsk_2,\leftarrow>$ can be paired with with the state $|-\vsk_2,\rightarrow>$ but not with the state $|-\vsk_2,\leftarrow>$. Consequently, for ${\bf k}_i$ taking the values ${\bf k}_1$ and ${\bf k}_2$, only linear combinations of the spin-singlet Cooper pair, $|\psi_-\rangle$, and only one of three possible spin-triplet Cooper pairs, $|\psi_+\rangle$, can be stable, with
\begin{equation}
|\psi_{\pm}({\bf k}_i)\rangle=\frac{1}{\sqrt{2}}\Big (|\vsk_i,\leftarrow\rangle|-\vsk_i,\rightarrow \rangle\pm |\bk_i,\rightarrow\rangle|-\bk_i,\leftarrow\rangle\Big ).
\end{equation} This provides an illustration of the first key idea in non-centrosymmetric superconductors: {\it that spin-singlet superconductivity is largely unaffected by broken inversion symmetry and that there is a single `protected' spin-triplet pairing state that survives the presence of the ASOC} \cite{CePt3SiMnSi}.

To gain a deeper understanding of the superconducting state, it is useful to re-express the usual spin-singlet Cooper pair $|\psi_s\rangle=(|\vsk,\uparrow>|-\vsk,\downarrow>-|\vsk,\downarrow>|-\vsk,\uparrow>)/\sqrt{2}$ in the helicity basis (ignoring pairing between fermions on different helicity bands)
\begin{equation}
|\psi_s \rangle =\frac{i}{\sqrt{2}}\Big (-e^{-i\phi_k}|\vsk,+\rangle|-\vsk,+\rangle+e^{i\phi_k} |\bk,-\rangle|-\bk,-\rangle\Big ).
\end{equation}
This expression implies that a spin-singlet superconductor will have the same gap magnitude on both helicity bands (the phase factors that appear stem from the operation of time-reversal operator on the helicity basis, $\mathcal{T}|\vsk,\pm>=ie^{\pm i\phi_k}|-\vsk,\pm >$ and the fact that the paired Fermions are partners under time-reversal \cite{ser04}). In general, there is no physical reason that both helicity bands should have the same gap magnitude. Indeed, different gap magnitudes on the two helicity bands can occur and signal the simultaneous existence of spin-singlet and spin-triplet pairing. Such a mixture is permitted since parity symmetry is broken. In particular, the Pauli exclusion principle implies that spin-triplet pairing is of odd parity and spin-singlet is of even parity. Consequently, when parity symmetry is present, these states cannot mix \cite{sig91}. However, once parity symmetry is broken, such mixing can occur \cite{NCSGorkov}. {\it Singlet-triplet mixing is the second key notion that is prevalent in non-centrosymmetric superconductors}. Generically, this mixing leads to two-gap physics for the Rashba case discussed here. The degree of the singlet-triplet mixing (and the resultant two-gap physics) is determined by the pairing interactions as well as the spin-orbit coupling. Indeed, if there is only spin-singlet $s$-wave pairing interactions, then $|\psi_s\rangle$ is the only state that appears, independent of the strength of the spin-orbit coupling, and there will only be a single gap. It is interesting to note that when considering more general forms of the ASOC,
singlet-triplet mixing can also lead to gap nodes that are not imposed by symmetry and have interesting topological properties. This is discussed in more detail later.

More insight can be gained by adding a Zeeman field $H_xS_x+H_yS_y+H_zS_z$ to the Hamiltonian. In this case, the quasi-particle energies become
$\epsilon_{\pm}=k^2/(2m)\pm\sqrt{(\alpha k_y+H_x)^2+(\alpha k_x-H_y)^2+H_z^2}$. In the limit that the spin-orbit coupling is much larger than the Zeeman field, this can be written as
\begin{equation}
\epsilon_{\pm}\cong k^2/(2m)\pm \alpha|k|\pm \hvg(\vsk)\cdot {\bf H}
\end{equation}
where $\hvg(\vsk)=(-ky,kx,0)/|k|$.  This latter equation reveals that if $\hvg$ is perpendicular to ${\bf H}$ (in the Rashba case this means for ${\bf H}$ along the $\hat{z}$ direction), then there is no coupling to the Zeeman field (in the large spin-orbit limit) and this suggests that there will be no Pauli paramagnetic suppression of the critical field (this is true even for a spin-singlet gap function).  This also suggests that the Pauli critical field will be anisotropic. Indeed, this turns out to be the case and the resultant anisotropy in the large spin-orbit coupling limit is largely independent of the degree of singlet-triplet mixing of the gap function, implying that this anisotropy cannot reveal any information about the relative size of the spin-singlet and spin-triplet components \cite{CePt3SiMnSi}. Similar arguments apply to spin susceptibility measurements \cite{NCSGorkov,FrigeriSusc}. This leads to the third key result: {\it in the strong spin-orbit coupling limit, Knight shift or critical field measurements cannot distinguish between spin-singlet and spin-triplet superconductivity}.

The above paragraph may suggest that there is no way to experimentally detect the difference between pairing states that are largely spin-singlet and those that are largely spin-triplet. However, this is not the case. There is an important difference between the two cases related to the topology of the bands. For the Rashba spin-orbit interaction discussed above, the protected spin-triplet superconductor has been examined in the context of superfluid $^3$He and is known to have Majorana edges states and Majorana modes in a vortex core \cite{vol88}. The pure spin-singlet superconductor has no such topological  states. {\it While magnetic response cannot distinguish spin-singlet or spin-triplet,  this topological difference survives even when the spin-singlet and spin-triplet mix and can be used to classify a non-centrosymmetric superconductor as predominantly spin-singlet or spin-triplet} \cite{lu08}.

Finally we turn to the last key concept that is discussed in this review, {\it the appearance of novel magneto-electric effects in the superconducing state}. These effects stem from a coupling of the supercurrent and magnetic fields that become allowed by symmetry once inversion symmetry is broken. This is discussed in detail later, but a hint at the origin of these effects can be seen by looking at the quasi-particle bands that result when a Zeeman field is applied in-plane. In particular, notice that when an in-plane Zeeman field is applied, say $H_y\ne 0$, the quasi-particle dispersions $\epsilon_{\pm}$  develop a linear term in $k_x$. This implies that the center of the band is shifted along the $k_x$-axis away from the $\Gamma$ point. If one considers creating Cooper pairs in this situation, it becomes clear that it is energetically favorable to form Cooper pairs through the new center of the band as opposed to pairing through the $\Gamma$ point. That is, for example, pairing the state $|\vsk,+\rangle$ with $|-\vsk+\vq,+\rangle$ with $\vq \ne 0$. This results in a superconducting condensate that has a spatial $e^{2i{\bf q}\cdot {\bf x}}$ dependence, much like a  Fulde-Ferrell-Larkin-Ovchinnikov  state. Since such a pairing state is often associated with a current carrying superconducting state (though in this case it does not carry a current), it is possible to see that non-centrosymmetric superconductors allow for a nontrivial coupling between  magnetic fields and supercurrents. This interplay has a series of new consequences on superconductivity.

\section{Experimental results}

\begin{table}[ptbh]
\caption{Properties of non-centrosymmetric superconductors. The nature of the superconducting gap from different measurements is indicated by F (fully gapped), P (point nodes) or L (line nodes), where 1 or 2 indicates evidence for single or multi-gap behaviour. The double lines separate materials with strong and weak correlations. Results of Knight shift (KS) measurements are displayed which are either constant (C) or reduced (R) below $T_c$. The presence or absence of time reversal symmetry breaking (TRSB) is denoted by Y and N respectively. Where available, the splitting of the band structure by the ASOC near $E_F$ is given by $E_{ASOC}$. References that are not given in the main text are displayed.}
\label{NCSTable}
\begin{center}
 \begin{tabular}{c  c  c  c  c  c  c  c  c  c  c  c}
\hline \\ [-2.0ex]
Compound&Structure&$T_{\rm c}$  & $\gamma$ &$H_{c2}$  &1/$T_1(T)$&KS&$C(T,H)$&TRSB&$\lambda(T)$& $E_{ASOC}$ & $E_{ASOC}/k_BT_c$ \\
&& (K) &(mJ/mol~K$^2$)&(T)   & &&&&&(meV)& \\
 \hline \\ [-1.5ex]

CePt$_3$Si &$P4mm$ & 0.75& 390&2.7$\parallel c$, 3.2$\parallel a$  & L&  C &L&&L&  200\cite{CePt3SiASOC}&3095\\
LaPt$_3$Si & &0.6&11& Type I \cite{LaPt3SiSpecH,LaPt3SiMuSR} &F & &F1&N&F1& 200&3868 \\
 \hline \\ [-1.5ex]
CeRhSi$_3$  & $I4mm$& 1.05&110 &$\sim30\parallel c$, $7\parallel a$ &  & &&&& 10&111 \\
CeIrSi$_3$ &&1.6& 100 &$\sim45\parallel c$, $9.5\parallel a$ & L &C,R  &&&&4&29\\
CeCoGe$_3$ & &0.64&  32&$>20\parallel c$, $3.1\parallel a$ & & &&&&9 \cite{CeCoGe3FS,CeCoGe3FSHF}&163 \\
CeIrGe$_3$&  &1.5&80& $>10\parallel c$ &&  & &&&\\
 \hline \\ [-1.5ex]
UIr & $P2_1$   &0.13 &49 &0.026&  & &&&&\\
\hline \hline \\ [-1.5ex]
Li$_2$Pd$_3$B& $P4_332$ & 7  & 9 &2&F  &R & F&&F2&30&50\\
Li$_2$Pt$_3$B & & 2.7 & 7 &5 &L   &C &F/L&&L2&200&860\\
Mo$_2$Al$_3$C & &9&17.8  &15& P & &&N&F1&\\
 \hline \\ [-1.5ex]
Y$_2$C$_3$ & $I\bar{4}3d$  &  18 & 6.3 &30& F2  & R &F&& L/F2&15&10 \\
La$_2$C$_3$ & &13&10.6& 19&  & C &F1&&F2&30&33\\
 \hline \\ [-1.5ex]
K$_2$Cr$_3$As$_3$ &$P\bar{6}m2$ & 6.1  & 70-75 &23 $\parallel$, 37$\perp$  & & &&&L&60&114\\
Rb$_2$Cr$_3$As$_3$ & & 4.8 & 55 &20&P & &&&&\\
Cs$_2$Cr$_3$As$_3$& & 2.2 & 39 &6.5 & & &&&&\\
 \hline \\ [-1.5ex]
BiPd & $P2_1$ & 3.8  & 4 &0.8 &F1&  &F1&&F2&50&153\\
 \hline \\ [-1.5ex]
Re$_6$Zr & $I\bar{4}3m$  & 6.75 &26 &12.2&  &  &&Y&F1&\\
Re$_3$W &  & 7.8 & 15.9&12.5 &  & &F1&N&F1 &  \\
Nb$_x$Re$_{1-x}$ &  &3.5-8.8  &3-4.8 &6-15 &  F &  R&F1/2&&F1& \\
Re$_{24}$Ti$_5$ & & 5.8&111.8&10.75  & & &F1&&&\\
 \hline \\ [-1.5ex]
Mg$_{10+x}$Ir$_{19}$B$_{16-y}$ & $I\bar{4}3m$  &2.5-5.7 & 52.6&0.8& F1 &R &F1&&F1/2& \\
 \hline \\ [-1.5ex]
 Ba(Pt,Pd)Si$_3$ & $I4mm$ & 2.3-2.8&4.9-5.7&0.05-0.10&  & & F1&&&\\
 La(Rh,Pt Pd,Ir)Si$_3$ & &0.7-2.7& 4.4-6&Type I/0.053&  &  & F1&N&F1&17(Rh)&93(Rh)\\
Ca(Pt,Ir)Si$_3$ & & 2.3-3.6&4.0-5.8&0.15-0.27&  &   & F1&N&&\\
 Sr(Ni,Pd,Pt)Si$_3$ & & 1.0-3.0&3.9-5.3&0.039-0.174&  & & F1&&& \\
  Sr(Pd,Pt)Ge$_3$  & & 1.0-1.5&4.0-5.0&0.03-0.05&  &  & F1&&&\\
 \hline \\ [-1.5ex]
Rh$_2$Ga$_9$ &$Pc$ & 1.95 &7.64 &Type I&  & & F1&&&5&30\\
Ir$_2$Ga$_9$ & & 2.25 &7.32 &Type I & &  & F1&&&25&129\\
 \hline \\ [-1.5ex]
Ru$_7$B$_3$ &$P6_3mc$ &3.3  &90 &1.1& F1  & &&&&20 &70\\
Re$_7$B$_3$ & &3.3  & &0.9& F1 &  &&&&\\
La$_7$Ir$_3$ & &2.25  & && &  &&Y&F1&\\
 \hline \\ [-1.5ex]
 Y$_3$Pt$_4$Ge$_{13}$ & $Cc$ &4.5 &19&3.8 & F1  & &&&&\\
 \hline \\ [-1.5ex]
LaIr(As,P) \cite{LaMPSC} &$I4_1md$ & 3.1, 5.3 & 8.8,9.1&0.64,1.64&  & &&&&\\
LaRhP \cite{LaMPSC} & & 2.5 & 7.0&0.27& & &&&&\\
 \hline \\ [-1.5ex]
PdBiSe \cite{PdBiSeRep} &$P2_13$ &1.8 &2.1 &  && &&&& 109 \cite{dHvaPdBiSe}&703\\
\hline \\ [-1.5ex]
LaPtSi & $I4_1md$  &3.35 &6.5 &0.4&  & &F1&&& \\
 \hline \\ [-1.5ex]
Li$_2$IrSi$_3$ \cite{Li2IrSi3Rep} & $P31c$ &3.8 & 5.3& 0.3 && &&&& \\
 \hline \\ [-1.5ex]
Cr$_2$Re$_3$B & $P4_132$ &4.8 & 11.2&10  && &F1&&& \\
(W,Mo)$_7$Re$_{13}$(B,C)&  &7.1-8.1 & 57.5-66.9&11.4-15.4  && &F1&&& \\
 \hline \\ [-1.5ex]
Ca$_3$Ir$_4$Ge$_4$ & $I\bar{4}3m$  & 1.8 & 25 & & &  &&&& \\
 \hline \\ [-1.5ex]
LaNiC$_2$ & $Amm2$ & 2.7  & 7.7 &0.5 &  &   &&Y&P/F2& 42&181 \\
ThCoC$_2$ \cite{ThCoC2Rep}&  &2.65    & 8.4&0.4&  &   &&&& \\
 \hline \\ [-1.5ex]
YPtBi &$F\bar{4}3m$ & 0.77  &&1.5  & & &&&&\\
LuPtBi & & 1.0  &  & 1.6&& &&&&\\
LaPtBi & & 0.9  &  & 1.5& F1& &&&&\\
LuPdBi & & 1.7  &11.9& 2.2/2.9 & &  &F1&&& \\
(Sm,Tb-Tm,Y)PdBi & & 0.6-1.6  & 0.5 (Y)  &0.5-2.7&& &&&&\\
 \hline \\ [-1.5ex]
PbTaSe$_2$ & $P\bar{6}m2$  &3.7 &6.9 &1.47&  & &F&&F1& \\

\hline \hline

\end{tabular}

\end{center}
\end{table}

In this section we  review experimental results from studies of superconductivity in systems with broken inversion symmetry. Here we put particular emphasis on evaluating what the effect of the ASOC is on the superconductivity and to what extent is there evidence for mixed-parity pairing states. A list of non-centrosymmetric superconductors (NCS) is displayed in Table~\ref{NCSTable}, which contains a summary of their properties and double lines separate strongly and weakly correlated materials. We begin with a discussion of non-centrosymmetric heavy fermion (strongly correlated) superconductors. The properties of non-centrosymmetric superconductors with weak electronic correlations are then reviewed. We include systems found to display evidence for parity mixing, those which show evidence for single-gap $s$-wave superconductivity and those where the presence of mixing remains a more open question. We also discuss those NCS found to break time reversal symmetry in the superconducting state and in particular, whether this is evidence in favor or against singlet-triplet mixing. We conclude our discussion of bulk materials by describing the properties of NCS with topologically non-trivial band structures, which have potential for realizing novel properties such as Majorana fermions. Finally we briefly review the properties of several low-dimensional superconducting systems, which provide different paths for breaking inversion symmetry. In many of these systems, the strength of the ASOC can be tuned more controllably, allowing for the detailed study of the effects of the ASOC on the superconductivity.

\subsection{Non-centrosymmetric heavy fermion superconductors}
\subsubsection{CePt$_3$Si}
\label{CePt3SiSec}

\begin{figure}[t]
\begin{center}
  \includegraphics[width=0.95\columnwidth]{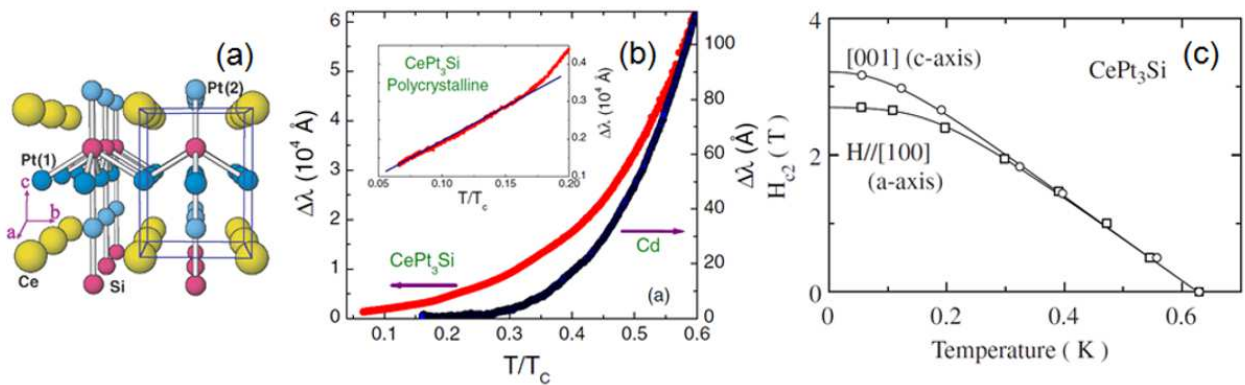}	
\end{center}
	\caption{(a) Crystal structure of CePt$_3$Si. Reprinted figure with permission from  Ref.~\onlinecite{CePt3Si2004}. Copyright 2004 by the American Physical Society. (b) Penetration depth of CePt$_3$Si  showing linear  behavior at low temperatures. Reprinted figure with permission from  Ref.~\onlinecite{CePt3Sinodepenpoly}. Copyright 2005 by the American Physical Society.  (c) Upper critical field of CePt$_3$Si  along the [001] and [100] directions.  Reprinted figure with permission from  Ref.~\onlinecite{CePt3SiPress1}. Copyright 2004 by the Physical Society of Japan.}
  \label{CePt3SiStruc}	
\end{figure}

CePt$_3$Si was the first reported noncentrosymmetric heavy fermion superconductor and is the only one confirmed to display superconductivity at ambient pressure \cite{CePt3SiCzech,CePt3Si2004,BauerNCS}. The noncentrosymmetric tetragonal crystal structure is shown in Fig.~\ref{CePt3SiStruc}(a) (space group $P4mm$), where a lack of a mirror plane perpendicular to the $c$~axis removes inversion symmetry, leading to a Rashba type ASOC. At  $T_{\rm N}$~=~2.2~K, the system orders antiferromagnetically, where the magnetic moments lie in the $ab$~plane and align ferromagnetically within the plane but antiferromagnetically between adjacent layers \cite{CePt3SiMagStruc}. Evidence for strong hybridization between $4f$ and conduction electrons due to the Kondo interaction arises from both the large electronic specific heat coefficient  \cite{CePt3Si2004} of  $\gamma~=~390$~mJ/mol~K$^2$ and a reduced ordered moment of 0.16$~\mu_{\rm B}$/Ce, compared to 0.65$~\mu_{\rm B}$/Ce expected for the deduced crystalline-electric field (CEF) scheme \cite{CePt3SiMagStruc,CePt3Sicrys1}.

Superconductivity is observed at ambient pressure with a bulk $T_c$ of around 0.7~K obtained from specific heat measurements of polycrystalline samples \cite{CePt3Si2004}, while transitions at lower temperatures are often reported in single crystals, at around 0.45~-~0.5~K \cite{CePt3Sicrys1,CePt3SinodesC,CePt3Sicrys2,CePt3SiPinning}. The reason for such a large discrepancy is not clear and many single crystals also show evidence for the inclusion of the high $T_c$ phase \cite{CePt3SiCrys2Tc,CePt3SiCrysInc}, which may lead to higher onset values of $T_c~\sim~0.7$~K in resistivity measurements \cite{CePt3Sicrys1,CePt3Sicrys2}.The presence of two superconducting transitions is not likely to be intrinsic and the lower transition has also been attributed to a magnetic transition arising from an impurity phase  \cite{CePt3SiSpur}. The upper critical field [$H_{c2}(T)$] displays only a small anisotropy as displayed in Fig.~\ref{CePt3SiStruc}(c) \cite{CePt3SiPress1}, reaching zero temperature values of $H_{c2}(0)\sim2.7~-~2.8$~T for $H\parallel[100]$ and $H_{c2}(0)\sim3.2~-~3.4$~T for $H\parallel[001]$, greatly exceeding the BCS Pauli paramagnetic limiting field of $0.8~-~1.4$~T in all directions. The relationship between antiferromagnetism and superconductivity in this compound is of particular interest. Muon spin relaxation ($\mu$SR) measurements reveal that all muons are implanted in magnetically ordered regions of the sample and therefore, there is microscopic coexistence between the magnetic and superconducting phases \cite{CePt3SiMuSR}, instead of the competition between the phases observed in some other heavy fermion superconductors such as CeCu$_2$Si$_2$ \cite{CeCu2Si2muSR,CeCu2Si2muSR2}. Both $T_N$ and $T_c$ are suppressed upon applying pressure \cite{CePt3SiPress1,CePt3SiPress2}. However, while the antiferromagnetic transition
 disappears around $0.6~-~0.8$~GPa, $T_c$ is suppressed to zero at a higher pressure around 1.5~GPa and there is no apparent enhancement of superconductivity upon the disappearance of magnetic order.

There are several pieces of evidence that the superconductivity of CePt$_3$Si is unconventional. Firstly the superconducting state is very sensitive to non-magnetic impurities. Superconductivity can be suppressed by the substitution   of just $4\%$ Th for Ce or $10\%~-~~15\%$ Ge for Si \cite{CePt3SiThdope,CePt3SiCzech,CePt3SiGedope}  and a pressure study of CePt$_3$Si$_{0.94}$Ge$_{0.06}$ \cite{CePt3SiGedope2} demonstrates that the suppression of $T_c$ cannot be accounted for by the negative chemical pressure of Ge substitution, but is due to the pair breaking effect of non-magnetic impurities on unconventional superconductivity. Secondly, penetration depth, thermal conductivity and  specific heat measurements all indicate the presence of line nodes in the superconducting gap \cite{CePt3Sinodepenpoly,CePt3SinodesCond,CePt3SinodesC}. The electronic contribution to the specific heat divided by temperature $C_{el}/T$ shows linear behaviour over a large temperature range, up to around 0.3~K, which is evidence for the presence of line nodes \cite{CePt3SinodesC}, instead of the exponentially activated behaviour of fully gapped superconductors. The change in the London penetration depth [$\Delta\lambda(T)$] measured using a tunnel diode oscillator based technique, shows linear behaviour of $\Delta\lambda(T)$ below $\sim0.16T_c$, for both polycrystalline [Fig.~\ref{CePt3SiStruc}(b)]\cite{CePt3Sinodepenpoly} and single crystal \cite{CePt3SiCryspen} samples, again indicating that there are line nodes in the superconducting gap. This behaviour is different from the isostructural non-magnetic LaPt$_3$Si, where the penetration depth is consistent with fully gapped superconductivity \cite{CePt3SiNodePen}. The thermal conductivity ($\kappa/T$) of CePt$_3$Si single crystals \cite{CePt3SinodesCond} shows a linear temperature dependence and nodal superconductivity is indicated by the large residual value of $\kappa/T$, which arises due to scattering from small amounts of non-magnetic impurities. The field dependence of $\kappa$ increases much more rapidly at low fields than expected for fully gapped superconductors and the behaviour can be explained by a gap structure with line nodes.

Since multiple experimental techniques provide evidence for unconventional superconductivity and line nodes in the superconducting gap, there was particular interest in determining the nature of pairing state. The large values of $H_{c2}(0)$, greatly exceeding the  Pauli paramagnetic limiting field in all directions suggested the possibility of triplet superconductivity, despite the requirement of Anderson's theorem that inversion symmetry is necessary for triplet superconductivity \cite{AndTrip}. While triplet states with a $\mathbf{d}$-vector with $\mathbf{d(k)}\parallel\hvg({\bf k})$ are protected in systems with broken inversion symmetry, the corresponding state for CePt$_3$Si has $\mathbf{d(k)}~=~ \hat{\mathbf{x}}k_y - \hat{\mathbf{y}}k_x$, which has point nodes instead of line nodes \cite{CePt3SiMnSi}. It was therefore suggested that line nodes may arise as a result of a mixed singlet-triplet pairing state,  where accidental line nodes arise on one of the split Fermi surface sheets due to the antisymmetric spin-orbit coupling \cite{NCSTheor1}. This possibility is discussed in more detail in Section~ \ref{general-apsects}.

\begin{figure}[t]
\begin{center}
\vspace{0.5cm}
  \includegraphics[width=0.45\columnwidth]{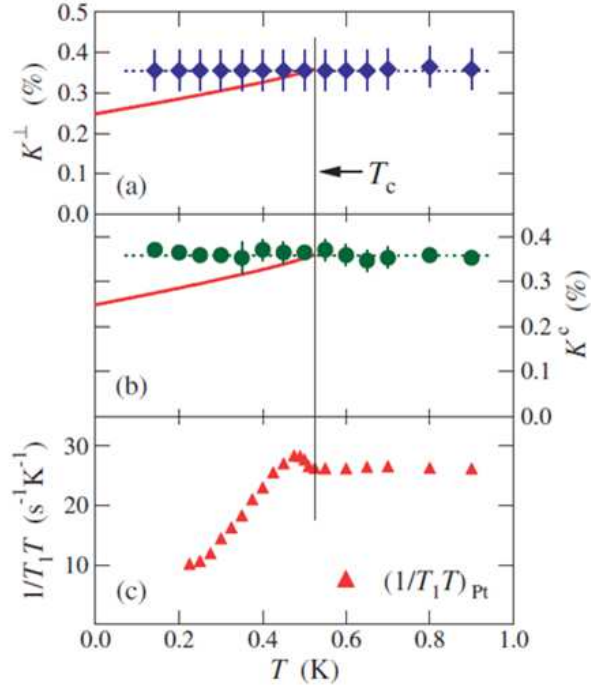}	
\end{center}
\caption{Temperature dependence of the Knight shift of CePt$_3$Si single crystals (a) perpendicular and (b) parallel to the $c$~axis. (c) Temperature dependence of 1/$T_1T$ of CePt$_3$Si showing a coherence peak just below $T_c$.  Reprinted figures with permission from  Ref.~\onlinecite{CePt3SiNMR}. Copyright 2006 by the Physical Society of Japan.}
  \label{CePt3SiNMRb}	
\end{figure}

Nuclear magnetic resonance (NMR) is a powerful tool for probing the nature of the superconducting pairing state and there have been several NMR studies of CePt$_3$Si (Fig.~\ref{CePt3SiNMRb}). The main results can be summarized as follows: (1) there is a peak in $1/T_1T$ below $T_c$ [Fig.~\ref{CePt3SiNMRb}(c)] \cite{CePt3SiNMR2,CePt3SiNMR,CePt3SiNMR3}; (2) the Knight shift remains constant upon crossing $T_c$ down to low temperatures for fields applied along all crystallographic directions, as displayed in Figs.~\ref{CePt3SiNMRb}(a) and (b); \cite{CePt3SiNMR} (3)  $1/T_1$ does not show the exponential behaviour of $s$-wave BCS superconductors \cite{CePt3SiNMR2}, with some measurements showing the $T^3$ dependence expected for line nodes in the superconducting gap.\cite{CePt3SiNMR3} Several aspects of these results remain quite puzzling.  A peak in $1/T_1T$ is usually associated with a Hebel Slichter coherence peak observed in isotropic BCS superconductors, and its observation in CePt$_3$Si was associated with a two component order parameter arising from singlet-triplet mixing \cite{NMRTheory}. However in Ref.~\onlinecite{CePt3SiNMR3}, the peak in $1/T_1T$ was also attributed to disordered regions with fully gapped behaviour, while the high quality single crystal with a low $T_c$ did not show any peak in $1/T_1T$ and a $T^3$ dependence was observed at low temperatures.

The constant Knight shift in all directions is difficult to explain from considering the effects of spin-orbit coupling, since it would be expected to remain constant only for fields applied along the $c$~axis, perpendicular to $\mbox{\boldmath$\gamma$}({\bf k})$. As a result, Pauli paramagnetic limiting would also be expected to be absent only for fields along this direction, which contradicts the nearly isotropic behaviour of $H_{c2}(T)$. Furthermore as discussed below, in the case of NCS with strong spin-orbit coupling, Knight shift measurements and the absence of Pauli limiting can not distinguish between singlet and triplet superconductivity and therefore these measurements can not determine the presence of singlet-triplet mixing. It was also pointed out by S.~Fujimoto that the effect of strong electronic correlations may lead to an enhancement of the spin susceptibility and therefore the Knight shift in the superconducting state \cite{SuscCorr}. In addition, he suggested that line nodes may arise when a fully gapped superconducting state coexists with antiferromagnetic order and since the gap does not change sign at these accidental nodes, this may explain the presence of a coherence peak in NMR results \cite{AFMNodes}.  As a result of the presence of these additional complications, as well as the experimental difficulties of producing high quality single crystals showing a single superconducting transition, unambiguously determining the nature of the pairing state of CePt$_3$Si remains a challenge and requires further experimental and theoretical investigations.

\subsubsection{Ce$TX_3$ ($T$~=~transition metal, $X$~=~Si or Ge)}
\label{CeTX3Sec}

\begin{figure}[t]
\begin{center}
\vspace{0.5cm}
  \includegraphics[width=0.9\columnwidth]{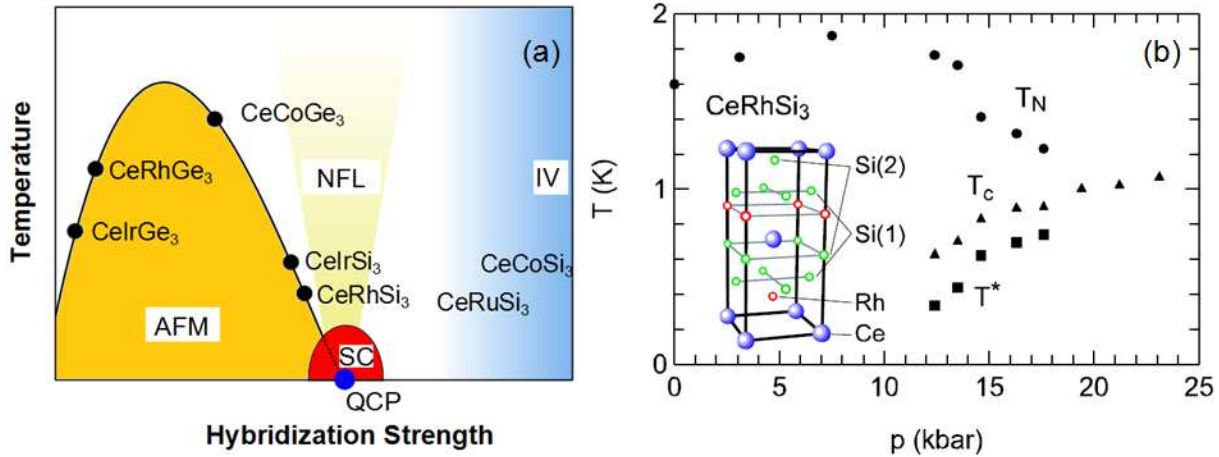}	
\end{center}
	\caption{(a) Schematic phase diagram of Ce$TX_3$ compounds, as a function of the hybridization strength between the $4f$ and conduction electrons. The labels AFM and SC denote the antiferromagnetic and superconducting phases respectively, while NFL and IV indicate the regions where non-Fermi liquid and intermediate valence behaviors occur. The position of the putative quantum critical point (QCP) is also labelled. (b) Temperature - pressure phase diagram of CeRhSi$_3$. The N{\'e}el temperature ($T_{\rm N}$), superconducting transition temperature ($T_{\rm c}$) are displayed as well as $T^*$, the position at which a kink in the resistivity is observed below $T_{\rm c}$. The crystal structure of CeRhSi$_3$ is shown in the inset.  Reprinted figure with permission from  Ref.~\onlinecite{CeRhSi3SC}. Copyright 2005 by the American Physical Society.}
  \label{CeRhSi3TP}	
\end{figure}

In addition to the superconductivity of CePt$_3$Si at ambient pressure, pressure-induced superconductivity was discovered in several compounds in the Ce$TX_3$ ($T~=~$transition metal, $X~=~$Si, Ge) series, which crystallize in the body centered  tetragonal crystal structure (space group $I4mm$) shown in the inset of Fig.~\ref{CeRhSi3TP}(b). Much like the crystal structure of CePt$_3$Si, there is no mirror plane perpendicular to the $c$~axis, which leads to broken inversion symmetry. Four of these compounds have been reported to display superconductivity,  CeRhSi$_3$ for $p~>$~1.2~GPa \cite{CeRhSi3SC}, CeIrSi$_3$ for $p~>$~1.8~GPa \cite{CeIrSi3SC}, CeCoGe$_3$  for $p~>$~4.3~GPa \cite{CeCoGe3SC}  and CeIrGe$_3$ for $p~>$~20~GPa \cite{CeIrGe3SC}. All of these compounds order antiferromagnetically at ambient pressure but the application of pressure suppresses $T_N$ leading to the observation of a superconducting dome.  The properties of the  Ce$TX_3$ have often been explained in the context of the Doniach phase diagram \cite{Doniach1977,CeTX32008}, where there is competition between the intersite Ruderman-Kittel-Kasuya-Yosida (RKKY) interaction which leads to magnetic ordering and the onsite Kondo interaction which favors the formation of a non-magnetic singlet. These results are summarized in the schematic phase diagram displayed in Fig.~\ref{CeRhSi3TP}(a), where the positions of various Ce$TX_3$ compounds are labelled based on the apparent strength of the hybridization between $4f$ and conduction electrons, which is enhanced by pressure. For the magnetically ordered compounds, there is evidence that those with stronger hybridization lie in closer proximity to superconductivity. For example CeRhGe$_3$ which orders at $T_N~=~14.6$~K \cite{CeRhSi3Rep}, shows no evidence of a reduced ordered magnetic moment due to Kondo screening \cite{CeRhGe32012} and no superconductivity is observed up to at least 8~GPa  \cite{CeTX32008}. Meanwhile, CeCoGe$_3$ which becomes superconducting for $p~>$~4.3~GPa \cite{CeCoGe3SC}, shows a reduced ordered moment compared to that predicted from the CEF scheme \cite{CeCoGe3SCND,CeCoGe3MS}, while CeRhSi$_3$ which becomes superconducting at a lower pressure of 1.2~GPa  \cite{CeRhSi3SC}, displays an even more significant reduction  \cite{CeRhSi3ND,CeRhSi3CEF}. Other Ce$TX_3$ compounds such as CeCoSi$_3$ and CeRuSi$_3$ are non-magnetic intermediate valence compounds with much stronger hybridization between $4f$ and conduction electrons \cite{CeTX32008,CeRuSi3IV}.

As an example, Fig.~\ref{CeRhSi3TP}(b) displays the temperature-pressure phase diagram of CeRhSi$_3$. The $T_N$ of CeRhSi$_3$ initially increases, before being suppressed with pressure \cite{CeRhSi3SC}, while for CeIrSi$_3$, $T_N$ continuously decreases \cite{CeIrSi3SC}. At certain pressures in  the two compounds, both antiferromagnetism and superconductivity are observed and it is of interest to determine whether $T_N$ disappears at the point where $T_N~=~T_c$ or whether it is suppressed to zero at a quantum critical point (QCP). Indeed  evidence for a QCP at $p~=~2.36$~GPa was reported from $\mu$SR measurements of  CeRhSi$_3$  \cite{CeRhSi3QCP}. On the other hand, although the resistivity in the paramagnetic state shows non-Fermi liquid behaviour with a linear temperature dependence, there is evidence for a lack of the enhancement of magnetic fluctuations expected near a QCP \cite{CeRhSi3NoQCP}. AC heat capacity measurements at low pressure for both compounds show a large anomaly for $T_N$ with a small anomaly at $T_c$ at lower temperatures \cite{CeRhSi3HCPress,CeIrSi3HC}. However in CeIrSi$_3$, once $T_c$ approaches $T_N$, the magnetic transition is no longer observed and the anomaly at $T_c$ is greatly enhanced, with a very large jump of  the ac heat capacity ($\Delta C/C$) of 5.7 at 2.58~GPa, indicating very strongly coupled superconductivity. These results suggest that at lower pressures, only a small fraction of the electrons form Cooper pairs, but once $T_c~=~T_N$, there is much greater number of electrons contributing to the superconducting state and a magnetic transition is no longer observed.

\begin{figure}[t]
\begin{center}
\vspace{0.5cm}
 \includegraphics[width=0.99\columnwidth]{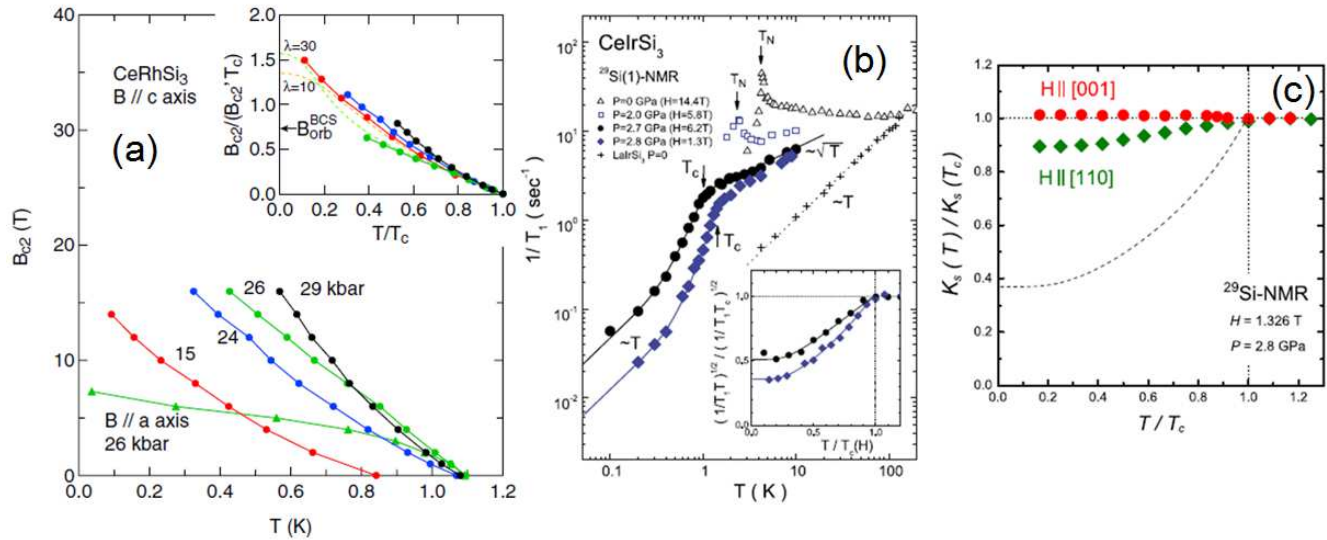}	
\end{center}
	\caption{(a) Upper critical field of CeRhSi$_3$ at various pressures along the [001] and [100] directions. The inset shows $B_{c2}$ along [001], normalized by the slope of the curve near $T_c$, as well as the calculated results of a strong coupling model.  Reprinted figure with permission from  Ref.~\onlinecite{CeRhSi3Hc2}. Copyright 2007 by the American Physical Society. (b) Temperature dependence of $1/T_1$ of polycrystalline CeIrSi$_3$ which shows good agreement with the calculated  behavior for a gap model with line nodes.  Reprinted figure with permission from  Ref.~\onlinecite{CeIrSi3NMR1}. Copyright 2008 by the American Physical Society. (c) Temperature dependence of the Knight shift of single crystal CeIrSi$_3$, which shows anisotropic  behavior, being constant for applied fields along the $c$ axis but decreasing below $T_c$ for fields applied perpendicular. From Mukuda~\textit{et al.}\cite{CeIrSi3NMR2}.  Reprinted figure with permission from  Ref.~\onlinecite{CeIrSi3NMR2}. Copyright 2010 by the American Physical Society.}
\label{CeIrSi3NMRfig}	
\end{figure}

A remarkable feature of the Ce$TX_3$ superconductors are the extremely large and anisotropic values of $H_{c2}(T)$ \cite{CeRhSi3Hc2,CeIrSi3Hc2,CeCoGe3Hc2b}. As shown in  Fig.~\ref{CeIrSi3NMRfig}(a), $H_{c2}(0)$ of CeRhSi$_3$  reaches about 7~T for fields applied in the $ab$~plane but is much larger for fields applied along the $c$~axis, where  $H_{c2}(T)$ shows an upward curvature reaching  up to $\sim30$~T at $T~=~0$ for $p$~=~2.9~GPa \cite{CeRhSi3Hc2}. As shown in the inset, the behaviour along the $c$~axis clearly deviates from that of the  Werthamer Helfand Hohenberg (WHH) model generally used to describe conventional BCS superconductors.  These values demonstrate that like in the case of CePt$_3$Si, the BCS Pauli limiting field is exceeded in all directions, but unlike CePt$_3$Si which only shows a small anisotropy in $H_{c2}(T)$, the values for fields applied along the $c$~axis in Ce$TX_3$ are much larger. This is compatible with calculations of the spin susceptibility for the permitted triplet state for this structure, \cite{FrigeriSusc}  which show the absence of Pauli limiting in one direction and an enhanced Pauli limiting field along others. In this scenario, $H_{c2}(T)$ would be limited by paramagnetic pair breaking for $H\parallel ab$ but would be determined solely by the orbital limiting field for $H\parallel c$, which may be enhanced in proximity to a QCP \cite{CeTX3Hc2,CeTX3Hc2b}. In this scenario, a magnetization perpendicular to the applied field $H\parallel ab$ is predicted in the vortex state, this is discussed in more detail in Section~\ref{vortex}.

To further characterize the pairing state, NMR measurements were also performed on CeIrSi$_3$ [Figs.~\ref{CeIrSi3NMRfig}(b) and (c)]. NMR measurements on polycrystalline samples reveal a $T^3$ dependence of $1/T_1$ with no coherence peak just below $T_c$, indicating an unconventional pairing state with line nodes in the superconducting gap [Fig.~\ref{CeIrSi3NMRfig}(b)]\cite{CeIrSi3NMR1}, similar to several other heavy fermion superconductors. These measurements were performed at pressures of $2.7~-~2.8$~GPa, where antiferromagnetism is completely suppressed and in the normal state  $1/T_1$ shows a $\sim T/\sqrt{T+\theta}$ dependence with $\theta\sim0$ at 2.7~GPa, indicating a close proximity to a QCP. NMR measurements were also performed on single crystals at a pressure of 2.8~GPa \cite{CeIrSi3NMR2} and as shown in Fig.~\ref{CeIrSi3NMRfig}(c), while there is no change in the Knight shift below $T_c$ for fields applied along [001], a decrease is observed for fields along [110], which is markedly different from the isotropic constant value observed in CePt$_3$Si. Such a decrease in the Knight shift for $H\perp c$ is more in line with the behaviour expected in the strong spin-orbit coupling limit. However, the decrease in the Knight shift is still relatively small, reaching just $\sim90\%$ of the normal state value, compared to expected drop of $\sim50\%$, which may again suggest an enhancement due to strong electronic correlations.

\subsubsection{UIr}

Pressure induced superconductivity has also been reported in the noncentrosymmetric U-based heavy fermion compound UIr \cite{UIrSC,UIr2}, which has a monoclinic structure with space group $P2_1$. At ambient pressure the system is an itinerant ferromagnet with a relatively high ordering temperature of 46~K. \cite{UIrSC,UIr2}. With increasing pressure this ordering temperature decreases and two additional pressure induced ferromagnetic phases are detected \cite{UIr3}. The magnetic ordering temperature is finally suppressed to zero around 2.6~GPa, where there is superconductivity in a narrow pressure range with $T_c~=~0.14$~K. This material may be a good candidate for a triplet superconductivity, since superconductivity appears on the border of ferromagnetism in common with many other U-based superconductors. However, the very low value of $T_c$ makes characterizing the superconducting state extremely challenging.

\subsection{Nodal superconductivity in weakly correlated systems}
\subsubsection{Li$_2$(Pd$_{1-x}$Pt$_x$)$_3$B}

Given the novel behaviour of Ce based NCS along with both the experimental difficulties in measuring the superconducting properties and the theoretical challenges in separating the effects of spin orbit coupling and strongly correlated phenomena, the superconducting properties of many weakly correlated NCS have been studied. Perhaps the best example of a weakly correlated system where unconventional superconductivity arises from tuning the spin-orbit coupling is  Li$_2$(Pd$_{1-x}$Pt$_x$)$_3$B. This system crystallizes in a cubic non-centrosymmetric antiperovskite structure with space group $P4_332$, which consists of distorted (Pd,Pt)$_6$ octahedra, with a B atom at the center \cite{Li2PtPd3BStruc}. Both Li$_2$Pd$_3$B and Li$_2$Pt$_3$B are superconductors with $T_c$ of around 7.6~K and 2.7~K respectively \cite{Li2Pd3BRep,Li2Pt3BRep}. Superconductivity exists across the entire compositional range and $T_c$  decreases with increasing $x$. The weakly correlated nature of these systems can be deduced from the low respective values  of $\gamma$ of 9 and 7~mJ/mol~K$^2$  \cite{Li2Pt3BHCGap}.

\begin{figure}[t]
\begin{center}
\vspace{0.5cm}
  \includegraphics[width=0.7\columnwidth]{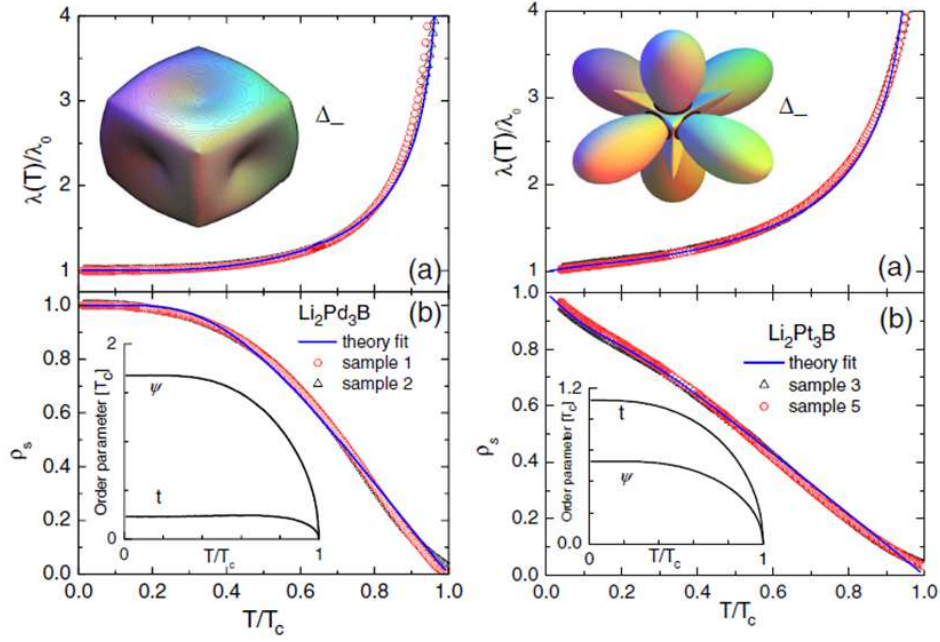}	
\end{center}
\caption{Penetration depth and superfluid density of Li$_2$Pd$_3$B and Li$_2$Pt$_3$B. The superfluid density has been fitted by a model of mixed singlet-triplet pairing and the relative strengths of the singlet and triplet components of the order parameter are shown in the inset. Also shown is the structure of one of the gap functions $\Delta_-$, which is anisotropic but fully gapped for Li$_2$Pd$_3$B but has line nodes for Li$_2$Pt$_3$B. Reprinted figures with permission from  Ref.~\onlinecite{Li2Pt3BNode}. Copyright 2006 by the American Physical Society.}
  \label{LPBTDO}	
\end{figure}

The properties of Li$_2$(Pd$_{1-x}$Pt$_x$)$_3$B evolve from fully gapped, spin singlet superconductivity in Li$_2$Pd$_3$B to nodal superconductivity in  Li$_2$Pt$_3$B with a significant spin triplet component, which corresponds to an increasingly strong ASOC as the heavier $5d$ Pt atoms are substituted for $4d$ Pd atoms. The significant number of (Pd,Pt) atoms in the crystal structure makes this a good system for studying the effect of tuning the ASOC. The evidence for this change in behaviour mainly arises from thermodynamic measurements which probe the gap structure and NMR measurements. The first evidence for nodal superconductivity and the tuning of singlet-triplet mixing by the ASOC came from penetration depth measurements of Li$_2$Pd$_3$B and Li$_2$Pt$_3$B by Yuan~\textit{et al.}\cite{Li2Pt3BNode}. The tell-tale feature of line nodes in the superconducting gap of  Li$_2$Pt$_3$B, as shown in Fig.~\ref{LPBTDO} is the linear temperature dependence of $\lambda(T)$ at low temperatures, while the exponential behaviour observed in  Li$_2$Pd$_3$B  indicates a fully open superconducting gap. The temperature dependence of the normalized superfluid density $\rho_s$ can be obtained from $\lambda(T)$ and this was fitted using a model with two gaps, which are an admixture of singlet ($\psi$) and triplet ($t$) components, $\Delta_\pm~=~\psi\pm t|\mbox{\boldmath$\gamma$}({\bf k})|$. The physical basis for this model is discussed in Section~\ref{general-apsects} (for this model it is assumed that the ASOC is sufficiently large that the spin-triplet ${\bf d}(\vk)$ vector is parallel to the ASOC vector $\vg(\vk)$). If $t$ is sufficiently large, line nodes may arise ``accidentally" on  $\Delta_-$ at certain nodal values $\mathbf{k_n}$, where $\psi-t|\hvg({\bf k_n})|$=0 and $\mathbf{k_n}$ is not determined by symmetry but by the relative strengths of the triplet and singlet channels. Consequently, while the observation of line nodes is strong evidence for a singlet-triplet admixture, their absence does not rule out, a priori, a mixed-parity gap. As shown in Fig.~\ref{LPBTDO}, $\rho_s$ of both compounds can be fitted by such a mixed-parity model and in  Li$_2$Pd$_3$B, the singlet component dominates and  $\Delta_-$ is anisotropic and fully open, but in  Li$_2$Pt$_3$B the larger triplet component leads to a significantly different temperature dependence of  $\rho_s$ and line nodes in $\Delta_-$. It should be noted that early specific heat measurements \cite{Li2Pt3BHCGap} and $\mu$SR measurements \cite{Li2Pd3BmuSR,Li2Pt3BmuSR} have been interpreted in terms of single band fully gapped superconductivity for both compounds. However, subsequent electronic specific heat measurements of  Li$_2$Pt$_3$B showed a $T^2$ dependence at low temperatures \cite{Li2Pt3BHCnode}, which also supports the presence of line nodes.

\begin{figure}[t]
\begin{center}
\vspace{0.5cm}
  \includegraphics[width=0.7\columnwidth]{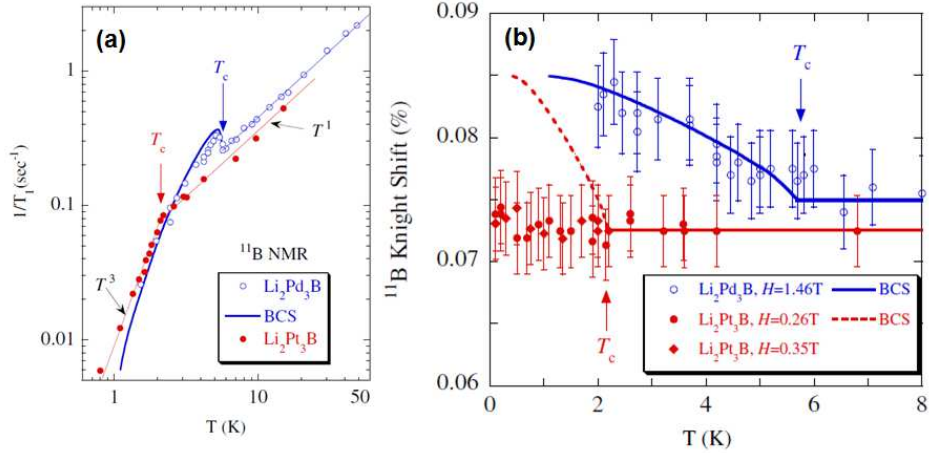}	
\end{center}
\caption{NMR measurements of Li$_2$Pd$_3$B and Li$_2$Pt$_3$B showing the temperature dependence of (a) $1/T_1$ and (b) the Knight shift. The $1/T_1$ measurements of Li$_2$Pd$_3$B  are fitted well to a BCS model and there is a Hebel-Slichter coherence peak just below $T_c$, while no coherence peak is observed in  Li$_2$Pt$_3$B and there is a $T^3$ dependence at low temperatures, indicating the presence of line nodes. Reprinted figures with permission from  Ref.~\onlinecite{Li2Pt3BNMR}. Copyright 2007 by the American Physical Society.}
  \label{LPBNMR1}	
\end{figure}

NMR measurements  \cite{Li2Pt3BNMR} also give evidence for a change in gap structure upon substituting Pt for Pd, as shown in Fig.~\ref{LPBNMR1}(a). The temperature dependence of $1/T_1$ for  Li$_2$Pd$_3$B  is consistent with fully gapped $s$-wave superconductivity, namely a Hebel-Slichter coherence peak is present below $T_c$ and the temperature dependence is well described by an isotropic $s$-wave model with a reduced gap size of $\Delta_0~=~1.1k_BT_c$. However, the temperature dependence of $1/T_1$ of   Li$_2$Pt$_3$B does not have a coherence peak and instead of an exponential temperature dependence, there is a $T^3$ dependence, which also indicates the presence of line nodes. Furthermore, Knight shift measurements are shown in Fig.~\ref{LPBNMR1}(b). An increase in the $^{11}$B Knight shift corresponds to a decrease in the spin susceptibility $\chi_s$ and this is exactly what is seen below $T_c$ in Li$_2$Pd$_3$B. However, the  Knight shift of  Li$_2$Pt$_3$B remains constant below $T_c$ and is unchanged from the normal state value. Similar results are observed in $^{195}$Pt NMR measurements \cite{Li2Pt3BNMR,Li2Pt3BNMR}, indicating that upon increasing the strength of the ASOC, the system changes from a decrease in $\chi_s$ below $T_c$, to a constant  $\chi_s$.

Since these strikingly different properties have often been interpreted as being due to different ASOC strengths, there has been great interest in studying the evolution of the singlet-triplet mixing by studying  Li$_2$(Pd$_{1-x}$Pt$_x$)$_3$B with intermediate values of $x$. Due to the large number of (Pd,Pt) atoms in the crystal structure, along with the significant differences in the atomic masses of Pd and Pt, it is expected that the strength of the ASOC can be effectively tuned by varying $x$ and indeed the ASOC splitting near the Fermi level has been calculated to reach 30~meV for $x~=~0$ but as high as 200~meV for $x~=~1$  \cite{Li2Pd3BASOC}. From penetration depth measurements \cite{Li2Pd3BDopePen}, linear behaviour of $\lambda(T)$ and therefore nodal superconductivity is reported for Pt rich samples, which is consistent with specific heat measurements \cite{Li2Pt3BHCnode,Li2Pd3BDopeHC} showing evidence for either line nodes or significant anistropy.  From NMR measurements \cite{Li2Pd3BDopeNMR}, a coherence peak and fully gapped  behaviour of $1/T_1$ was found for  $x\leq0.8$, along with a decrease of the Knight shift below $T_c$, whereas evidence for nodal behaviour with a constant Knight shift below $T_c$ was observed for $x$~=~0.9. The authors argue that this change is correlated with both distortions of the  (Pd,Pt)$_6$ octahedra and a reduction in the angle between adjacent corner sharing octahedra. These structural changes in turn lead to significant enhancements of the ASOC, beyond that expected for simply substituting Pt for Pd. It is suggested that the identification of such structural features provide a further guide for finding NCS which show strong singlet-triplet mixing, beyond simply the presence of heavy atoms. Therefore, the studies of intermediate compositions of the Li$_2$(Pd$_{1-x}$Pt$_x$)$_3$B system, support the scenario that the unconventional behaviour evolves as the ASOC is enhanced upon increasing Pt doping.

However, other explanations have also been proposed to explain the evolution of a nodal gap structure upon increasing the Pt content. Indeed electronic calculations show that the bands which cross the Fermi level of Li$_2$Pt$_3$B  have an enhanced $d$~character and significant nesting \cite{Li2Pt3BNMRElec1,Li2Pt3Bspm}. In Ref.~\onlinecite{Li2Pt3Bspm}, the pairing symmetry is proposed to be  a singlet $s_\pm$ state with accidental line nodes. While the difference in the behaviour of the Knight shift between the Pd and Pt rich samples gives evidence for the effects of tuning the ASOC on the superconductivity, as discussed below Knight shift measurements can not be used to distinguish between singlet and triplet superconductivity in noncentrosymmetric superconductors. As a result,  a wider range of measurements is desirable to directly measure the presence of a triplet component, to detect whether there is evidence for singlet-triplet mixing.

\subsubsection{Y$_2$C$_3$ and La$_2$C$_3$ }

\begin{figure}[t]
\begin{center}
\vspace{0.5cm}
  \includegraphics[width=0.7\columnwidth]{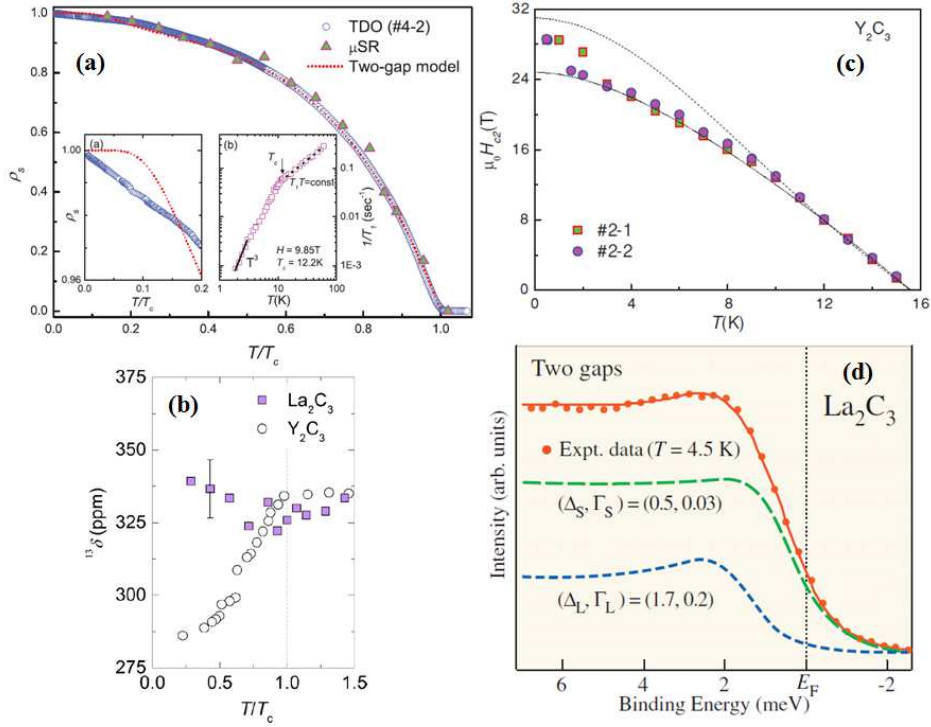}	
\end{center}
\caption{Normalized superfluid density $\rho_s$ derived from TDO measurements of Y$_2$C$_3$. The $\rho_s$ obtained from $\mu$SR measurements is also shown, digitized from Kuroiwa \textit{et~al.}  \cite{Y2C3TwoGap} and the dotted lines show a fit to a two-gap model. A close up of the low temperature region is displayed in the inset (a), which shows the clear deviation from fully gapped  behavior at low temperatures. The other inset (b) shows $1/T_1$ from NMR measurements by Harada \textit{et~al.} \cite{Y2C3NMR} and the reanalysis by  Chen \textit{et~al.} \cite{Y2C3node} hints at the onset of $T^3$  behavior. Reprinted figure with permission from  Ref.~\onlinecite{Y2C3node}. Copyright 2011 by the American Physical Society. (b) Temperature dependence of the $^{13}$C NMR shift of La$_2$C$_3$ and Y$_2$C$_3$, where the data for Y$_2$C$_3$ was adapted from Harada \textit{et~al.} \cite{Y2C3NMR}. Reprinted figure with permission from  Ref.~\onlinecite{La2C3NMR}. Copyright 2014 by the American Physical Society. (c) $H_{c2}(T)$ of Y$_2$C$_3$ with $T_c~=~$15~K. Reprinted figure with permission from  Ref.~\onlinecite{Y2C3node}. Copyright 2011 by the American Physical Society. (d) Photoemission spectrum of La$_2$C$_3$. The solid lines show fits to a two-gap model, whereas the two components are shown by dashed lines. Reprinted figure with permission from  Ref.~\onlinecite{La2C3Photo}. Copyright 2007 by the American Physical Society.}
  \label{R2C3Fig}	
\end{figure}

Evidence for unconventional superconducting properties have also  been found in the NCS Y$_2$C$_3$ \cite {Y2C3Rep} and La$_2$C$_3$ \cite{La2C3Tc}, both of which crystallize in a body-centered non-centrosymmetric structure with space group $I\bar{4}3d$. Both compounds display a wide range of $T_c$ values which depends strongly on the sample synthesis conditions, with a relatively large maximum $T_c$ of 18~K for  Y$_2$C$_3$ \cite{Y2C3HighTc1,Y2C3HighTc2} and 13.2~K for La$_2$C$_3$ \cite{La2C3Tc}. The main evidence for unconventional superconductivity in  Y$_2$C$_3$ arises from penetration depth measurements of polycrystalline samples with $T_c~=~15$~K \cite{Y2C3node}, which show a linear temperature dependence of $\Delta\lambda(T)$ at low temperatures. This is indicative of line nodes in the superconducting gap and given that this is a weakly correlated electron system, a natural conclusion would be that this arises from mixed singlet-triplet pairing. However, there are reports of specific heat, \cite{Y2C3SpecH} $\mu$SR \cite{Y2C3TwoGap} and NMR measurements \cite{Y2C3NMR} which are consistent with fully gapped superconductivity. Both the NMR and  $\mu$SR data are also consistent with multiband superconductivity that is dominated by a large gap with $\Delta_1(0)/k_BT_c\sim2.5$ but also has a smaller contribution of around $15-25\%$ from a smaller gap  $\Delta_2(0)/k_BT_c\sim1$. In fact, there may not be a particularly significant contradiction between these measurements and at higher temperatures, as shown in Fig.~\ref{R2C3Fig}(a) there is good agreement between the values of $\rho_s$ obtained from tunnel diode oscillator and $\mu$SR measurements. Furthermore the TDO data  at higher temperatures can also be well fitted by a similar two gap model. However the TDO measurements were performed down to 90~mK, whereas the $\mu$SR, NMR and specific heat measurements were all carried out above 1.8~K and it can be seen that in the inset of Fig.~\ref{R2C3Fig}(a) there is a clear deviation in TDO measurements of $\rho_s$ from the saturated behaviour expected for fully gapped superconductivity. Furthermore, a reanalysis of the temperature dependence of $1/T_1$ from Ref.~\onlinecite{Y2C3NMR}[inset (b) of Fig.~\ref{R2C3Fig}(a)], hints at the onset of $T^3$ behaviour at low temperatures \cite{Y2C3node}, which would also suggest the presence of line nodes. Although the authors report the presence of a tiny coherence peak just below $T_c$, its magnitude is significantly smaller than that expected for BCS superconductors. It should be noted that electron tunneling spectroscopy \cite{STSY2C3} and field dependent specific heat measurements \cite{Y2C3CvH} are also consistent with fully gapped superconductivity and therefore further measurements are required to determine whether there is a nodal gap structure.

NMR measurements of the Knight shift of  Y$_2$C$_3$ were also performed, which are shown in Fig.~\ref{R2C3Fig}(b). There is a decrease in the Knight shift below $T_c$, which was attributed to a decrease in $\chi_s$ and therefore spin singlet superconductivity. However, this behaviour does not necessarily exclude an enhancement of the Knight shift due to the ASOC, since these measurements were performed on polycrystalline samples and as discussed previously, the Knight shift is only expected to remain constant along crystallographic directions perpendicular to $\mbox{\boldmath$\gamma$}({\bf k})$. The upper critical field $H_{c2}(T)$ may also be affected by the presence of ASOC. In the aforementioned  Li$_2$(Pd$_{1-x}$Pt$_x$)$_3$B, $H_{c2}(0)$ is considerably lower than  $H_p$ \cite{Li2Pd3BRep,Li2Pt3BRep} and therefore no information about the presence of Pauli paramagnetic limiting can be deduced. However, as shown in Fig.~\ref{R2C3Fig}(c),  $H_{c2}(0)$  of Y$_2$C$_3$ is large, reaching $28-30$~T when $T_c~=~15$~K \cite{Y2C3node,Y2C3Hc2}. This suggests that $H_{c2}(0)$ may slightly exceed the weak coupling value of $H_p$~=~1.86$T_c~\sim$28~T. There are however other explanations for increases of  $H_p$ beyond the Clogston-Chandrasekhar limit of 1.86$T_c$, such as an energy gap larger than the BCS value \cite{ClogstonHp} and therefore this does not as of yet provide clear evidence for a suppression of paramagnetic limiting as a result of ASOC.

An interesting contrast is provided by NMR measurements \cite{La2C3NMR} of  La$_2$C$_3$, where as shown in Fig.~\ref{R2C3Fig}(b), the Knight shift is nearly constant through $T_c$. While this may be due to the effects of the ASOC, these measurements were performed on polycrystalline samples and therefore the same caveat as discussed previously applies, that $\chi_s$ is only expected to be constant perpendicular to $\hvg({\bf k})$. Furthermore, instead of a decrease $1/T_1$ below $T_c$, there is an anomalous increase of $1/T_1$ with decreasing temperature. The origin of the highly unusual behaviour is not known but various sources for the increase can be ruled out and it was suggested that it may arise as a result of collective modes of spin-triplet Cooper pairs  \cite{La2C3NMR}. Two gap superconductivity has also been deduced from  $\mu$SR \cite{Y2C3TwoGap,La2C3MuSR} and photoemission spectroscopy measurements  \cite{La2C3Photo}. The photoemission spectrum in  Fig.~\ref{R2C3Fig}(d) deviates significantly from the behaviour of single band $s$- and $d$-wave models but is well fitted by two bands. However, the specific heat of La$_2$C$_3$ could be accounted for by a single band $s$-wave model with strong electron-phonon coupling \cite{La2C3SpecH}.

These results pose a question about whether the superconducting properties of $R_2$C$_3$ are governed by strong singlet-triplet mixing. In this scenario, the two gaps would arise on the bands split due to the ASOC and if there are nodes in Y$_2$C$_3$ these can arise accidentally when the triplet component exceeds the singlet. Since La is heavier than Y, it might be expected that the ASOC strength and therefore the triplet component is stronger in  La$_2$C$_3$ than Y$_2$C$_3$, which is qualitatively consistent with the comparison between the Knight shifts and also the larger differences between the magnitudes of the two gaps in La$_2$C$_3$. However, as discussed in the previous section, the ASOC strength upon atomic substitution may be governed by more than simply the relative atomic weights and differences may arise due to detailed structural differences. It should also be noted that the presence of multiple Fermi surface sheets in both compounds  \cite{La2C3SpecH,Y2C3Elec} means that if both compounds are nodeless, the two band superconductivity can also be  naturally explained by conventional multiband $s$-wave superconductivity. Along with the difficulties in interpreting NMR data and confirming nodal behaviour, the nature of the pairing symmetry is not well understood and further measurements are required to confirm whether there is mixed-parity pairing.

\subsubsection{$A_2$Cr$_3$As$_3$ ($A$~=~K, Rb, Cs)} \label{ACrAs}

The recently discovered superconductors $A_2$Cr$_3$As$_3$ ($A$~=~K, Rb or Cs) \cite{K2Cr3As3Rep,Rb2Cr3As3Rep,Cs2Cr3As3Rep} also show evidence for unconventional behaviour. These compounds crystallize in a noncentrosymmetric hexagonal crystal structure (space group $P\bar{6}m2$) consisting of concentric chains of [(Cr$_3$As$_3$)$^{2-}]^\infty$  units, with Cr on the inside and As on the outside, which run along the $c$~axis. Despite these quasi-one-dimensional structural features, electronic structure calculations reveal that there are both two quasi-one dimensional and one anisotropic three-dimensional Fermi surface sheets \cite{K2Cr3As3Elec}. The most striking of the initially reported properties was the very high values of $H_{c2}(T)$. From low field measurements of polycrystalline samples, of K$_2$Cr$_3$As$_3$ ($T_c~=~6.1$~K), $H_{c2}(0)$ could be extrapolated to $\sim32$~T, significantly in excess of the weak coupling Pauli limiting field \cite{K2Cr3As3Rep,K2Cr3As3Crys}, which was taken as an indication of possible spin-triplet superconductivity. The presence of line nodes  in the superconducting gap of  K$_2$Cr$_3$As$_3$ was  suggested from penetration depth measurements \cite{K2Cr3As3Pen,K2Cr3As3Pen2,K2Cr3As3MuSR}, which show a linear temperature dependence. Unconventional superconductivity in   K$_2$Cr$_3$As$_3$  and   Rb$_2$Cr$_3$As$_3$   is also indicated by NMR measurements \cite{K2Cr3As3NMR,Rb2Cr3As3NMR}, where there is a lack of a coherence peak in $1/T_1$.

 Despite the noncentrosymmetric crystal structure, the possibility of mixed singlet-triplet pairing has not been the focus of theoretical work on this compound and spin triplet superconductivity has been predicted on the basis of a three band Hubbard model driven by spin fluctuations \cite{K2Cr3As3Theor1,K2Cr3As3Theor2}(this was calculated without including the ASOC). On the other hand, other work has explained the properties as resulting from  strong electron-phonon coupling and suggested that the nodes may arise from the anisotropy of this coupling \cite{K2Cr3As3Phonon}. One interesting result in particular is the measurement of $H_{c2}(T)$ of single crystals in high magnetic fields. In applied fields up to 14~T, $H_{c2}(T)$ only has a small anisotropy and its value parallel to the direction of Cr chains ($H_{c2\parallel}(T)$) is greater than in the perpendicular direction ($H_{c2\perp}(T)$)\cite{K2Cr3As3Crys}. However, at lower temperatures the curves cross and the behaviour of $H_{c2\parallel}(T)$ indicates the presence of Pauli paramagnetic limiting with   $H_{c2\parallel}(0)\sim$23~T, while  $H_{c2\perp}(T)$  appears not to show Pauli limiting with $H_{c2\perp}(0)\sim$37~T. This behavior is consistent with triplet superconducting states where the order parameter $\mathbf{d(k)}$  lies along the $c$-axis, which means Pauli limiting is absent perpendicular to this direction. Indeed this is the case for the $p_z$ pairing state given in  Refs.~\onlinecite{K2Cr3As3Theor1,K2Cr3As3Theor2,K2Cr3As3Theor3} . Furthermore, since there are pseudo-spin representations corresponding to the point group $D_{3h}$ (see Table~\ref{Table1}) which do not couple to an in-plane Zeeman field, the effects of the ASOC may also account for this behaviour. However, detailed characterization of the superconducting properties are further complicated by the extreme air sensitivity of these materials and sample decomposition can affect the low temperature properties \cite{K2Cr3As3Rep,K2Cr3As3Pen,K2Cr3As3Pen2}.

\subsection{Fully gapped, weakly correlated superconductors}

\begin{figure}[t]
\begin{center}
\vspace{0.5cm}
  \includegraphics[width=0.7\columnwidth]{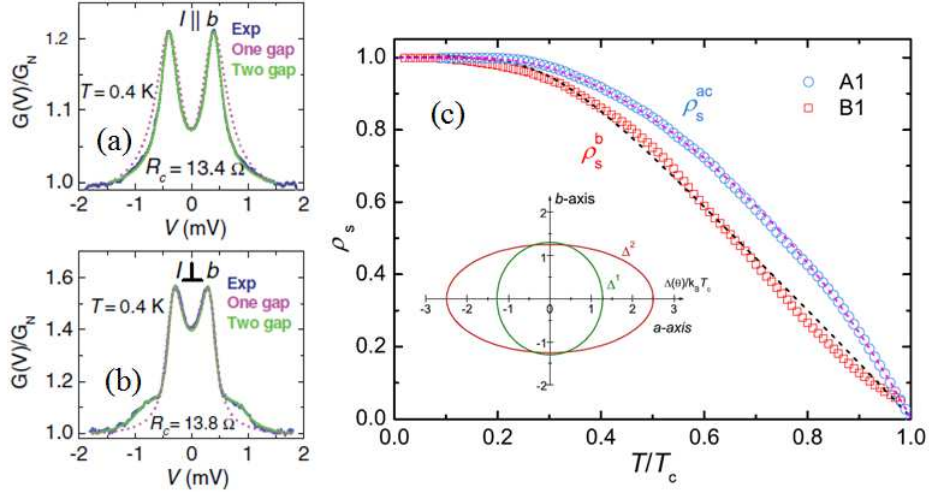}	
\end{center}
\caption{Conductance spectra from point contact spectroscopy measurements at 0.4~K of BiPd with (a) $I\parallel b$ and (b) $I\perp b$. Also shown are fits to a one-gap model (dotted lines), which can not account for the data and a two-gap model (solid lines) which fit the data well. Reprinted figure with permission from  Ref.~\onlinecite{BiPdPCAR}. Copyright 2012 by the American Physical Society. (c) The temperature dependence of the superfluid density of BiPd along two directions, obtained from penetration depth measurements of BiPd from. The superfluid density was fitted with a two band model, with one isotropic and one anisotropic gap. A cross section of the fitted gap structure is shown in the inset. Reprinted figure with permission from  Ref.~\onlinecite{BiPdYuan}. Copyright 2014 by the American Physical Society. }
  \label{PCSFig}	
\end{figure}

\subsubsection{BiPd}

The NCS $\alpha$-BiPd  crystallizes in a monoclinic structure (space group $P2_1$) \cite{BiPdStruc} and becomes superconducting at $T_c~=~3.8$~K  \cite{OldBiPdSC,BiPdRep}. The presence of multiple superconducting gaps in BiPd has been proposed on the basis of point contact Andreev reflection (PCAR) \cite{BiPdPCAR} and penetration depth measurements  \cite{BiPdYuan}.

 The conductance as a function of the bias voltage from PCAR spectroscopy experiments is shown for BiPd,  deep in the superconducting state  at 0.4~K, with $I\parallel b$ and  $I\perp b$ in Figs.~\ref{PCSFig}(a) and (b) respectively. A superconducting gap is revealed by a peak in the conductance and the voltage at which this occurs depends on the gap size. The dominant feature in  all the spectra for both directions is a peak which corresponds to $\Delta_1\sim0.4$~meV. In addition, it is reported that some of the spectra for $I\parallel b$ show a second smaller gap around  $\Delta_2\sim0.1$~meV, while some with $I\perp b$ show second larger gap $\Delta_2\sim0.8$~meV, as clearly shown in  Fig~\ref{PCSFig}(b). The data for both orientations can be accounted for by a two gap Blonder-Thinkham-Klapwijk (BTK) model but not one with a single gap, indicating the presence of multiple superconducting gaps.
In addition, some spectra show a peak at zero voltage bias which is ascribed to the presence of Andreev bound states, suggesting a sign change of the superconducting gap on the Fermi surface. On this basis the authors propose that the two gaps arise from singlet-triplet mixing, where one gap $\Delta_1$ is fully open, while the second gap $\Delta_2$  is highly anistropic and changes sign across the Fermi surface.  On the other hand, measurements of the differential conductance using scanning tunneling spectroscopy \cite{BiPdNatComm} show different results to the PCAR data, where no peak at zero bias is observed and the data can be fitted by a single band $s$-wave model.

\begin{figure}[t]
\begin{center}
\vspace{0.5cm}
  \includegraphics[width=0.3\columnwidth]{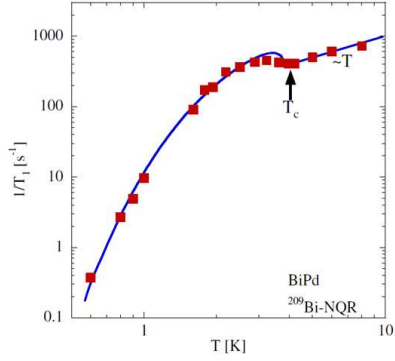}	
\end{center}
\caption{Temperature dependence of $1/T_1$ of BiPd from NQR measurements from Matano~\textit{et al.} \cite{BiPdNQR}, showing a reduced coherence peak below $T_c$. Reprinted figure with permission from  Ref.~\onlinecite{BiPdNQR}. Copyright 2013 by the Physical Society of Japan.}
  \label{PCSFig2a}	
\end{figure}

The superfluid density obtained from penetration depth measurements \cite{BiPdYuan} can also be fitted with a two-gap model [Fig.~\ref{PCSFig}(c)], with one isotropic gap  $\Delta_1\sim0.41$~meV and an anisotropic gap  $\Delta_2$ which has a maximum amplitude of 0.8~meV. A cross section of the structure of the fitted gaps is shown in the inset of Fig.~\ref{PCSFig}(c). This shows reasonable agreement with the PCAR data and the results indicate that there are no nodes on either of the gaps.  The presence of an anisotropic two-gap structure is also consistent with a splitting due to ASOC of around 50~meV \cite{BiPdNQR} which is slightly higher than the similarly nodeless Li$_2$Pd$_3$B, but smaller than the nodal Li$_2$Pt$_3$B. However, the spin relaxation rate $1/T_1$ obtained from nuclear quadrupole measurements (NQR)  \cite{BiPdNQR} could be fitted by a single band $s$-wave model, although curiously the size of the coherence peak below $T_c$ is greatly reduced (Fig.~\ref{PCSFig2a}), which may also indicate anisotropy of the superconducting gap.

\begin{figure}[t]
\begin{center}
\vspace{0.5cm}
  \includegraphics[width=0.3\columnwidth]{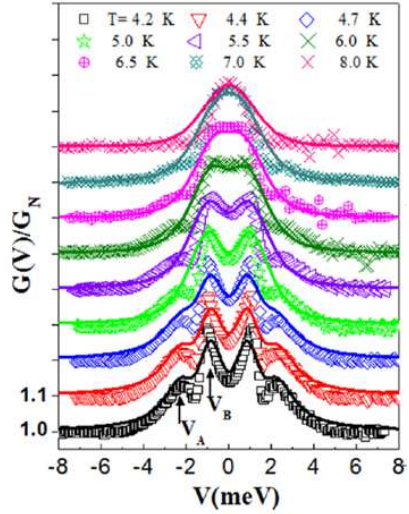}	
\end{center}
\caption{Conductance spectra of  Nb$_{0.18}$Re$_{0.82}$  showing a clear two peak structure, which has been fitted by a two-gap model.  Reprinted figure with permission from  Ref.~\onlinecite{NbRePCS}. Copyright 2015 by the American Physical Society.}
  \label{PCSFig2b}	
\end{figure}

The fact that in BiPd, the two band behaviour from PCAR is corroborated by thermodynamic measurements gives further evidence for multi-band behaviour, but the lack of two gap features in the scanning tunneling spectroscopy spectra needs to be accounted for. In addition it is clear that while two band behaviour is consistent with the presence of significant singlet-triplet mixing, it by no means demonstrates its presence.  Electronic structure calculations \cite{BiPdNatCommSupp} reveal that there are multiple Fermi surface sheets and therefore the multi-band superconductivity could also be of the conventional $s$-wave type.

\subsubsection{Re based superconductors with the $\alpha$-Mn structure}

Several superconductors are found to form in the cubic noncentrosymmetric $\alpha$-Mn structure (space group $I\bar{4}3m$),   many of which contain Re and other heavy elements. The Nb$_x$Re$_{1-x}$ system forms in the $\alpha$-Mn structure \cite{KnaptonNbRe} for $0.13\leq x\leq0.38$, where the unit cell is large, containing 58 atoms occupying four sites. Given that all of the heavy Re atoms occupy noncentrosymmetric positions, it was considered a good candidate for looking for the effects of strong ASOC and in particular, it might be expected that the ASOC could be tuned by varying $x$. Despite this, thermodynamic measurements such as the penetration depth and specific heat \cite{YuanNbRe,NbRe2011} as well as NMR measurements \cite{NbReNMR} for $x\sim0.17-0.18$ were all explained in terms of single band, isotropic $s$-wave superconductivity with either a moderately enhanced gap size, or a gap very close to the BCS value in the case of NMR.  In addition, measurements of other isostructural superconductors Re$_3$W \cite{Re3WRep,BiswasRe3W1} and Re$_{24}$Ti$_5$ \cite{Re24Ti5} were also shown to be well fitted by a single band model,  while $\mu$SR measurements of Re$_6$Zr suggest fully gapped behaviour and time reversal symmetry breaking \cite{Re6ZrTRS} (this is discussed in more detail in Section~\ref{TRSBSec}).

On the other hand,  point contact spectroscopy and specific heat measurements of Nb$_{0.18}$Re$_{0.82}$ also provide evidence for two band superconductivity \cite{NbRePCS}. As shown in Fig.~\ref{PCSFig2b}, two peaks are observed in the conductance spectra, without evidence for a zero bias voltage peak. The spectra could be fitted by an isotropic two-gap BTK model, with gap amplitudes of 1.99~meV and 1.0~meV at 4.2~K. Evidence for the presence of two energy scales in the superconducting state of Nb$_{0.18}$Re$_{0.82}$ was also found from an analysis of the derivative of the specific heat and the data  could also be accounted for by a two-gap model \cite{NbRePCS}. This is therefore another system where two-gap superconductivity has been suggested from point contact spectroscopy, while some other measurements are consistent with single-gap behaviour. As in the case of BiPd, if there is multi-gap behaviour in  Nb$_{0.18}$Re$_{0.82}$, this would not necessarily imply singlet triplet mixing since more than one band crosses the Fermi level \cite{NbReElec}. Further systematic  measurements as a function of Re content may help clarify the gap structure of  the Nb$_x$Re$_{1-x}$ system and determine the role played by the ASOC in determining the superconducting properties.

\begin{figure}[t]
\begin{center}
\vspace{0.5cm}
  \includegraphics[width=0.4\columnwidth]{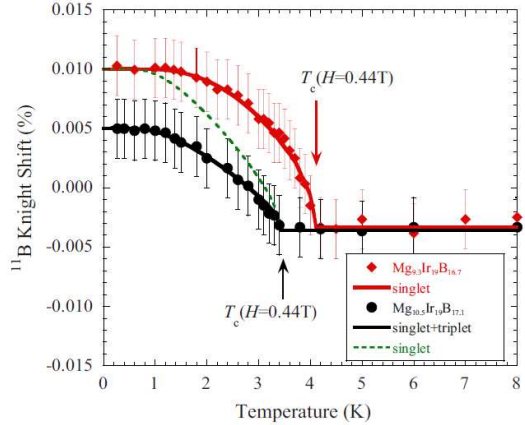}	
\end{center}
\caption{The $^{11}$B Knight shift of  Mg$_{10+x}$Ir$_{19}$B$_{16-y}$  ($x~=~-0.7, y~=~-0.7$ and $x~=~-0.5, y~=~-1.1$). The red solid and the green dashed lines show the purely singlet  behavior calculated for the two compounds, while the black solid line shows the calculation of mixed singlet and triplet pairing. Reprinted figure with permission from  Ref.~\onlinecite{MgBIrNMR}. Copyright 2009 by the American Physical Society.}
  \label{MgIrBFig}	
\end{figure}

\subsubsection{ Mg$_{10+x}$Ir$_{19}$B$_{16-y}$}

The  Mg$_{10+x}$Ir$_{19}$B$_{16-y}$ series of superconductors \cite{MgBIrRep} form in the same  noncentrosymmetric space group as Nb$_x$Re$_{1-x}$, but with a different crystal structure. These materials  have been widely studied without firm conclusions about the nature of the pairing state. Evidence for two gaps was reported on the basis of penetration depth measurements \cite{MgBIrPen}. Converting the data to superfluid density, they found a mixed singlet-triplet model with two anistropic but fully open gaps fits the data at least as well, or possibly even better than a conventional isotropic two-gap picture. Although specific heat \cite{MgBIrHH,MgBIrKlim} and transverse field $\mu$SR data \cite{MgBIrmuSR} were well fitted by a single-gap model, point contact spectroscopy \cite{MgBIrKlim} measurements were better understood in terms of  two-gaps due to the presence of two features in the conductance spectra. NMR measurements of the Knight shift are shown in Fig.~\ref{MgIrBFig} for two compositions ($x=-0.7, y=-0.7$ and $x=-0.5, y=-1.1$) \cite{MgBIrNMR}. An increase in the $^{11}$B Knight shift corresponds to a decrease in the spin susceptibility and in the Mg deficient compound this follows the expected behaviour for spin singlet superconductivity, while the increase for Mg rich sample is not as large as the expected curve for singlet behaviour shown by the green dashed line. This suggests that there remains a finite spin susceptibility and they argue that the higher $T_c$ of the $x~=~0.7$ sample suggests fewer defects. The residual spin susceptibility is attributed to defects which enhance the ASOC, leading to a smaller reduction in the Knight shift than expected for BCS superconductivity.

\subsubsection{Other fully gapped noncentrosymmetric systems}
Another focus for  the study of weakly correlated NCS was to compare the superconducting properties of Ce based NCS with the non-magnetic compounds LaPt$_3$Si and $RTX_3$ ($R$~=~La, Sr or Ba;  $T$~=~transition metal; $X$~=~Si or Ge), which are isostructural to  CePt$_3$Si  and  CeRhSi$_3$ respectively. Unlike CePt$_3$Si, LaPt$_3$Si is a weakly correlated material with $\gamma~=~11$~mJ/mol~K$^2$ and becomes superconducting below $T_c~=~0.6$~K \cite{CePt3SinodesC}. Measurements of the specific heat and penetration depth demonstrate that instead of the nodal superconductivity of CePt$_3$Si,  LaPt$_3$Si is a fully gapped superconductor \cite{CePt3SinodesC,CePt3SiNodePen} and these results, along with NMR measurements are consistent with BCS superconductivity \cite{LaPt3SiNMR}. A considerably larger number of $RTX_3$ superconductors have been studied and thermodynamic measurements of Ba(Pt,Pd)Si$_3$ \cite{BaPtSi32009,Kneidinger2014}, La(Rh,Pt Pd,Ir)Si$_3$ \cite{LaRhSi32011,LaPdSi3Rep,LaPdPtSi3,LaIrSi3SC,LaIrSi32014}, Ca(Pt,Ir)Si$_3$ \cite{CaMSi32011,CaTSi3Singh}, Sr(Ni,Pd,Pt)Si$_3$ \cite{Kneidinger2014} and  Sr(Pd,Pt)Ge$_3$  \cite{RTX32011,Kneidinger2014} are consistent with fully gapped, $s$-wave superconductivity with energy gaps similar to or slightly below the isotropic, weak coupling BCS value, although in the case of CaIrSi$_3$, which has the highest $T_c$ (3.6~K) of the aforementioned compounds, a larger gap value is inferred from $\mu$SR measurements \cite{CaTSi3Singh}. Unlike the Ce$TX_3$ superconductors, which have extremely large upper critical fields, the weakly correlated $RTX_3$ superconductors either have moderate values of $H_{c2}(0)$,  typically less than 0.3~T obtained from bulk measurements \cite{BaPtSi32009,CaMSi32011,LaPdPtSi3,Kneidinger2014} or are type-I superconductors, as in the case of La(Rh,Pd,Ir)Si$_3$ \cite{LaRhSi32011,LaPdPtSi3,LaIrSi32014}. Interestingly, the resistivity of several of these compounds show significantly different behaviour under magnetic fields and the (upper) critical field derived from resistivity measurements is significantly larger and curves upwards at low temperatures  \cite{BaPtSi32009,CaMSi32011,LaPdPtSi3,LaIrSi32014}, unlike the conventional behaviour of $H_{c2}(T)$ described by the WHH model. While most of the reports have been on polycrystalline material, similar behaviour was also observed in single crystals of  LaRhSi$_3$ \cite{LaRhSi3Crys}. It would be interesting to determine whether the much more robust superconductivity in the resistivity is an intrinsic feature or whether it is an extrinsic effect, possibly due to the presence of regions with a larger number of defects, leading to a reduced mean free path \cite{CaMSi32011}. Weakly correlated $RTX_3$ superconductors have also been studied using detailed electronic  structure calculations \cite{BaPtSi32009,BauerNCS,Kneidinger2014}. In all the compounds, the bands are split by the ASOC and there is  significant splitting at some positions. However in the vicinity of the Fermi level $E_F$,  the splitting is small and mainly discernable near the Brillouin zone boundaries. It is suggested that since the band splitting at $E_F$ is of primarily importance for the superconducting properties, the weak effect of the ASOC near $E_F$ explains the agreement between the physical properties of weakly correlated $RTX_3$ superconductors and single band BCS superconductivity \cite{BaPtSi32009,Kneidinger2014}.

The properties of other NCS have also been found to be consistent with single band BCS superconductivity. Some examples include (Rh,Ir)$_2$Ga$_9$ \cite{T2Ga9Rep,T2Ga92009}, (Re,Ru)$_7$B$_3$ \cite{Re7B3Rep,BiPdNQR,Ru7B3SC}, LaPtSi \cite{LaPtSiRep}, Cr$_2$Re$_3$B \cite{Cr2Re3BRep}, (W,Mo)$_7$Re$_{13}$(B,C) \cite{BMnSC} and Y$_3$Pt$_4$Ge$_{13}$ \cite{Y3Pt4Ge13Rep}. Although early measurements of Mo$_3$Al$_2$C hinted at unconventional superconductivity \cite{Mo3Al2CRep,Mo3Al2CRep2}, this was later attributed to the presence of superconducting impurity phases and the behaviour was described by single band $s$-wave behaviour \cite{Mo3Al2CConv,Mo3Al2CMuSR}. While all these NCS show evidence for  single band $s$-wave superconductivity and therefore an absence of significant singlet-triplet mixing, in some instances the properties have only been characterized by a few measurements and as demonstrated in the preceding sections, in several systems evidence for two-band behaviour is not readily discernable using some techniques, particularly if the two gaps are a similar size. For some compounds mainly consisting of light elements, the lack of significant ASOC and singlet-triplet mixing is readily understandable. However, for others such as Re$_3$W and Re$_{24}$Ti$_5$ which mainly consist of heavy elements, there would be expected to be an appreciable ASOC and evidence for two-gap behaviour is only seen in some measurements of the isostructural Nb$_x$Re$_{1-x}$.  It is clear that despite the intrinsic mixed-parity pairing expected when inversion symmetry is broken, many weakly correlated NCS show evidence for entirely singlet $s$-wave pairing. The reasons for these differences and the lack of singlet-triplet mixing remain to be determined, but it may be the case that in addition to the presence of a significant ASOC, there also needs to be pairing interactions in both the singlet and triplet channels for there to be significant parity mixing.

\subsection{Time reversal symmetry breaking: both evidence for and against parity mixing.}
\label{TRSBSec}

\begin{figure}[t]
\begin{center}
\vspace{0.5cm}
  \includegraphics[width=0.99\columnwidth]{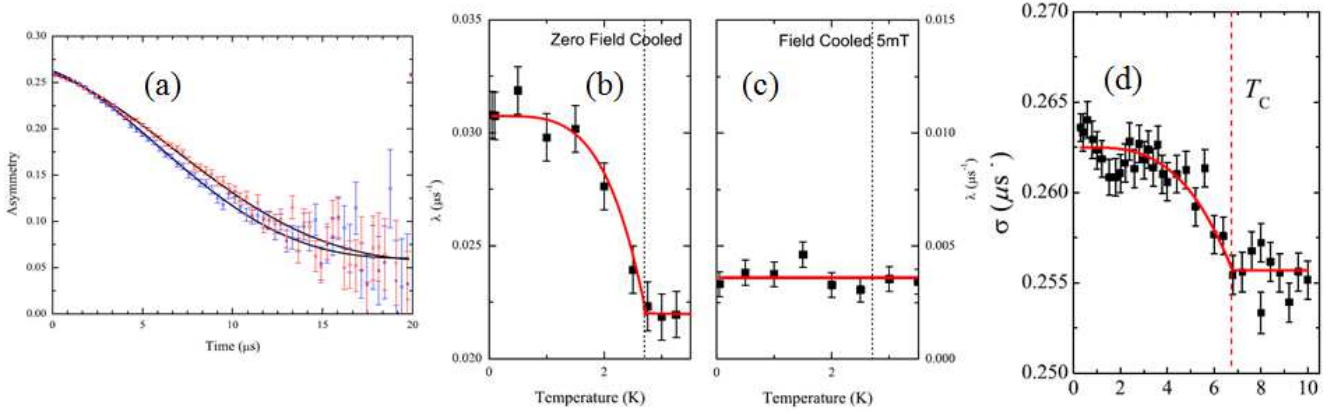}	
\end{center}
\caption{(a) Zero field $\mu$SR measurements of LaNiC$_2$ measured at 0.054~K (blue) and 3~K (red), where a faster decay of the asymmetry below $T_c$ is clearly observed. The temperature dependence of the exponential relaxation rate $\lambda$ is shown for (b) zero field, where there is a clear increase below $T_c$ and (c) a field of 5~mT, where there is no change. Reprinted figures with permission from  Ref.~\onlinecite{LaNiC2TRS}. Copyright 2009 by the American Physical Society. (d) Temperature dependence of the Kubo-Toyabe relaxation rate $\sigma$ from zero field $\mu$SR measurements of Re$_6$Zr, which also increases below $T_c$. Reprinted from Ref.~\onlinecite{Re6ZrTRS}, available under a Creative Commons Attribution 3.0 License.}
  \label{TRSFig}	
\end{figure}

One hallmark of unconventional superconductivity is the breaking of additional symmetries in the superconducting state and in certain triplet states time reversal symmetry (TRS) is broken. As a result spontaneous magnetic moments arise below $T_c$ and these can be detected using muon spin relaxation ($\mu$SR) measurements, which can probe the local magnetic field distribution of a sample, without an applied field. The asymmetry of the emitted positrons as a function of time gives information about the field distribution at the muon stopping site and in particular, a faster decay of the asymmetry below $T_c$ indicates the formation of spontaneous magnetic fields \cite{BlundellMuon}.

Time reversal symmetry breaking has been detected using $\mu$SR in the NCS, LaNiC$_2$ \cite{LaNiC2TRS}, Re$_6$Zr \cite{Re6ZrTRS} and La$_7$Ir$_3$ \cite{La7Ni3TRS}. Figure~\ref{TRSFig}(a) shows zero-field $\mu$SR measurements of LaNiC$_2$ performed above and below $T_c~=~2.7$~K. Below $T_c$ as shown by the blue points, there is a noticeably increased decay rate of the asymmmetry. The data were fitted with the product of two decay functions and the exponential decay rate $\lambda$, associated with a Lorentzian field distribution, shows a sudden increase at $T_c$ (Fig.~\ref{TRSFig}(b)). This is taken as an indication of the formation of spontaneous magnetic fields at $T_c$ and therefore time reversal symmetry breaking. The effect is not seen in an applied field of 5~mT [Fig.~\ref{TRSFig}(c)], which is sufficient to decouple the muon from these local fields. TRS breaking was also observed in the superconducting state of Re$_6$Zr \cite{Re6ZrTRS}, as shown in Fig.~\ref{TRSFig}(d), although in this instance the exponential relaxation rate remains constant, but there is an increase of the Kubo-Toyabe relaxation rate $\sigma$ below $T_c$.

If the superconducting states of LaNiC$_2$ and  Re$_6$Zr do break TRS, two further questions arise; does TRS breaking necessarily imply spin triplet superconductivity and do the results support the presence or absence of singlet triplet mixing? Insights into these questions have been provided by means of group theoretical analysis  \cite{LaNiC2TRS,LaNiC2Sym,Re6ZrTRS}. LaNiC$_2$ has a noncentrosymmetric orthorhombic crystal structure (space group $Amm2$) \cite{LaNiC2Rep}.  Due to the low symmetry of the point group of the crystal structure ($C_{2v}$) of LaNiC$_2$, all the allowed TRS breaking states correspond to non-unitary triplet pairing  \cite{LaNiC2TRS}, where there is spin polarization of the Cooper pairs \cite{SigristUnCon}. Furthermore, when there is significant spin-orbit coupling, the non-unitary triplet states are no longer allowed to form at $T_c$ \cite{LaNiC2Sym}. Therefore the observation of TRS breaking at $T_c$ implies that the effect of the ASOC on superconductivity is weak, which is evidence against significant singlet-triplet mixing in LaNiC$_2$. It should be noted that it would be possible for TRS to be broken at a second transition below $T_c$, but no evidence for this has been observed. The fact that TRS breaking is also observed in the similar but \textit{centrosymmetric} LaNiGa$_2$ \cite{LaNiGa2TRS} further indicates that in this instance, the breaking of TRS in LaNiC$_2$ is unlikely to be a result of the lack of inversion symmetry. In the case of Re$_6$Zr, the crystal structure is the noncentrosymmetric cubic $\alpha$-Mn type, which has a much higher symmetry. As a result, a TRS breaking mixed singlet-triplet state can be permitted \cite{Re6ZrTRS}. However, another result of the high symmetry of the point group is  that broken TRS in Re$_6$Zr may not necessarily imply the presence of a spin triplet component and it has been suggested that TRS breaking multiband $s$-wave superconductivity with a conventional pairing mechanism can arise in some cubic systems \cite{AgterConv}.

Measurements of the gap structure of LaNiC$_2$ have produced conflicting results, with evidence for nodal behaviour suggested from early specific heat \cite{LaNiC2Rep} and penetration depth measurements \cite{LaNiC2PenNodes}. However, later specific heat \cite{LaNiC2SpecH1,LaNiC2Yuan} and penetration depth measurements \cite{LaNiC2Yuan} as well as NQR experiments \cite{LaNiC2NQR} indicate fully gapped behaviour, with two-band superconductivity being proposed in Ref.~\onlinecite{LaNiC2Yuan}. All the allowed TRS breaking triplet states for LaNiC$_2$ have nodes  \cite{LaNiC2TRS} and how to reconcile the TRS breaking and fully gapped superconductivity is currently an important issue. Furthermore, evidence for two gap superconductivity has also been found in centrosymmetric LaNiGa$_2$, indicating that this is another common feature of these compounds \cite{LaNiGa2HQ}. Meanwhile transverse field $\mu$SR measurements \cite{Re6ZrTRS} of the penetration depth indicate that Re$_6$Zr is fully gapped, but there is as of yet no evidence for multiband superconductivity. In the NCS La$_7$Ir$_3$ \cite{La7Ni3TRS}, where TRS was also recently reported to be broken, the penetration depth obtained from transverse field $\mu$SR measurements was fitted with a single fully gapped model. Further progress may be made from detailed measurements of single crystals. The recent successful growth of single crystals of LaNiC$_2$ may allow for the superconductivity to be probed using a wider range of techniques \cite{LaNiC2FS}.  The synthesis of Re$_6$Zr single crystals has yet to be reported and the high melting point of Re rich alloys presents a challenge for crystal growth. However, the successful growth of Nb$_{0.18}$Re$_{0.82}$ single crystals using the optical floating zone technique \cite{NbReCrys} may be a promising method for obtaining Re$_6$Zr crystals.

These results indicate that zero field $\mu$SR can be a powerful tool for looking for TRS breaking and therefore triplet superconductivity.  Where evidence has been obtained from  $\mu$SR measurements, it is particularly important to find other experimental evidence which supports the presence of broken TRS and triplet superconductivity. This has been achieved for example in centrosymmetric Sr$_2$RuO$_4$ which is widely believed to be a spin triplet superconductor \cite{Sr2RuO4Trip}, where evidence for TRS breaking has been observed using $\mu$SR \cite{Sr2RuO4TRS1} and polar Kerr effect \cite{Sr2RuO4TRS2} measurements. Indeed further evidence has been found in the case of LaNiC$_2$ from measurements of a small spontaneous magnetization along the $c$~axis below $T_c$ \cite{LaNiC2Mag}. In many instances the presence or absence of TRS breaking can not definitively demonstrate or rule out the presence of triplet pairing, since there are triplet states where TRS is not broken and there can be other explanations for TRS breaking other than triplet superconductivity \cite{SigristUnCon}. However in the case of LaNiC$_2$, the results of the symmetry analysis should perhaps provide a cautionary note that the observation of unconventional behaviour in NCS can not necessarily be taken as evidence for singlet-triplet mixing, but may in fact be evidence for its absence.

\subsection{Non-centrosymmetric superconductivity in topological systems}

The half Heusler alloys are a series of compounds which have recently attracted particular attention. These compounds lack inversion symmetry and many have been predicted to have topologically non-trivial surface states, due to the presence of a band inversion in the electronic structure \cite{HHBandInv}. One such compound is YPtBi  which becomes superconducting at $T_c~=~0.77$~K  \cite{YPtBiRep}. Measurements of the  transverse field magnetoresistance  at low temperatures reveal that the spin-orbit coupling splits the Fermi surface into two similar sized sheets and the carrier density $n$ is surprisingly low compared to most other superconducting systems  \cite{YPtBiRep}. Calculations on the basis of the BCS theory of superconductivity indicate that much larger values of $n$ are necessary to account for the value of $T_c$, which was taken to suggest that the superconductivity of YPtBi can not be explained within the framework of the conventional electron-phonon pairing mechanism, but rather that there is unconventional pairing \cite{YPtBiCalc}. Meanwhile $H_{c2}(T)$ remains linear to low temperatures, in excess of the orbital limiting field of 0.85~T calculated from WHH theory, but it can be extrapolated to a zero temperature value of $\sim1.5$~T, close to the weak coupling Pauli limit. Although $\mu$SR measurements were unable to provide $\lambda(T)$ of sufficient precision \cite{YPtBiPressMuSR}, recent measurements of $\Delta\lambda(T)$ showed linear behaviour at low temperatures, indicating the presence of line nodes \cite{YPtBiPen}. Due to the band inversion and the four fold degeneracy of the bands near the Fermi level, it was suggested that the Cooper pairs may form between fermions with  $j=3/2$, instead of  the commonly considered $j=1/2$ \cite{YPtBiPen,YPtBiTheor}. As a result, pairing states with higher angular momentum such as quintet ($J~=~2$) and septet ($J~=~3$)  states may occur, in addition to the singlet ($J~=~0$)  and triplet ($J~=~1$)  states arising from $j=1/2$ pairing. It was suggested that the superconducting state may correspond to mixed-parity pairing involving these higher angular momentum states, where the line nodes may arise from a mixture of singlet-septet pairing \cite{YPtBiPen,YPtBiTheor}.

Other half Heusler alloys displaying both band inversion and superconductivity are $R$PdBi ($R$~=~Ho, Er, Tm or Lu) \cite{ErPdBiRep,LuPdBiRep,RPdBirep} ($T_c \sim1.2-1.7$~K),  LuPtBi \cite{LuPtBiRep} ($T_c \sim1.0$~K) and LaPtBi \cite{LaPtBiRep} ($T_c \sim 0.9$~K). Meanwhile other half heusler compounds showing superconductivity include $R$PdBi ($R$~=~Sm, Dy, Tb or Y)\cite{RPdBirep}, although for these compounds, the  electronic structures are predicted not to show band inversion, but to be topologically trivial \cite{HHBandInv,RPdBirep}. Another interesting feature of the $R$PdBi superconductors is that when $R$ is a magnetic element, the compound displays both magnetic order and superconductivity \cite{ErPdBiRep,RPdBirep}. The  relationship between the superconducting and magnetic phases in these compounds is yet to be determined and requires further detailed measurements. For most of the superconducting half Heusler alloys, the superconducting gap structure has yet to be characterized. Besides  the case of YPtBi discussed previously, the specific heat below $T_c \sim 1.7$~K of LuPdBi  was fitted by an isotropic, fully gapped single band model with a gap larger than the weakly coupled BCS value \cite{LuPdBiRep} and $1/T_1$ obtained from NQR measurements of LaPtBi shows the coherence peak just below $T_c$ expected for $s$-wave superconductivity . It is therefore important for the gap structure of a wider range of half Heusler alloys to be measured, which is made more challenging  in some instances by a lack of a superconducting transition observed in specific heat measurements \cite{RPdBirep}.

Another promising candidate for topological superconductivity is the NCS PbTaSe$_2$ \cite{PbTaSe2Rep}, which has a $T_c$ of 3.7~K. It was previously proposed that NCS may be ideal systems for observing Majorana modes, as long as the triplet component is larger than the singlet one \cite{Majoran1}. This is unlikely to be the case for PbTaSe$_2$, where the penetration depth and superfluid density are well accounted for by a single-gap $s$-wave model \cite{PbTaSe2TDO}, indicating that any triplet component should be very small. However it is reported in Ref.~\onlinecite{PbTaSe2Topo} that PbTaSe$_2$  has topologically non-trivial Dirac surface states and as a result, fully gapped superconductivity induced in these states is sufficient to allow for Majorana bound states. Since there is evidence for fully gapped superconductivity in  PbTaSe$_2$ obtained from a variety of measurements \cite{PbTaSe2TDO,PbTaSe2SpecH,PbTaSe2Cond}, it is highly desirable to look for signatures of Majorana fermions in this system. Indeed the requirement that the superconducting gap is fully open is significantly less stringent than the need for a dominant triplet component and therefore the identification of NCS with such non-trivial surface states may be a promising route for observing such novel behaviour.

\subsection{Two dimensional superconductivity: Interfaces, monolayers and heterostructures}

\begin{figure}[t]
\begin{center}
\vspace{0.5cm}
  \includegraphics[width=0.99\columnwidth]{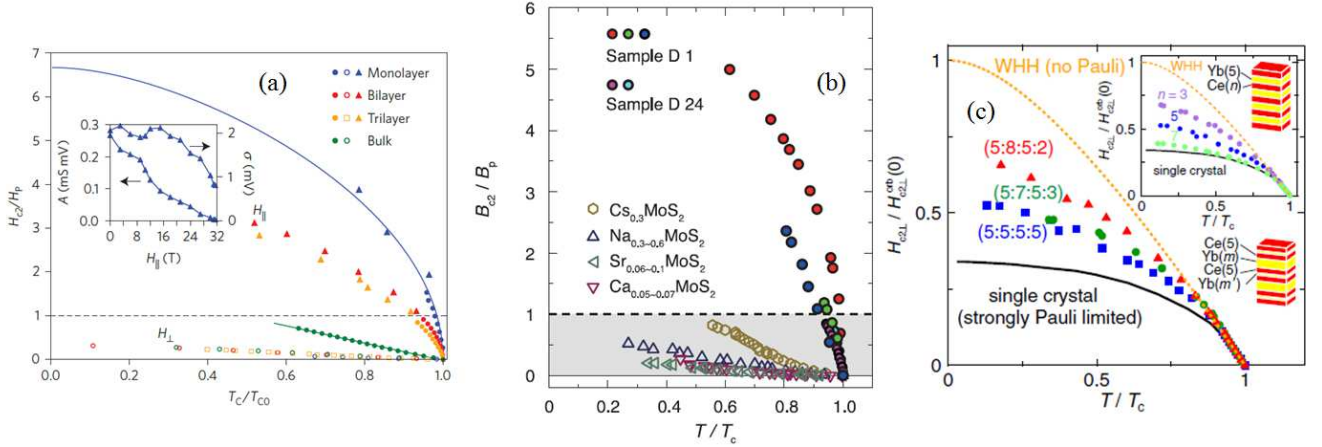}	
\end{center}
\caption{Temperature dependence of the ratio of the in-plane upper critical and the Pauli limiting field for atomically thin and bulk (a) NbSe$_2$, reprinted by permission from Macmillan Publishers Ltd: Nature Physics (Ref.~\onlinecite{IsingNbSe2}), copyright 2016; (b) MoS$_2$, from Ref.~\onlinecite{IsingMoS2}. Reprinted with permission from AAAS. (c) Out of plane upper critical field of  CeCoIn$_5$/YbCoIn$_5$ heterostructures, normalized by the orbital limiting value at zero temperature. The solid line shows the bulk value while the dashed line shows the values for purely orbital limiting. The main panel shows data for heterostructures with five CeCoIn$_5$ layers with different thicknesses of YbCoIn$_5$, while the inset shows   the values for five YbCoIn$_5$ layers with different numbers $n$ of CeCoIn$_5$ layers. Reprinted figure with permission from  Ref.~\onlinecite{SuperLattHc2b}. Copyright 2014 by the American Physical Society.}
  \label{IsingFig}	
\end{figure}

The properties of low-dimensional systems has been an important topic in condensed matter research in recent years, which show a wide range of unusual phenomena. For example, a two dimensional conducting layer can be formed at the interface of two bulk insulators \cite{InterfaceRep1,InterfaceRep2} and superconductivity has also been observed at the interface of some systems such as LaAlO$_3$/SrTiO$_3$ \cite{LO3STO3SCRep} and LaTiO$_3$/SrTiO$_3$ \cite{LTO3STO3SCRep}. Another route to realizing two-dimensional superconductivity is via single layers of materials, which may be non-superconducting in the bulk, as  in the case of the semiconductor MoS$_2$ \cite{MoS2SC1,MoS2SC2} or also have bulk superconductivity as with  NbSe$_2$ \cite{NbSe2SC}. Superconductivity in many of these systems is either induced or enhanced upon the application of a gate voltage \cite{LOSTOGateV,LTOSTOGateV,NbSe2GateV}. An important property of these interfaces and surfaces is that inversion symmetry is broken and the electrons experience a Rashba type spin-orbit interaction, the strength of which can be tuned by varying the applied gate voltage \cite{InterfaceRashba1,InterfaceRashba2}. These systems therefore provide a valuable opportunity to study the effects of ASOC on the superconducting properties.

An example of this phenomena is the recent evidence for Ising superconductivity in both monolayer NbSe$_2$ \cite{IsingNbSe2}, and MoS$_2$ with an applied gate voltage \cite{IsingMoS2}. For single layers of both these systems,  inversion symmetry is broken within the plane, so instead of a Rashba type interaction, the predominant effect of the resulting ASOC is to give rise to an effective magnetic field which locks the spins perpendicular to the plane. This is described later in more detail in Section~\ref{ASOC}. As discussed previously, an applied field greater than the Pauli paramagnetic limiting value can induce a Zeeman splitting which breaks singlet Cooper pairs. However, since the fermion spins in these systems are locked perpendicular to the plane, they show an unusual resilience to in-plane magnetic fields. The temperature dependences of the in-plane upper critical field ($H_{c2\parallel}$) are displayed in Fig.~\ref{IsingFig}. For NbSe$_2$ (Fig.~\ref{IsingFig}(a)) \cite{IsingNbSe2}, the values for the atomically thin samples are compared to the values from the bulk material as well as the out of plane upper critical field  ($H_{c2\perp}$). In  Fig.~\ref{IsingFig}(b) \cite{IsingMoS2} the values from the thin samples are compared to the bulk behaviour of MoS$_2$ intercalated with alkali metals. In both cases the bulk $H_{c2}$ has a linear temperature dependence and remains lower than $H_p$. Upon reducing the dimensionality there are two effects which enhance $H_{c2\parallel}$. Firstly, once the thickness becomes less than $\lambda$, in-plane orbital pair breaking significantly weakens. Secondly, $H_{c2\parallel}$ greatly exceeds $H_p$, reaching more than $6H_p$ for monolayer NbSe$_2$ and more than $4H_p$ in MoS$_2$. Instead of a linear temperature dependence, square root behaviour of $H_{c2\parallel}(T)$ is observed. When the applied field is sufficiently large compared to the effective field from the ASOC, the spins will tilt towards the plane and paramagnetic pair breaking can occur. From fitting the data (solid line  in Fig.~\ref{IsingFig}(a)), the effective field is estimated to be $\sim660$~T for monolayer NbSe$_2$ and $\sim114$~T for MoS$_2$, which are consistent with theoretical calculations of the band splitting. These results therefore provide strong evidence for a significant effect of ASOC on low dimensional superconducting systems. A range of further novel superconducting properties have been proposed to result from the enhanced ASOC, including a Fulde-Ferrell-Larkin-Ovchinikov  (FFLO) -type state \cite{InterfaceFFLO} and topological superconductivity. \cite{TopolRashba}

Superconductivity has also been observed in two-dimensional Kondo lattices via the creation of heterostructures consisting of layers of CeCoIn$_5$ and YbCoIn$_5$ \cite{2DHFSCRep}. CeCoIn$_5$ is a centrosymmetric heavy fermion superconductor \cite{CeCoIn5SC} and superlattices can be grown consisting of $n$ layers of CeCoIn$_5$ interspaced with non-superconducting  YbCoIn$_5$  \cite{2DHFSCRep}. Superconducting transitions are observed in resistivity measurements for all $n$, but zero resistivity is only seen for $n\geq3$ and the large difference in the Fermi velocity between the two materials means that two-dimensional superconductivity can be realized. Since inversion symmetry is more strongly broken at the interface of the two types of layer, the ASOC can be tuned by varying the layer thickness. The angular dependence $H_{c2}(\theta)$ was measured for a different number of CeCoIn$_5$ layers $n=3-5$ \cite{SuperLattHc2a}. For $n~=~3$, a cusp is observed near $T_c$ in $H_{c2}(\theta)$ at $\theta~=~0$, corresponding to the in-plane direction. This feature becomes rounded and less pronounced for larger $n$ and for $n~=~3$ it also weakens upon reducing the temperature. Such a cusp is expected for a two-dimensional superconductor with orbital pair breaking and therefore the weakening with increasing $n$ was interpreted as an increased contribution of Pauli paramagnetic limiting, which dominates $H_{c2}$ of the bulk material. The change in shape of $H_{c2}(\theta)$ upon reducing the temperature, which is only seen in the $n~=~3$ sample, was interpreted as a signal of an FFLO-like state, where the superconducting order parameter has a real space modulation. In fact this has been proposed to be a mixed-parity pair density wave state \cite{PDW1,PDW2} where there is a spatially uniform spin triplet component, but the sign of the singlet component varies between layers.

In the inset of Fig.~\ref{IsingFig}(c),  $H_{c2}(\theta=\pi/2)$ is plotted for different $n$, normalized by the zero temperature orbital limiting field \cite{SuperLattHc2b}. It can be seen that for all $n$, the data lie between the bulk behaviour, which is strongly Pauli limited (solid line), and the purely orbital limiting case (dashed line). With decreasing $n$ the data get closer to the orbital limiting line, indicating the weakening of Pauli limiting. More complex superlattice structures have also been fabricated, where the YbCoIn$_5$ layers are no longer all of the same thickness but alternate between $m$ and $m'$ layers, giving a ($n:m:n:m'$) arrangement \cite{SuperLattHc2b}. As displayed in the main panel of  Fig.~\ref{IsingFig}(c), Pauli limiting is also weakened with increasing $|m-m'|$ when $n$ is fixed. These results give evidence for the reduction of paramagnetic limiting upon varying the degree of inversion symmetry breaking and demonstrate the potential of using heterostructures as a means of controllably tuning the ASOC.

\section{Theory: antisymmetric spin-orbit coupling} \label{ASOC}

In the remainder of the article, we provide the theoretical basis for understanding experimental results. The key ingredient is the antisymmetric spin-orbit coupling (ASOC). In this Section we present an overview of this interaction. From a microscopic point of view, the ASOC stems from the spin-orbit interaction, which is  proportional to ${\bf S}\cdot (\nabla U({\bf r})\times {\bf P})$ where ${\bf S}$ is spin, $U({\bf r})$ is the lattice potential experienced by the electrons and ${\bf P}$ is the momentum operator. When the spin-orbit interaction is expressed in a basis of Block spinors, the broken inversion symmetry reveals itself though $\nabla U({\bf r})$ and leads to the second term in the Hamiltonian of Eq.~\ref{H_01}. In the following we rely on symmetry arguments to construct the form of the ASOC.

Prior to discussing the non-centrosymmetric case, it is useful to review the situation when parity ($\mathcal{P}$) and time-reversal $\mathcal{T}$ symmetries are present (here we adopt a formalism and notation similar to Ref.~\onlinecite{yip14}). In this case, at each momentum ${\bf k}$, there are two degenerate states, the state $|{\bf k},\alpha\rangle$ and the state $ \mathcal{P}\mathcal{T}|{\bf k},\alpha\rangle$. In particular, we can define a pseudo-spin basis through $|{\bf k},\uparrow\rangle$ and $|{\bf k},\downarrow\rangle=\mathcal{P}\mathcal{T}|{\bf k},\uparrow\rangle$. Since they are fermions, these states satisfy $\mathcal{T}^2|{\bf k},s\rangle=-|{\bf k},s\rangle$ and we take $\mathcal{T}=-i\sigma_y \mathcal{K}$ where $\mathcal{K}$ is the complex conjugation operator. The two states at momentum ${\bf k}$ are degenerate with two states at momentum $-{\bf k}$, these we take to be $|-{\bf k},\uparrow\rangle=\mathcal{P}|{\bf k},\uparrow\rangle$ and $|-{\bf k},\downarrow\rangle=\mathcal{T}|{\bf k},\uparrow\rangle$. It is important to understand, particularly with strong spin orbit coupling, that the pseudo-spin basis defined this way is not necessarily the same as the usual spin 1/2 basis.

From the two-dimensional pseudo-spin basis at momentum ${\bf k}$, single particle operators can be constructed from the following Pauli-matrix like Hermitian operators
\begin{eqnarray}
\tilde{\sigma}_x({\bf k})=&&|{\bf k},\uparrow\rangle\langle{\bf k},\downarrow|+|{\bf k},\downarrow\rangle\langle{\bf k},\uparrow|\\
\tilde{\sigma}_y({\bf k})=&&-i|{\bf k},\uparrow\rangle\langle{\bf k},\downarrow|+i|{\bf k},\downarrow\rangle\langle{\bf k},\uparrow|\\
\tilde{\sigma}_z({\bf k})=&&|{\bf k},\uparrow\rangle\langle{\bf k},\uparrow|-|{\bf k},\downarrow\rangle\langle{\bf k},\downarrow|.
\label{sigma}
\end{eqnarray}
Note that under $\mathcal{T}$, these operators transform as $\tilde{\sigma}_i({\bf k})\rightarrow -\tilde{\sigma}_i(-{\bf k})$ and under $\mathcal{P}$, $\tilde{\sigma}_i({\bf k})\rightarrow \tilde{\sigma}_i(-{\bf k})$.
 In many cases, a unitary transformation can be found such that $\tilde{\sigma}_x,\tilde{\sigma}_y,\tilde{\sigma}_z$ transform as the usual Pauli matrices $\sigma_x,\sigma_y,\sigma_z$ under spatial rotations. In the following, when such a transformation exists, we will use the notation $\sigma_x,\sigma_y,\sigma_z$ but will keep the notation $\tilde{\sigma}_x,\tilde{\sigma}_y,\tilde{\sigma}_z$ when such a transformation does not exist. The single particle Hamiltonian expressed in these operators takes the form $\xi({\bf k})\sigma_0$ (where $\sigma_0$ is the identity matrix) because: i) the Hamiltonian is Hermitian; and ii) under $\mathcal{PT}$, $\tilde{\sigma}_i({\bf k})\rightarrow -\tilde{\sigma}_i({\bf k})$ and the Hamiltonian must be invariant under this operation.

We now consider the case when parity symmetry is broken. In particular, with just $\mathcal{T}$ symmetry, the single particle Hamiltonian takes the form $\xi({\bf k})\sigma_0+\sum_i \gamma_i({\bf k})\tilde{\sigma}_i$; requiring that the Hamiltonian is Hermitian imposes $\xi({\bf k})=\xi^*({\bf k})$ and $\gamma_i({\bf k})=\gamma_i^*({\bf k})$  and the condition of time reversal invariance further implies $\xi({\bf k})=\xi(-{\bf k})$ and $\gamma_i({\bf k})=-\gamma_i(-{\bf k})$. This leads to the following time-reversal invariant single-particle Hamiltonian in a crystal without parity symmetry
\begin{equation}
\label{H_01}
    H_0=\sum\limits_{{\bf k}}\sum_{\alpha\beta=\uparrow,\downarrow}
[(\xi({\bf k})-\mu)\delta_{\alpha\beta}+\mbox{\boldmath$\gamma$}({\bf k})
   \cdot \tilde{\mbox{{\boldmath$\sigma$}}}
 _{\alpha\beta}]
    a^\dagger_{{\bf k}\alpha}a_{{\bf k}\beta}
\end{equation}
where $ a^\dagger_{{\bf k}\alpha} $ ($a_{{\bf k}\alpha}$) creates (annihilates) an electronic state $ | {\bf k} \alpha \rangle $, $\alpha,\beta=\uparrow,\downarrow$ are pseudo-spin indices, the sum over ${\bf k}$
is restricted to the first Brillouin zone, and $\mu$ is the chemical potential.  The second term in Eq.
(\ref{H_01}) is the {\it antisymmetric} spin-orbit (ASOC) coupling. The main role of the ASOC is fix the direction of the single-particle spin at each ${\bf k}$ so that the spin is no longer able to rotate freely. It does this by lifting the pseudo-spin degeneracy at each ${\bf k}$ point, leaving only a degeneracy between states related by time-reversal (that is, between a state at momentum ${\bf k}$ and its time reversed partner at momentum $-{\bf k}$). In particular, the resultant quasi-particle energies become $\xi_{\pm} ({\bf k})=\xi({\bf k})\pm|\mbox{\boldmath$\gamma$}({\bf k})|$. The basis in which the single-particle Hamiltonian takes this diagonal form is often called the helicity basis.

The detailed form of $\mbox{\boldmath$\gamma$}({\bf k})$ depends upon the space group of the crystal and the particular pseudo-spin representation for the single-particle states. In many cases the pseudo-spin representation has the same rotation properties as usual spin $1/2$ operators. This implies that for states near the $\Gamma$ point, ${\bf k}=(0,0,0)$, the following symmetry condition can be imposed (note that this is not true in general)
\begin{equation}
\mbox{\boldmath$\gamma$}({\bf k})=\tilde{R}_g\mbox{\boldmath$\gamma$}(R_g^{-1}{\bf k})
\label{half}
\end{equation}
 where, for proper rotations, $\tilde{R}_g=R_g$ and $R_g$ is the $3\times 3$ rotation matrix corresponding to the point group operation $g$, and for improper rotations (for which the determinant of $R_g$ is $-1$) $\tilde{R}_g=-R_g$. In most previous discussions of non-centrosymmetric superconductors, Eq.~\ref{half} has been used to find the structure of $\mbox{\boldmath$\gamma$}({\bf k})$ for different point group symmetries near the $\Gamma$ point in the Brillouin zone (specifically through an expansion in powers of $k_i$). However, there exist pseudo-spin representations for which \ref{half} does not hold. Here, in Table II, we list the symmetry allowed form for all non-centrosymmetric point groups and for all possible pseudo-spin representations. Table II was found using the spinor representations of the double groups listed in Ref.~\onlinecite{kos} and the labelling of the representations is taken from there. To aid in understanding the representation labels, we have provided the basis function for the state $|\uparrow\rangle$, the state $|\downarrow\rangle$ is found by applying time-reversal symmetry.  The listed basis functions for these representations are expressed in terms of the usual $|j,m>$ angular momentum basis functions (with total angular momentum $j$ and with $J_z|j,m>=m\hbar |j,m>$). In some cases, the relevant representation is expressed as a direct sum, for example for the point group $C_2$ we have $\Gamma_3\oplus \Gamma_4$. This is a consequence of two different (or two of the same) point group representations that become degenerate due to time reversal symmetry. If time-reversal symmetry is not present, then the representations in the direct sum would not be degenerate.  Also, as mentioned above, the use of $\sigma_i$ specifies that this operator transforms as the usual spin $1/2$ Pauli matrix operator under rotations and the $\tilde{\sigma}_i$ specifies that this is not the case. Finally, we did not include any 4-dimensional spinor representations  for the  groups $T$, $T_d$, $O$. The theory in these cases is more involved and is treated in Ref.~\onlinecite{YPtBiTheor} in the context of half-Heulser superconductors.

\begin{table}
 \begin{tabular}{|c|c|c|c|}
   \hline
   Point Group & Representation  & $|\uparrow\rangle$ &$\mbox{\boldmath$\gamma$}({\bf k})\cdot\vec{\tilde{\sigma}}$\\
   \hline
   $C_1$ & $\Gamma_2$ & $|1/2,1/2\rangle$& $\sum_{i,j=x,y,z}a_{i,j}k_i\sigma_j$\\
   $C_2$ & $\Gamma_3\oplus\Gamma_4$&$|1/2,1/2\rangle$ & $\alpha_{zz}k_z\sigma_z+\alpha_{xx}k_x\sigma_x+\alpha_{yy}k_y\sigma_y+\alpha_{xy}k_x\sigma_y+\alpha_{yx}k_y\sigma_x$\\
   $C_s$ & $\Gamma_3\oplus\Gamma_4$ &$|1/2,1/2\rangle$  &$\alpha_{xz}k_x\sigma_z+\alpha_{yz}k_y\sigma_z+\alpha_{zx}k_z\sigma_x+\alpha_{zy}k_z\sigma_y$\\
    $D_2$& $\Gamma_5$ & $|1/2,1/2\rangle$&$\alpha_{xx}k_x\sigma_x+\alpha_{yy}k_yS_y+\alpha_{zz}k_z\sigma_z$\\
   $C_{2v}$ &$\Gamma_5$  & $|1/2,1/2\rangle$ & $\alpha_{xy}k_x\sigma_y+\alpha_{yx}k_y\sigma_x+\alpha_3k_xk_yk_z\sigma_z$\\
   $C_4$ & $\Gamma_5\oplus\Gamma_6$ & $|1/2,1/2\rangle$ &$\alpha_{xx}(k_x\sigma_x+k_y\sigma_y)+\alpha_{xy}(k_x\sigma_y-k_y\sigma_x)+\alpha_{zz}k_z\sigma_z$ \\
    & $\Gamma_7\oplus\Gamma_8$ & $|3/2,3/2\rangle$ &$\alpha_{xx}(k_x\sigma_x+k_y\sigma_y)+\alpha_{xy}(k_x\sigma_y-k_y\sigma_x)+\alpha_{zz}k_z\sigma_z$ \\
    $S_4$ & $\Gamma_5\oplus\Gamma_6$ & $|1/2,1/2\rangle$ &$\alpha_{xx}(k_x\sigma_x+k_y\sigma_y)+\alpha_{xy}(k_x\sigma_y-k_y\sigma_x)+\beta_1k_z(k_x^2-k_y^2)\sigma_z+\beta_2k_zk_xk_y\sigma_z$ \\
    & $\Gamma_7\oplus\Gamma_8$ & $|3/2,3/2\rangle$ &$\alpha_{xx}(k_x\sigma_x+k_y\sigma_y)+\alpha_{xy}(k_x\sigma_y-k_y\sigma_x)+\beta_1k_z(k_x^2-k_y^2)\sigma_z+\beta_2k_zk_xk_y\sigma_z$ \\
    $D_4$ & $\Gamma_6$ & $|1/2,1/2\rangle$ &$\alpha_{xx}(k_x\sigma_x+k_y\sigma_y)+\alpha_{zz}k_z\sigma_z$ \\
    & $\Gamma_7$ & $(x^2-y^2)|1/2,1/2\rangle$ &$\alpha_{xx}(k_x\sigma_x+k_y\sigma_y)+ \alpha_{zz}k_z\sigma_z$\\
    $C_{4v}$ & $\Gamma_6$ & $|1/2,1/2\rangle$ &$\alpha_{xy}(k_x\sigma_y-k_y\sigma_x)+\beta k_zk_xk_y(k_x^2-k_y^2)\sigma_z$ \\
    & $\Gamma_7$ & $(x^2-y^2)|1/2,1/2\rangle$ &$\alpha_{xy}(k_x\sigma_y-k_y\sigma_x)+\beta k_zk_xk_y(k_x^2-k_y^2)\sigma_z$\\
    $D_{2d}$ & $\Gamma_6$ & $|1/2,1/2\rangle$ &$\alpha_{xx}(k_x\sigma_x-k_y\sigma_y)+\beta k_z(k_x^2-k_y^2)\sigma_z$ \\
    & $\Gamma_7$ & $(x^2-y^2)|1/2,1/2\rangle$ &$\alpha_{xx}(k_x\sigma_x-k_y\sigma_y)+\beta k_z(k_x^2-k_y^2)\sigma_z$\\
    $C_3$ & $\Gamma_4\oplus\Gamma_5$ & $|1/2,1/2\rangle$ &$\alpha_{xx}(k_x\sigma_x+k_y\sigma_y)+\alpha_{xy}(k_x\sigma_y-k_y\sigma_x)+\alpha_{zz}k_z\sigma_z$ \\
    & $\Gamma_6\oplus \Gamma_6$ & $|3/2,3/2\rangle$ &$\alpha_{x}k_z\tilde{\sigma}_x+\alpha_{y}k_z\tilde{\sigma}_y+\alpha_{zz}k_z\sigma_z$ \\
     $D_3$ & $\Gamma_4$ & $|1/2,1/2\rangle$ &$\alpha_{xx}(k_x\sigma_x+k_y\sigma_y)+\alpha_{zz}k_z\sigma_z$ \\
    & $\Gamma_5\oplus \Gamma_6$ & $|3/2,3/2\rangle-i|3/2,-3/2\rangle$ & $\alpha_{x}k_z\tilde{\sigma}_x+\alpha_{y}k_z\tilde{\sigma}_y+\alpha_{zz}k_z\sigma_z$\\
     $C_{3v}$ & $\Gamma_4$ & $|1/2,1/2\rangle$ &$\alpha_{xy}(k_x\sigma_y-k_x\sigma_y)+\beta k_y(3k_x^2-k_y^2)\sigma_z$ \\
    & $\Gamma_5\oplus \Gamma_6$ & $|3/2,3/2\rangle-i|3/2,-3/2\rangle$ &$\beta_xk_y(3k_x^2-k_y^2)\tilde{\sigma}_x+\beta_yk_y(3k_x^2-k_y^2)\tilde{\sigma}_y+ \beta_zk_y(3k_x^2-k_y^2)\sigma_z$\\
   $C_6$ & $\Gamma_7\oplus \Gamma_8$ & $|1/2,1/2\rangle$ &$\alpha_{xx}(k_x\sigma_x+k_y\sigma_y)+\alpha_{xy}(k_x\sigma_y-k_y\sigma_x)+\alpha_{zz}k_z\sigma_z$ \\
    & $\Gamma_9\oplus \Gamma_{10}$ & $|5/2,5/2\rangle$ & $\alpha_{xx}(k_x\sigma_x+k_y\sigma_y)+\alpha_{xy}(k_x\sigma_y-k_y\sigma_x)+\alpha_{zz}k_z\sigma_z$\\
    & $\Gamma_{11}\oplus \Gamma_{12}$ & $|3/2,3/2\rangle$ &$\beta_1k_y(3k_x^2-k_y^2)\tilde{\sigma}_x+\beta_2k_x(3k_y^2-k_x^2)\tilde{\sigma}_x+\beta_3k_y(3k_x^2-k_y^2)\tilde{\sigma}_y$\\ &&&$+\beta_4k_x(3k_y^2-k_x^2)\tilde{\sigma}_y+\alpha_{zz}k_z\sigma_z$ \\
     $C_{3h}$ & $\Gamma_7\oplus \Gamma_8$ & $|1/2,1/2\rangle$ &$\beta_1k_z[(k_x^2-k_y^2)\sigma_x+2k_xk_y\sigma_y]+\beta_2k_z[-2k_xk_y\sigma_x+(k_x^2-k_y^2)\sigma_y]$ \\&&&$+\beta_3k_x(3k_y^2-k_x^2)\sigma_z+\beta_4k_y(3k_x^2-k_y^2)\sigma_z$\\
    & $\Gamma_9\oplus \Gamma_{10}$ & $|5/2,5/2\rangle$ &$\beta_1k_z[(k_x^2-k_y^2)\sigma_x+2k_xk_y\sigma_y]+\beta_2k_z[-2k_xk_y\sigma_x+(k_x^2-k_y^2)\sigma_y]$ \\&&&$+\beta_3k_x(3k_y^2-k_x^2)\sigma_z+\beta_4k_y(3k_x^2-k_y^2)\sigma_z$\\
     & $\Gamma_{11}\oplus \Gamma_{12}$ & $|3/2,3/2\rangle$ & $\alpha_xk_z\tilde{\sigma}_x+\alpha_yk_z\tilde{\sigma}_y+\beta_1k_x(3k_y^2-k_x^2)\sigma_z+\beta_2k_y(3k_x^2-k_y^2)\sigma_z$\\
      $D_6$ & $\Gamma_7$ & $|1/2,1/2\rangle$ &$\alpha_{xx}(k_x\sigma_x+k_y\sigma_y)+\alpha_{zz}k_z\sigma_z$ \\
    & $\Gamma_8$ & $y(y^2-3x^2)|1/2,1/2\rangle$ &$\alpha_{xx}(k_x\sigma_x+k_y\sigma_y)+\alpha_{zz}k_z\sigma_z$ \\
    & $\Gamma_9$ & $|3/2,3/2\rangle$ & $\beta_1k_x(k_x^2-3k_y^2)\tilde{\sigma}_x+\beta_2k_y(k_y^2-3k_x^2)\tilde{\sigma}_y+\alpha_{zz}k_z\sigma_z$\\
      $C_{6v}$ & $\Gamma_7$ & $|1/2,1/2\rangle$ & $\alpha_{xy}(\sigma_xk_y-\sigma_yk_x)+\beta k_z(3k_x^5k_y-10k_x^3k_y^3+3k_xk_y^5)\sigma_z$\\
    & $\Gamma_8$ & $x(x^2-3y^2)|1/2,1/2\rangle$ &$\alpha_{xy}(\sigma_xk_y-\sigma_yk_x)+\beta k_z(3k_x^5k_y-10k_x^3k_y^3+3k_xk_y^5)\sigma_z$ \\
    & $\Gamma_9$ & $|3/2,3/2\rangle$ & $\beta_1k_y(k_y^2-3k_x^2)\tilde{\sigma}_x+\beta_2k_x(k_x^2-3k_y^2)\tilde{\sigma}_y+\beta_3k_z(3k_x^5k_y-10k_x^3k_y^3+3k_xk_y^5)\sigma_z$\\
      $D_{3h}$ & $\Gamma_7$ & $|1/2,1/2\rangle$ &$\beta_1k_z[(k_x^2-k_y^2)\sigma_x-2k_xk_y\sigma_y]+\beta_2k_x(k_x^2-3k_y^2)\sigma_z$\\
    & $\Gamma_8$ & $zx(x^2-3y^2)|1/2,1/2\rangle$ & $\beta_1k_z[(k_x^2-k_y^2)\sigma_x-2k_xk_y\sigma_y]+\beta_2k_x(k_x^2-3k_y^2)\sigma_z$\\
    & $\Gamma_9$ & $|3/2,3/2\rangle$ & $\alpha k_z\tilde{\sigma}_x+\beta_1k_z(3k_x^5k_y-10k_x^3k_y^3+3k_xk_y^5)\tilde{\sigma}_y+\beta_2k_x(k_x^2-3k_y^2)\sigma_z$\\
    $T$ & $\Gamma_5$&$|1/2,1/2\rangle$&$\alpha_{xx}(k_x\sigma_x+k_y\sigma_y+k_z\sigma_z)$\\
    $O$ &$\Gamma_6$&$|1/2,1/2\rangle$&$\alpha_{xx}(k_x\sigma_x+k_y\sigma_y+k_z\sigma_z)$\\
    &$\Gamma_7$& $xyz|1/2,1/2\rangle$&$\alpha_{xx}(k_x\sigma_x+k_y\sigma_y+k_z\sigma_z)$\\
     $T_d$ &$\Gamma_6$&$|1/2,1/2\rangle$&$\beta[k_x(k_y^2-k_z^2)\sigma_x+k_y(k_z^2-k_x^2)\sigma_y+k_z(k_x^2-k_y^2)\sigma_z]$\\
    &$\Gamma_7$& $f(x)|1/2,1/2\rangle$&$\beta[k_x(k_y^2-k_z^2)\sigma_x+k_y(k_z^2-k_x^2)\sigma_y+k_z(k_x^2-k_y^2)\sigma_z]$\\

   \hline
 \end{tabular}

\caption{Form of ASOC for all non-centrosymmetric point groups expanded for ${\bf k}$ near the $\Gamma$ point. The labels for the spinor representations are those used in Ref.~\onlinecite{kos} (as explained in more detail in the text, the direct sum of representations refers to two point group representations that become degenerate under time-reversal symmetry). The basis function $|\uparrow\rangle$ for these representations are given in the column with this name, the corresponding $|\downarrow\rangle$ is found by applying time-reversal symmetry. The form of the spin-orbit coupling is given in the column labelled $\mbox{\boldmath$\gamma$}({\bf k})\cdot\vec{\tilde{\sigma}}$, the notation $\sigma_i$ refers to pseudo-spin operators that share the same transformation properties as the usual spin 1/2 Pauli matrices and $\tilde{\sigma}_i$ denote pseudo-spin operators for which this is not the case. Some detailed notes on constructing this table: for the group $D_{2d}$, we took the two-fold rotation element about $x$ (usually denoted $C_{2x}$) as a rotation element; for the group $C_{3v}$, we took the mirror reflection with normal to the $\hat{y}$ axis (usually denoted $\sigma_y$ and not to be confused with the Pauli matrices discussed elsewhere) as an element; for the group $D_{3h}$,  the mirror reflection with normal to the $\hat{x}$ axis (usually denoted $\sigma_x$ and not to be confused with the Pauli matrices discussed elsewhere) as an element; and $f(x)=[(y^2-z^2)x^4+(z^2-x^2)y^4+(x^2-y^2)z^4]$. }
\label{Table1}
\end{table}

While the form for the ASOC is useful for understanding the physics near the $\Gamma$ point, in many cases the ASOC is required throughout the first Brillouin zone. This can be found by imposing \begin{equation}
\mbox{\boldmath$\gamma$}({\bf k})=\sum_{n,m,l}{\bf a}_{n,m,l}\sin[{\bf k}\cdot (n{\bf a}_1+m{\bf a}_2+l{\bf a}_3)]
\end{equation}
where ${\bf a}_i$ are the primitive translation vectors of the lattice. This expansion ensures that $\mbox{\boldmath$\gamma$}({\bf k})=\mbox{\boldmath$\gamma$}({\bf k}+{\bf G}_i)$ where ${\bf G}_i$ is a reciprocal lattice vector.

As a specific example, consider the space group $P\overline{6}m2$ with point group $D_{3h}$ in two-dimensions. This applies to single layer NbSe$_2$, which is superconducting \cite{IsingNbSe2} and to MoS$_2$ which is superconducting under electric field \cite{IsingMoS2}. Both these materials have been discussed earlier in this article. This group can be visualized  as the direct product of the point group and the translation group spanned by the two translation vectors ${\bf a}_1=a(1/2,-\sqrt{3}/2,0)$, ${\bf a}_2=a(1/2,\sqrt{3}/2,0)$.  Explicitly, we will consider pseudo-spin 1/2 bands, which near the $\Gamma$ point correspond to either the $\Gamma_7$ or the $\Gamma_8$ representations of $D_{3h}$. Consequently, we treat $\tilde{\sigma}_i=\sigma_i$ as the usual Pauli matrices (complete with the usual rotation properties). For our choice of basis functions, the point group $D_{3h}$ is generated by the elements $\sigma_h$ (mirror plane reflection with normal  $\hat{z}$), $C_3$ (a three-fold rotation about the $z$-axis), and $C_{2y}$ (a two-fold rotation about the $y$ axis). For this point group $\sigma_z$ and $\{\sigma_x,\sigma_y\}$ belong to different representations, so we consider them separately. Let us consider $\mbox{\boldmath$\gamma$}({\bf k})=\gamma_z\hat{z}$. Then under the point group generators: $C_{2y}$ $\sigma_z\rightarrow -\sigma_z$, $\sigma_h$ $\sigma_z\rightarrow -\sigma_z$, and $C_3$ $\sigma_z\rightarrow \sigma_z$.  To construct $\mbox{\boldmath$\gamma$}({\bf k})$, we also need the action of the point group elements on the vector $n{\bf a}_1+m{\bf a}_2$, which can be characterized by the action on $(n,m)$. Under $\sigma_h$ $(n,m)\rightarrow (n,m)$, under $C_3$ $(n,m)\rightarrow (-m,n-m)$ and under $C_{2y}$ $(n,m)\rightarrow(-m,-n)$. Consequently, writing $\mbox{\boldmath$\gamma$}({\bf k})=\sum_{n,m} a_{z,n,m}\sin[{\bf k}\cdot(n{\bf a}_1+m{\bf a}_2)]$, yields the symmetry relations $a_{z,n,m}=a_{z,n,m}$ (from $\sigma_h$), $a_{z,n,m}=a_{z,-m,n-m}$ (from $C_3$) and $a_{z,n,m}=-a_{z,-m,-n}$ (from $C_{2y}$). Starting from nearest neighbors, that is assuming $a_{z,1,0}\ne 0$, yields $a_{z,1,0}=a_{z,0,1}=-a_{z,1,1}$ or
\begin{equation}
\mbox{\boldmath$\gamma$}_1({\bf k})=\hat{z}\Big( \sin[{\bf k}\cdot {\bf a}_1]+\sin[{\bf k}\cdot {\bf a}_2]-\sin[{\bf k}\cdot ({\bf a}_1+{\bf a}_2)]\Big)
\label{g-1}
\end{equation}
where $k_i$ is in units $2\pi/a$.
Near the $\Gamma$ point Eq.~\ref{g-1} becomes proportional to $\hat{z}k_x(3k_y^2-k_x^2)$. Note how $k_x$ and $k_y$ are switched with respect to Table II. This is a consequence of the choice $C_{2y}$ as an element of $D_{3h}$, while Table II has used $C_{2x}$ as the corresponding element (this was done intentionally to highlight that care must be taken in defining the relative orientation of the rotation axes and the translation vectors). Near the ${\bf K}$ point, Eq.~\ref{g-1} becomes a constant (a constant is allowed because the ${\bf K}$ point is not invariant under parity symmetry). As discussed eariler, in the NbSe$_2$ and MoS$_2$ literature, this term has been labeled an Ising spin-orbit coupling \cite{IsingNbSe2,IsingMoS2}. Note that Eq.~\ref{g-1} also applies to the recently discovered quasi-1D material K$_2$Cr$_3$As$_3$ \cite{K2Cr3As3Rep}.

As emphasized in Refs.~\onlinecite{min05,sam09}, the structure and  magnitude of $\mbox{\boldmath$\gamma$}({\bf k})$ at the Fermi surface can be measured through de-Haas-van Alphen (dHvA) measurements. The primary effect follows from the fact that  in the presence of ${\mbox{\boldmath$\gamma$}}({\bf k})$, the band energies become $\xi_{\pm} ({\bf k})=\xi({\bf k})\pm|\mbox{\boldmath$\gamma$}({\bf k})|$. This leads to a difference in the area of the extremal orbits for the two bands, leading to a difference in the dHvA frequencies. For example in LaRhSi$_3$, the spin-orbit coupling was measured this way to be approximately 200 K \cite{CeRhSi3FS2}. The interplay with a Zeeman field and the spin-orbit coupling and the subsequent distortion of the Fermi surface can lead to additional effects in dHvA measurements, including phase shifts in in the dHvA signal and a non-linear field dependence of the dHvA frequencies.

Finally, we note that ASOC also gives rise to a non-trivial topological Berry curvature in some cases and this plays a role in the context of topological superconductivity discussed later \cite{sam09,sam15}.

\section{Antisymmetric spin orbit coupling as a parity symmetry breaking field on superconductivity}

In the presence of the ASOC, which splits the spin-degeneracy at each momentum ${\bf k}$,  it follows that the correct description of the
superconducting state is a two-band theory (this theory is described in Section~\ref{general-apsects}). However, many consequences of
broken parity symmetry, most importantly with respect to the role of magnetic fields, can readily be seen by treating the ASOC as a symmetry breaking field to the superconducting state \cite{CePt3SiMnSi}. This is similar in spirit to the treatment of a Zeeman field in the theory of superconductivity. This treatment reveals a few key results that apply in the large ASOC limit. The first of which is presented in this subsection. In particular, it is found that only a single spin-triplet state survives in the limit of strong ASOC (where strong here means relative to the pairing gap). Prior to presenting the formal results, a qualitative explanation is given.

Within the pseudospin basis discussed in the previous section, the gap function can be written as \cite{sig91}
\begin{equation}
\Delta=i\tilde{\sigma}_y[\psi({\bf k})+{\bf d}({\bf k})\cdot\tilde{\mbox{\boldmath$\sigma$}}]=\left(
    \begin{array}{cc}
    -d_x({\bf k})+id_y({\bf k})& \psi({\bf k})+d_z({\bf k}) \\
     -\psi({\bf k})+d_z({\bf k})& d_x({\bf k})+id_y({\bf k}) \\
    \end{array}
  \right).
  \label{gap1}
\end{equation}
While the pseudo-spin singlet gap function $\psi({\bf k})$ has often been discussed, the spin-triplet gap function ${\bf d}({\bf k})$ is not as familiar. Under a pseudospin rotation, the  ${\bf d}({\bf k})$ rotates as a vector. Eq.~\ref{gap1} is useful in understanding which spin-triplet state is stable in the presence of the ASOC. In particular, if the pseudospin quantization axis is chosen to be along the ${\bf d}$-vector (so ${\bf d}=d_z\hat{z}$), then Eq.~\ref{gap1} shows there is only pairing between opposite spins. When inversion symmetry is removed, the ASOC chooses the  particular spin quantization axis $\hvg$. With this quantization axis, time-reversal symmetry ensures that the up state at ${\bf k}$ is degenerate with the down state at $-{\bf k}$. However, the up state at ${\bf k}$ is not degenerate with the up state at $-{\bf k}$ (an example of this is shown in Fig.~1). This implies that once the ASOC is sufficiently large, the only degenerate states available to create Cooper pairs must have opposite spin. Eq.~\ref{gap1} then implies that only the component of ${\bf d}({\bf k})$ that is parallel to $\hvg=\vg({\bf k})/|\vg({\bf k})|$ is non-zero (note that this argument ignores the small regions in momentum space where $\vg({\bf k})=0$). Hence ${\bf d}({\bf k})$ is parallel to $\hvg$ when the ASOC  is sufficiently strong. In the following, we develop this result more formally.

The approach used in this section is to consider a parity invariant superconductor and to break parity symmetry solely through the ASOC. For convenience we work in the weak-coupling limit. When parity (and time-reversal) symmetries are present, superconductivity can be included through the following pairing interaction (in this section we examine only spatially homogeneous superconducting states):
\begin{equation}
{\cal H}_{pair} =
\frac{1}{2} \sum_{\vk, \vk', s_i} V_{s_1s_2,s_2's_1'}(\vk,\vk')
 c^{\dag}_{\vsk,s_1} c^{\dag}_{-\vsk,s_2} c_{-\vsk',s_2'} c_{\vsk',s_1'}.
\end{equation}
Symmetry dictates that \cite{sig91} the pairing interaction can be broken up into separate pseudo-spin singlet  and  pseudo-spin triplet parts and that these parts can be characterized by interactions that quantify the $T_c$ for different irreducible representations of the point group of the crystal.

In particular,
\begin{equation}
V_{s_1s_2,s_2's_1'}(\vk,\vk')=\sum_{\Gamma,i} \Big \{v_{s,\Gamma}\psi_{\Gamma,i}(\vk)\psi^*_{\Gamma,i}(\vk')(i\tilde{\sigma_y})_{s_1,s_2}(i\tilde{\sigma_y})^{\dagger}_{s_2',s_1'}
+v_{t,\Gamma}[{\bf d}_{\Gamma,i}(\vk)\cdot (i \vec{\tilde{\sigma}})_{s_1,s_2}][{\bf d}^*_{\Gamma,i}(\vk')\cdot (i \vec{\tilde{\sigma}})^{\dagger}_{s_2',s_1'}]\Big \}
\end{equation}
where $\Gamma,i$ describes the particular irreducible representation $\Gamma$ and $i$ the basis of that representation, $v_{s,\Gamma}$ represents singlet pairing with gap function basis functions $\psi_{\Gamma,i}({\bf k})$, $v_{t,\Gamma}$ represents triplet pairing with gap basis functions ${\bf d}_{\Gamma,i}(\bf k)$. Of importance to note is that the Pauli matrices $\tilde{\sigma}$ that appear in the interaction are the same as those that appeared in Eq.~\ref{sigma} in the context of the ASOC. This follows by noting that Cooper pairs are constructed by pairing single electron states with momenta ${\bf k}$ and the time-reversed momenta $-{\bf k}$ and that $\mathcal{T}|{\bf k},\alpha>$ has the same transformation properties under symmetry operations as $<{\bf k}, \alpha|$.

To determine the role of the ASOC, we begin with a calculation of $T_c$ for the singlet and triplet states. A usual BCS weak coupling approach yields the following linearized gap equation (this approach is nicely presented in Ref.~\onlinecite{sig91})
\begin{equation}
\Delta_{ss'} (\vk) = - k_B T \sum_{n, \vsk'} \sum_{s_1,s_2}
V_{\vsk,\vsk'} G^0_{ss_1}(\vk', i\omega_n)  \Delta_{s_1,s_2}
(\vk') G^0_{s's_2} ( - \vk', - i\omega_n)
\end{equation}
where $ G^0(\vk, i \omega_n)$ is the normal state Greens function and is given by \begin{equation}
G^0(\vk, i \omega_n) = G_{+} (\vk, i \omega_n) \sigma_{0} +
  \hvg({\bf k}) \cdot   \vsig G_-(\vk, i \omega_n)
\end{equation}
where $ \sigma_0 $ is the unit matrix and
\begin{equation}
G_{\pm} (\vk, i \omega_n) = \frac{1}{2} \left[ (i \omega_n -
    \epsilon_{\vsk,+})^{-1}  \pm  (i \omega_n - \epsilon_{\vsk,-} )^{-1}
    \right] \; ,
\end{equation}
$ \epsilon_{\vsk,\pm} = \xi_{\vsk} \pm | \vg({\bf k})| $ and $
\hvg({\vsk}) = \vg({\vsk}) / | \vg({\vsk}) | $ ($ | \vg | =
\sqrt{\vg^2} $) \cite{Edel96,BauerNCS}.

The gap function is expressed as $\Delta(\vk) = \{ \psi(\vk)
\sigma_0 + \vd (\vk) \cdot \tilde{\vsig} \} i \tilde{\sigma}_y $ and we assume in this Section that the gap magnitudes are the same on both the spin-split Fermi surfaces (which is consistent with the assumption that the ASOC is much smaller than the chemical potential).
The linearized gap
equations become
\begin{equation}
\psi (\vk) = - k_B T \sum_{n, \vsk'} V_{\vsk, \vsk'} \Big\{
[G_+ G_+ + G_- G_- ]\psi (\vk')  + [G_+ G_- + G_- G_+] \hvg({\vsk'}) \cdot \vd ( \vk') \Big \}
\end{equation}
and
\begin{equation}
\vd (\vk) = - k_B T \sum_{n, \vsk'} V_{\vsk, \vsk'} \Big\{
[G_+ G_+ +
 G_- G_- ] \vd (\vk')  + 2 G_- G_- [\hvg({\vsk'}) (\hvg({\vsk'})
\cdot \vd ( \vk') ) - \vd(\vk') ]
+ [G_+ G_- + G_- G_+] \hvg({\vsk'}) \psi(\vk')
\Big\}
\end{equation}
where $ G_a G_b =
G_a (\vk, i \omega_n) G_b (-\vk, - i \omega_n ) $ with $ a,b=\pm $. In general, since $\vg({\bf k})\ne 0$,
the spin-singlet and triplet channels are
coupled \cite{NCSGorkov}. This
coupling vanishes when there is particle-hole asymmetry and when the density of states on the two spin-split Fermi surface sheets are equal. In particular, the singlet-triplet coupling is of the order $ \alpha / \epsilon_F  \ll
1 $ where we have set $\alpha$ as the typical energy scale of the ASOC, that is $\alpha=\sqrt{\langle \vg({\bf k})^2\rangle_{{\bf k}}}$, the magnitude of the ASOC averaged over the Fermi surface. In this Section we ignore this coupling and consider
the singlet and triplet pairing channels separately. This precludes the mixing of spin-singlet and spin-triplet superconductivity and this possibility is discussed in more detail later in this article. By expanding the interaction, $\psi({\bf k})$, and $\vd(\vk)$ in a basis of irreducible representations and using the orthogonality of different irreducible representations, the equation for the transition temperature can be found \cite{sig91}. In the following expressions for $T_c$, we drop the label specifying which irreducible representation $T_c$ and the gap functions belong to.

For spin-singlet pairing the transition temperature
($T_c$) is given by
\begin{equation}
\ln  \left( \frac{T_c}{T_{cs}}\right) = O \left(
\frac{\alpha^2}{\epsilon_F^2}  \right) \;.
\end{equation}
The main conclusion is that $T_c$ is essentially from $T_{cs}$, which is the transition temperature when $\vg=\alpha=0$. For triplet pairing $T_c$ is given by
\begin{equation}
\ln \left( \frac{T_c}{T_{ct}}\right) = 2 \langle \{ |\vd (\vk)|^2
- | \hvg({\vsk})  \cdot \vd (\vk) |^2  \} f( \rho_{\vk})
  \rangle_{\vsk} + O \left(
  \frac{\alpha^2}{\epsilon_F^2} \right)
\label{tc-triplet}
\end{equation}
where $T_{ct}$ is the transition temperature when $\vg=\alpha=0$,  $\rho_{\vk}=|\vg(\vk)|/(\pi k_B T_c)$, and the function $ f (\rho) $ is defined as
\begin{equation}
f(\rho) = Re \; \sum_{n=1}^{\infty} \left( \frac{1}{2 n -1 + i \rho} -
\frac{1}{2n -1} \right) \; .
\label{tct}
\end{equation}
In contrast to the spin-singlet case, $T_c$ for spin-triplet pairing is generally strongly suppressed  once
$\alpha\gtrapprox k_B T_c$. However, there remains a single spin-triplet pairing channel  for which $T_c$ is not suppressed, that is when $ \vd(\vk) \parallel \hvg({\vsk})$, such a pairing channel is often called the protected spin-triplet pairing state. Remarkably, as will be made clear in the following, many of the magnetic properties of this spin-triplet pairing state are adopted by the spin-singlet state, even when no spin-triplet order co-exists with the spin-singlet order. We note that $ \vd(\vk) \parallel \hvg({\vsk})$ applies not just for the isotropic pairing channel, but also to pairing in other symmetry channels. In the latter case, $\vd(\vk) =f(\vk)\hvg({\vsk})$ where $f(\vk)$ can have nodes that are dictated by the pairing symmetry (and are unrelated to the form of $\vg(\vk)$).

\subsection{Role of Zeeman field} \label{Zeeman-sec}

It is useful to consider the usual spin-singlet case in which the Zeeman field separates the energies of the up and down spins and hence pairing (with zero momentum Cooper pairs) is no longer between degenerate states. This implies that the formation of Cooper pairs, while still gaining potential energy, now also have an energy cost. Once this energy cost becomes greater than the gap, superconductivity is unstable, leading to paramagnetic pair breaking. When inversion symmetry is broken, the single particle spins are no longer free to rotate but are pinned by the ASOC, with spins oriented along $\pm \hvg(\vk)$. If this pinning energy is much larger than the Zeeman energy, then when the applied field is perpendicular to this spin, there will be no energy cost in forming Cooper pairs. In this case, we expect that superconductivity will be more stable than when inversion symmetry is present. This is indeed what the detailed calculations below reveal in the strong ASOC limit.

Before proceeding, it is worthwhile pointing out that if $\tilde{\sigma}_i$ does not belong to the same representation as $\sigma_i$ (this is discussed in detail in Section III), then the usual Zeeman coupling will not exist. For example, consider the  point group $D_{3h}$.
The coupling to the Zeeman field is different for the different spinor representations at the $\Gamma$ point. For $\Gamma_7$ and $\Gamma_8$ this coupling is the usual (we have ignored for simplicity the possibility of anisotropy between in-plane and c-axis fields)
\begin{equation}
H_{Zeeman}=\mu_B \sum_k {\bf H}\cdot \vec{\sigma}({\bf k})
\label{eqZeeman}
\end{equation}
while for the $\Gamma_9$ representation it is
\begin{equation}
H^{\Gamma_9}_{Zeeman}=\mu_B \sum_k H_z \tilde{\sigma}_z({\bf k}).
\end{equation}
Consequently, for the $\Gamma_9$ representation, we do not expect an in-plane Zeeman field to play a role. This may play a role in understanding the upper critical field behavior observed in A$_2$Cr$_3$As$_3$ superconductors discussed in Section~\ref{ACrAs}.

In the following we assume that the coupling to the Zeeman field is the usual coupling given by Eq.~\ref{eqZeeman}, the results for when this is not the case can be found by setting the corresponding components of the magnetic field to zero (for example in the case of the $\Gamma_9$ representation, $H_x=H_y=0$). Typically, a Zeeman field suppresses spin-singlet superconductivity, but does not necessarily suppress spin-triplet superconductivity. The following shows that breaking parity symmetry qualitatively changes this picture. In particular, spin-singlet superconductivity can be robust against Zeeman fields, depending on the field orientation. Here we do not consider the orbital effect of the field, which generically suppresses superconductivity. We also do not consider any finite momentum pairing states, these are discussed later in Section~\ref{helical}.

We replace $ \vg({\vk})
\to \tilde{\vg}({\vk}) =\vg({\vk}) - \vh $ with $ \vh
= \mu_B \vH $ (note: $ \tilde{\vg}({-\vsk}) \neq -
\tilde{\vg}({\vsk}) $). For
spin-singlet pairing, $ \psi (\vk) $,  (ignoring any induced
spin-triplet pairing) $T_c$ becomes \cite{CePt3SiMnSi}
\begin{equation}
{\rm ln} \left( \frac{T_c}{T_{cs}} \right) =\left\langle
|\psi(\vk)|^2 \left\{\left[f( \rho_{\vsk}^- )+f(
\rho_{\vsk}^+)\right]+ \frac{ \vg({\vsk})^2 -
  \vh^2}{[(\vg({\vsk}) + \vh)^2 (\vg({\vsk}) - \vh)^2]^{1/2} }
  \left[f( \rho_{\vsk}^- )-f( \rho_{\vsk}^+)\right] \right\}\right\rangle_{\vsk}
  \label{para}
\end{equation}
with $ \rho_{\vk}^{\pm} = |\vg({\vsk}) + \vh |/ 2\pi k_B T_c
\pm|\vg({\vsk}) - \vh |/ 2\pi k_B T_c $. In the limit that $\vg(\vk)=0$, Eq.~\ref{para} describes the usual paramagnetic pair breaking. However, this changes dramatically when $\vg(\vk)\ne 0$. For example, if it is possible
to choose $ \vh \perp \vg({\vsk}) $ for all $ \vk$ then in the limit $T_c(h)/T_{cs}<<1$, the
critical field obeys $h'^2\ln h'=-\langle \vg(\vk)^2\rangle_{\bf k}\ln
(T_c/T_{cs})$; with $ h'= |\vh|/\pi k_B T_{cs} $. This implies that
the critical field diverges as $T\rightarrow 0$. In general, this divergence is suppressed by impurity scattering.

For spin-triplet pairing \cite{CePt3SiMnSi}
\begin{equation} \begin{array}{ll}
\displaystyle {\rm ln} \left( \frac{T_c}{T_{ct}} \right) =
&\displaystyle  \left\langle | \vd (\vk ) |^2 \left\{
\left[f(\rho_{\vk}^+ )+f(\rho_{\vk}^- )\right]+\frac{
\vg({\vsk})^2 - \vh^2}{[ (
     \vg({\vsk}) + \vh)^2 ( \vg({\vsk}) - \vh)^2 ]^{1/2}}
 \left[
f(\rho_{\vk}^+ )-f(\rho_{\vk}^- )\right]\right\} \right\rangle_{\vsk} \\
& \displaystyle + 2 \left\langle \left[f(\rho_{\vsk}^+
)-f(\rho_{\vsk}^- ) \right]\frac{ | \vh \cdot
  \vd(\vk) |^2 -
   | \vg({\vsk}) \cdot \vd(\vk) |^2}{[ (
   \vg({\vsk}) + \vh)^2 (  \vg({\vsk}) - \vh)^2 ]^{1/2}}
\right\rangle_{\vsk}
\end{array}
\label{d}
\end{equation}
When $\vg(\vk)=0$, then Eq.~\ref{d} reproduced the known results for superconductors with parity symmetry. That is there is no suppression of superconductivity, provided $
\vd(\vk) \cdot \vh = 0 $ can be found for all $ \vk $. When $\vg(\vk)\ne 0$, then according
to Eq.(\ref{d}) there is no suppression of $T_c$, if for all $ \vk
$ $ \vh \perp \vd(\vk) $ and $ \vd(\vk)
\parallel \vg({\vsk}) $.

It is instructive to compare the singlet and triplet cases in the large spin-orbit limit $\alpha=\sqrt{\langle \vg(\vk)^2\rangle_{\bf k}}>>\mu_B H$ (where we ignore the small regions in momentum space for which the ASOC vanishes), for spin-singlet superconductivity
\begin{equation}
\ln \frac{T_c}{T_c^0}=2\langle |\psi({\bf k})|^2f(\frac{\hvg({\vsk}) \cdot{\bf h}}{\pi T_c}\rangle_{k}
\end{equation}
For the spin-triplet case, if ${\bf d}$ and $\vg$ are parallel, and the ASOC is much larger than the Zeeman field, then the paramagnetic limiting field is given by
\begin{equation}
\ln \frac{T_c}{T_c^0}=2\langle |{\bf d}({\bf k})|^2f(\frac{\hvg({\vsk}) \cdot{\bf h}}{\pi T_c}\rangle_{k}
\end{equation}
where it has been assumed that $\langle |\psi({\bf k})|^2\rangle_{\bf k}=\langle |{\bf d}({\bf k})|^2\rangle_{\bf k}=1$. The key point is that the equation for both the singlet and triplet pairing state become the same, indicating that
the observation of a critical field exceeding the Pauli critical field does not provide an indication of spin-triplet superconductivity.

Qualitatively the above predictions are in agreement with experimental results on MoS$_2$ \cite{IsingMoS2}, NbSe$_2$ \cite{IsingNbSe2}, and CeRhSi$_3$ \cite{CeRhSi3Hc2}. We also note that the predicted critical field behavior also applies to materials that are {\it locally non-centrosymmetric} . Locally non-centrosymmetric refers to structures that have subunits that do not have inversion symmetry, and these subunits are related by inversion symmetry. An example is a bi-layer material where each individual layer does not contain inversion symmetry, but the bi-layer as a whole has an inversion center between the two layers.  If the ASOC of the individual subunit is much stronger than the coupling between these subunits, the behavior predicted in this Section will apply. A recent short review of such materials is given in \cite{sig14}.

\subsection{Spin susceptibility}

Spin susceptibility is a popular probe to identify spin-singlet or spin-triplet pairing. However, as this Section shows, when strong ASOC is present, this identification can no longer be made. Indeed, as discussed in the last section of this review, the most reliable means of identifying whether a superconductor is predominantly spin-singlet or spin-triplet is through the observation of topological edge states. The main conclusion drawn here is that in the large ASOC limit and even when pairing is pure spin-singlet, the spin susceptibility resembles that expected for the protected spin-triplet pairing state found above (with ${\bf d}({\bf k})$ parallel to $\hvg(\vk)$). Below, a qualitative understanding of this result is given followed by a more detailed formal analysis.

To gain an understanding of the spin susceptibility, it is useful to contrast the normal state susceptibility with and without the ASOC \cite{Y02}. In particular, when a Zeeman field is applied to a centrosymmetric material, the usual argument finds the magnetic moment resulting from the change in the population of $n_{\uparrow}$ and $n_{\downarrow}$ due to the energy difference between these two states. This gives the usual Pauli susceptibility. However, when the ASOC is present, the usual argument cannot be readily applied. In particular, the ASOC locks the spin along the $\hvg$ direction, so the spin is not free to rotate to be entirely aligned with the field. In the limit that the Zeeman energy is much smaller than the ASOC, it is useful to consider ${\bf H}=H_{\parallel}\hvg+{\bf H}_{\bot}$ with $H_{\parallel}={\bf H}\cdot \hvg$. The contribution of $H_{\parallel}\hvg$ to the susceptibility is similar to that in the usual case. However, the contribution from  ${\bf H}_{\bot}$ is not. Specifically, the energies of the up, $|\hvg,+\rangle$ and down, $|\hvg,-\rangle$, spins are not changed by ${\bf H}_{\bot}$. However, just as there is a matrix element for the $\sigma_x$ operator between the different eigenstates of $\sigma_z$, ${\bf H}_{\bot}$ yields a non-zero matrix element between $|\hvg,+\rangle$ and $|\hvg,-\rangle$. Consequently, while there is no first order energy shift from ${\bf H}_{\bot}$, there is a first order shift in the wavefunctions $|\hvg,\pm \rangle$, this leads to a Van Vleck (or interband) contribution to the susceptibility. The key physical point that matters here is that, in a spin-singlet superconducting state, the Pauli susceptibility is strongly affected by the pairing due to the formation of the gap since a spin-singlet Cooper pair carries no spin. However, if the ASOC is much larger than the gap, the Van Vleck contribution is not affected by the gap, since it stems from an interband process. Consequently, the zero temperature susceptibility for a spin-singlet superconductor will be strongly dependent on the relative orientation of $\hvg$ and ${\bf H}$. Indeed, if these two vectors are orthogonal, then we expect no change in the spin susceptibility for a spin-singlet superconductor. Note that this is what would be expected for a spin-triplet superconductor with a ${\bf d}(\vk)$ vector perpendicular to the applied field when parity symmetry is present. This justifies the statement that in the large ASOC limit, the response of a spin-singlet superconductor and the protected spin-triplet superconductor will exhibit the same spin susceptibility. These qualitative considerations are born out by the more formal approach adopted below.

Here we follow ~\cite{FrigeriSusc}, which generalizes earlier related work found in \cite{bul76,NCSGorkov,Y02}.  We note that our results apply when the ASOC is much smaller  than the Fermi energy. If the ASOC becomes comparable to the chemical potential, then it is not possible to separate the spin and orbital contributions and it becomes reasonable to ask about the physical meaning of the spin susceptibility.

As shown in Ref~\cite{abr62}, the expression for the spin susceptibility is
\begin{equation}
\chi_{ij}^s=-\mu_B^2k_BT\sum_k\sum_{\omega_n}tr\Big \{\hat{\sigma}_i\hat{G}({\bf k},\omega_n)\hat{\sigma_j}\hat{G}({\bf k},\omega_n)- \hat{\sigma}_i\hat{F}({\bf k},\omega_n)\hat{\sigma_j}^{T}\hat{F}^{\dagger}({\bf k},\omega_n)\Big \}
\end{equation}
where $\hat{G}$ and $\hat{F}$ are the normal and anomalous Greens functions. For a spin-singlet gap function $\hat{\Delta}({\bf k})=i\psi({\bf k})\hat{\sigma}_y$ or a spin-triplet superconductor with  $\hat{\Delta}({\bf k})=i{\bf d}({\bf k})\cdot \vsig \hat{\sigma}_y$ with $\vd({\vk})$ constrained to be parallel to $\vg(\vk)$, $\hat{G}$ and $\hat{F}$ have the same form \cite{FrigeriSusc}:
\begin{equation}
\hat{G}(\vk, i \omega_n) = G_{+} (\vk, i \omega_n) \sigma_{0} +
  \hvg({\bf k}) \cdot   \vsig G_-(\vk, i \omega_n)
\end{equation}
where $ \sigma_0 $ is the unit matrix,
\begin{equation}
G_{\pm} (\vk, i \omega_n) = -\frac{1}{2} \left[ \frac{i \omega_n +
    \epsilon_{\vsk,+}}{\omega_n^2+|\Delta({\bf k})|^2+ \epsilon^2_{\vsk,+}} \pm  \frac{i \omega_n +
    \epsilon_{\vsk,-}}{\omega_n^2+|\Delta({\bf k})|^2+ \epsilon^2_{\vsk,-}}
    \right] \; ,
\end{equation}
 \begin{equation}
\hat{F}(\vk, i \omega_n) = [F_{+} (\vk, i \omega_n) \sigma_{0} +
  \hvg({\bf k}) \cdot   \vsig F_-(\vk, i \omega_n)]\hat{\Delta}(\vk),
\end{equation}

\begin{equation}
F_{\pm} (\vk, i \omega_n) = \frac{1}{2} \left[ \frac{1}{\omega_n^2+|\Delta({\bf k})|^2+ \epsilon^2_{\vsk,+}} \pm  \frac{1}{\omega_n^2+|\Delta({\bf k})|^2+ \epsilon^2_{\vsk,-}}
    \right] \; ,
\end{equation}
and $|\Delta({\bf k})|^2=|\psi(\vk)|^2$ for the spin-singlet case and $|\Delta({\bf k})|^2=|\vd(\vk)|^2$ for the spin-triplet case.

In the normal state, this expression yields the usual normal state spin susceptibility $\chi_n=2\mu_B 2 N(0)$ when electron-hole asymmetry is ignored. In the superconducting state, for a spin-singlet superconductor, this expression yields
\begin{equation}
\chi_{ii}^s=\chi_n\left\{1-k_BT\pi\sum_{\omega_n}\left \langle \frac{1-\hvg_i^2(\vk)}{\omega_n^2+|\psi({\bf k})|^2+|\vg({\bf k})|^2}\frac{|\psi({\bf k})|^2}{\sqrt{\omega_n^2+|\psi({\bf k})|^2}} +\frac{\hvg_i^2(\vk)|\psi({\bf k})|^2}{\sqrt{\omega_n^2+|\psi({\bf k})|^2}}\right\rangle_k\right\}
\end{equation}
while for the spin-triplet case, this yields
\begin{equation}
\chi_{ii}^s=\chi_n\left\{1-k_BT\pi\sum_{\omega_n}\left \langle \frac{\hvg_i^2(\vk)|{\bf d}({\bf k})|^2}{\sqrt{\omega_n^2+|{\bf d}({\bf k})|^2}}\right\rangle_k\right\}
\end{equation}
in these expressions $\hvg_i(\vk)=\hvg(\vk)\cdot \hat{x}_i$ where $\hat{x}_i$ is a Cartesian normal vector with direction specified by the index $i$ in $\chi_{ii}^s$.
The role of the Van-Vleck term is apparent in the singlet case, here it is the term multiplied by $1-\hvg_i^2(\vk)$, once $|\psi({\bf k})|^2<<\vg^2(\vk)$, this term drops out and the expressions for the spin-singlet and spin-triplet superconducting states become the same as qualitatively discussed in the opening paragraph. The factor $1-\hvg_i^2(\vk)$ in the spin-singlet case projects out the component of ${\bf H}$ that is perpendicular $\hvg$. Note, as was the case when the critical field was calculated in Section~\ref{Zeeman-sec}, these results are also relevant for locally non-centrosymmetric superconductors \cite{sig14}.

The results presented here apply in the clean limit when correlations are not included and there is some qualitative agreement with measurements on CeIrSi$_3$ (see Fig. 5c) \cite{CeIrSi3NMR2} if the ASOC is assumed to be a Rashba spin-orbit interaction. However, the measured spin-susceptibility is more isotropic than that calculated above. In Ref.~\onlinecite{sam07}, the role of disorder has been considered, and it was shown that this tends to increase the residual value of the spin susceptibility in the limit $T=0$. In Ref.~\onlinecite{fuj07}, the role of correlations in the context of heavy fermion superconductivity have been considered. Here it was found that the Van Vleck contribution to the spin-susceptibility can develop a strong temperature dependence due to the strong energy dependence of the density of states (in the calculation presented above, there is no such energy dependence). This may account for the difference between the theory presented above and the measured results in CeIrSi$_3$.

\section{General aspects of broken parity superconductivity} \label{general-apsects}

While treating broken parity symmetry as a symmetry breaking field provides useful insight and also leads to good qualitative agreement with experiment, there are situations when a more general analysis is needed, for example when considering the mixing of spin-singlet and spin-triplet superconductivity. This Section provides the basis of this more general analysis and follows Ref.~\onlinecite{ser04}. To be explicit, we consider the typical situation in which at a particular ${\bf k}$ point in the BZ, no degeneracies exist (this applies at all ${\bf k}$ for which $\vg$ is not zero). In addition, we consider the realistic situation that the band splitting is sufficiently large so that only intra-band pairing can occur.
We consider a basis in which the single-particle Hamiltonian is diagonal and the eigenstates are labelled by $|\vk,n\rangle$. Since time-reversal symmetry  exists, the state  $|\vk,n\rangle$ is degenerate with its time-reversal partner $\mathcal{T}|\vk,n\rangle$. Time reversal takes $\vk$ to $-\vk$ and since we assume there are no degeneracies at momentum $\vk$, this implies
\begin{equation}
\mathcal{T}|\vk,n\rangle>=t_n(\vk)|-\vk,n\rangle,
\end{equation}where $t_n(\vk)$ is a phase factor. Since these states are fermions $\mathcal{T}^2|\vk,n\rangle=-|\vk,n\rangle$, leading to
\begin{equation}
t_n(\vk)=-t_n(-\vk).
\end{equation}
In general the phase factors $t_n(\vk)$ are not related for different bands. However, using the pseudospin description given in the first section, and ignoring any other possible bands, the following relationship can be found
\begin{equation}
t_{\pm}(\vk)= \mp\frac{\gamma_x(\vk)-i\gamma_y(\vk)}{\sqrt{\gamma_x^2(\vk)+ \gamma_y^2(\vk)}}
\label{phase}
\end{equation}

In the context of superconductivity, the phase factors $t_n(\vk)$ play a fundamental role, since pairing occurs between a state and its time-reversed partner. In particular, the pairing Hamiltonian can be written as
\begin{equation}
H_{\Delta}=\frac{1}{2}\sum_{n,k}[\Delta_n(\vk)c^{\dagger}_{\vk}c^{\dagger}_{-\vk}+\Delta^*_n(\vk)c_{-\vk}c_{\vk}].
\end{equation}
Anticommutation of the fermions implies $\Delta_n(\vk)=-\Delta_n(-\vk)$. In addition, using the transformation properties of the single particle states under a point group operation  reveals that the function $\Delta_n(\vk)$ does not transform simply under such a rotation \cite{ser04}. However, if a new function $\Delta_n(\vk)=t_n(\vk)\tilde{\Delta}_n(\vk)$ is defined, then under a point group operation $g$,  $\tilde{\Delta}_n(\vk)$ becomes $\tilde{\Delta}_n(g^{-1}\vk)$, which can be used to classify $\tilde{\Delta}_n(\vk)$ basis functions according to irreducible representations of the point group. One simplifying feature of this analysis is that $\tilde{\Delta}_n(\vk)=\tilde{\Delta}_n(-\vk)$, so that only even functions of $\vk$ need to be considered. A list of these functions can be found in Table III and Table IV.  The reason that the usual group theoretical analysis works for $\tilde{\Delta}(\vk)$ is that physically, this is the gap that results from pairing time-reversed partners, that is
\begin{equation}
H_{\Delta}=\frac{1}{2}\sum_{n,k}[\tilde{\Delta}_n(\vk)c^{\dagger}_{\vk}\tilde{c}^{\dagger}_{\vk}+\tilde{\Delta}^*_n(\vk)\tilde{c}_{\vk}c_{\vk}].
\label{pairing}
\end{equation}
where $\tilde{c}^{\dagger}_{\vk}$ is the raising operator for the time-reversed partner of $|\vk,n\rangle$.

It is informative to relate the above description in terms of eigenstates of the single-particle Hamiltonian to the description in the original spin basis. In particular, simplifying to the pseudospin basis that gives rise to the two-band helicity basis (that is ignoring other bands, so that the phase factors are given by Eq.~\ref{phase}), the resulting singlet and triplet gap functions become
\begin{eqnarray}
\psi(\vk)=&&\frac{\tilde{\Delta}_+(\vk)+\tilde{\Delta}_-(\vk)}{2} \nonumber \\
\vd(\vk)=&&\hvg(\vk)\frac{\tilde{\Delta}_+(\vk)-\tilde{\Delta}_-(\vk)}{2}.
\label{gaps}
\end{eqnarray}
 This reveals the result found above that the $\vd(\vk)$ vector describing spin-triplet pairing is parallel to $\vg$, which describes the single particle spin-orbit coupling. It further reveals how spin-singlet and spin-triplet mixing occur in the helicity basis, if the two gaps are unequal, then there is singlet-triplet mixing. Note that this relationship can be reversed to yield
\begin{equation}
\tilde{\Delta}_{\pm}=\psi({\bf k})\pm \hvg(\vk)\cdot \vd(\vk).
\label{mix}
\end{equation}
This relationship has often been used to show how singlet-triplet mixing can lead to line nodes (which are topologically protected). One illustrative example that has been applied to CePt$_3$Si \cite{NMRTheory} is a Rashba spin orbit coupling $\vg(\vk)=\hat{x}k_y-\hat{y}k_x$ with a spherical Fermi surface with radius $k_F$. In this case, for pairing in the same symmetry channel as isotropic $s$-wave ($\psi(\vk)=\Delta_s$), the protected spin-triplet state is $\vd(\vk)=\Delta_t(\hat{x}k_y-\hat{y}k_x)/(\sqrt{2}k_F)$.  Eq.~\ref{mix} gives $\tilde{\Delta}_{\pm}(\theta,\phi)=\Delta_s\pm \Delta_t|\sin\theta|$, where $\theta$ and $\phi$ are the spherical angles that specify the position on the Fermi surface. Taking $\Delta_s$ and $\Delta_t$ to be positive, we see that $\Delta_+$ is nodeless while $\Delta_-$ can have nodes for $\theta=\theta_n$ with $|\sin\theta_n|=\Delta_s/\Delta_t$. This is only possible if $\Delta_s/\Delta_t<1$, meaning that the spin-triplet component dominates. When this occurs, there are two circular line nodes on the spherical Fermi surface given by $\theta=\theta_n$ and $\theta=\pi-\theta_n$. As discussed in Section~\ref{toposec}, these nodes lead to interesting topological edge states. Singlet-triplet mixing has also been argued to give rise to topological line nodes in Li$_2$Pt$_3$B \cite{Li2Pt3BNode}.

It was discussed in previous sections that when there is no spin-triplet superconductivity, that is $\vd(\vk)=0$ in Eq.~\ref{gaps}, the superconductor exhibits properties of a spin-triplet superconductor (for example a similar spin susceptibility and enhanced Pauli critical fields). Another perspective on why this occurs follows from the pairing Hamiltonian in Eq.~\ref{pairing}. In particular, in the response functions that determine the magnetic field response, the anomalous averages $\langle c^{\dagger}_{\vk}\tilde{c}^{\dagger}_{\vk} \rangle=\Delta_n/E_n$ enter. For the response to resemble a spin-singlet superconductor, it is required that when these are re-expressed in the spin basis, only the spin-singlet $\langle c^{\dagger}_{\vk,\uparrow}\tilde{c}^{\dagger}_{\vk,\downarrow}-c^{\dagger}_{\vk,\downarrow}\tilde{c}^{\dagger}_{\vk,\uparrow} \rangle$ contribution is non-zero. In the pseudospin basis this implies $\tilde{\Delta}_-/E_-=\tilde{\Delta}_+/E_+$, revealing that non-spin singlet behavior arises not only from $\tilde{\Delta}_-\ne \tilde{\Delta}_+$, but also from $E_+\ne E_-$. Hence if ASOC is large, spin-triplet like behavior can occur.

There have been a few microscopic studies on the interplay between pairing interactions and the ASOC to generate mixed spin-singlet and spin-triplet states \cite{vaf11,Li2Pt3Bspm,sch15}. More research to develop a deeper understanding of this important issue is desirable.

\begin{table}
\begin{tabular}{|c|c|c|}
\hline
Point Group & Representation & Basis Functions\\
  \hline
  $O$ & $\Gamma_1$ ($A_1$)  & 1, $k_x^2+k_y^2+k_z^2$\\
  &$\Gamma_2$ ($A_2$)&$(k_x^2-k_y^2)(k_y^2-k_z^2)(k_x^2-k_z^2)$\\
  &$\Gamma_3$ ($E$) &$\{2k_z^2-k_x^2-k_y^2,\sqrt{3}(k_x^2-k_y^2)\}$\\
  &$\Gamma_4$ ($F_1$) &$\{k_yk_z(k_y^2-k_z^2),k_zk_x(k_z^2-k_x^2),k_xk_y(k_x^2-k_y^2)\}$\\
  &$\Gamma_5$ ($F_2$) &$\{k_yk_z,k_zk_x,k_xk_y\}$\\ \hline
  $T_d$ & $\Gamma_1$ ($A_1$)  & 1, $k_x^2+k_y^2+k_z^2$\\
  &$\Gamma_2$ ($A_2$)&$(k_x^2-k_y^2)(k_y^2-k_z^2)(k_x^2-k_z^2)$\\
  &$\Gamma_3$ ($E$) &$\{2k_z^2-k_x^2-k_y^2,\sqrt{3}(k_x^2-k_y^2)\}$\\
  &$\Gamma_4$ ($F_1$) &$\{k_yk_z(k_y^2-k_z^2),k_zk_x(k_z^2-k_x^2),k_xk_y(k_x^2-k_y^2)\}$\\
  &$\Gamma_5$ ($F_2$) &$\{k_yk_z,k_zk_x,k_xk_y\}$\\ \hline
  $T$ & $\Gamma_1$ ($A_1$) & 1, $k_x^2+k_y^2+k_z^2$\\
  &$\Gamma_2$ ($A_2$)&$2k_z^2-k_x^2-k_y^2-\sqrt{3}(k_x^2-k_y^2)$\\
  &$\Gamma_3$ ($B_1$) &$2k_z^2-k_x^2-k_y^2+\sqrt{3}(k_x^2-k_y^2)$\\
  &$\Gamma_4$ ($F_1$) &$\{k_yk_z,k_zk_x,k_xk_y\}$\\\hline
  $D_6$ & $\Gamma_1$ ($A_1$) & 1, $k_x^2+k_y^2$,$k_z^2$\\
  &$\Gamma_2$ ($A_2$)&$(k_x^3-3k_xk_y^2)(k_y^3-3k_yk_x^2)$\\
  &$\Gamma_3$ ($B_1$) &$k_z(k_x^3-3k_xk_y^2)$\\
  &$\Gamma_4$ ($B_2$) &$k_z(k_y^3-3k_yk_x^2)$\\
  &$\Gamma_5$ ($E_1$) &$\{k_zk_x,k_zk_y\}$\\
  &$\Gamma_6$ ($E_2$)&$\{k_x^2-k_y^2,2k_xk_y\}$\\\hline
  $C_{6v}$ & $\Gamma_1$ ($A_1$) & 1, $k_x^2+k_y^2$,$k_z^2$\\
  &$\Gamma_2$ ($A_2$)&$(k_x^3-3k_xk_y^2)(k_y^3-3k_yk_x^2)$\\
  &$\Gamma_3$ ($B_1$) &$k_z(k_y^3-3k_yk_x^2)$\\
  &$\Gamma_4$ ($B_2$) &$k_z(k_x^3-3k_xk_y^2)$\\
  &$\Gamma_5$ ($E_1$) &$\{k_zk_x,k_zk_y\}$\\
  &$\Gamma_6$ ($E_2$)&$\{k_x^2-k_y^2,2k_xk_y\}$\\\hline
  $D_{3h}$ & $\Gamma_1$ ($A_1'$) & 1, $k_x^2+k_y^2$,$k_z^2$\\
  &$\Gamma_2$ ($A_2'$)&$(k_x^3-3k_xk_y^2)(k_y^3-3k_yk_x^2)$\\
  &$\Gamma_3$ ($A_1''$) &$k_z(k_x^3-3k_xk_y^2)$\\
  &$\Gamma_4$ ($A_2''$) &$k_z(k_y^3-3k_yk_x^2)$\\
  &$\Gamma_5$ ($E''$) &$\{k_zk_x,k_zk_y\}$\\
  &$\Gamma_6$ ($E'$)&$\{k_x^2-k_y^2,2k_xk_y\}$\\\hline
  $C_6$ & $\Gamma_1$ ($A_1$) & 1, $k_x^2+k_y^2$,$k_z^2$\\
  &$\Gamma_2$ ($A_2$)&$(k_x-ik_y)^2$\\
  &$\Gamma_3$ ($B_1$) &$(k_x+ik_y)^2$\\
  &$\Gamma_4$ ($B_2$) &$k_z(k_y^3-3k_yk_x^2),k_z(k_x^3-3k_xk_y^2)$\\
  &$\Gamma_5$ ($C_1$) &$k_zk_x$\\
  &$\Gamma_6$ ($C_2$)&$k_zk_x$\\\hline
  $C_{3h}$ & $\Gamma_1$ ($A_1$) & 1, $k_x^2+k_y^2$,$k_z^2$\\
  &$\Gamma_2$ ($A_2$)&$(k_x-ik_y)^2$\\
  &$\Gamma_3$ ($B_1$) &$(k_x+ik_y)^2$\\
  &$\Gamma_4$ ($B_2$) &$k_z(k_y^3-3k_yk_x^2),k_z(k_x^3-3k_xk_y^2)$\\
  &$\Gamma_5$ ($C_1$) &$k_zk_x$\\
  &$\Gamma_6$ ($C_2$)&$k_zk_x$\\\hline
  $D_{3}$ & $\Gamma_1$ ($A_1$) & 1, $k_x^2+k_y^2$, $k_z^2$\\
  &$\Gamma_2$ ($A_2$)&$(k_x^3-3k_xk_y^2)(k_y^3-3k_yk_x^2)$\\
  &$\Gamma_3$ ($E$) &$\{k_x^2-k_y^2,2k_xk_y\}$,$\{k_zk_x,k_zk_y\}$ \\
  \hline
  $C_{3v}$ & $\Gamma_1$ ($A_1$) & 1, $k_x^2+k_y^2$, $k_z^2$\\
  &$\Gamma_2$ ($A_2$)&$(k_x^3-3k_xk_y^2)(k_y^3-3k_yk_x^2)$\\
  &$\Gamma_3$ ($E$) &$\{k_x^2-k_y^2,2k_xk_y\}$,$\{k_zk_x,k_zk_y\}$ \\
  \hline
  $C_{3}$ & $\Gamma_1$ ($A_1$) & 1, $k_x^2+k_y^2$, $k_z^2$\\
  &$\Gamma_2$ ($A_2$)&$k_x^2-k_y^2$,$k_zk_x$\\
  &$\Gamma_3$ ($B_1$) &$2k_xk_y$,$k_zk_y$ \\
  \hline

\end{tabular}
\caption{Gap basis functions for different cubic, hexagonal, and trigonal non-centrosymmetric point groups. The $\Gamma_i$ irreducible representation labels are from Ref.~\onlinecite{kos}, the labels in brackets are also commonly used. The functions in brackets $\{f_1,f_2,...\}$ are the degenerate basis functions for irreducible representations of more than 1 dimension. }
\end{table}
\begin{table}
\begin{tabular}{|c|c|c|}
\hline
Point Group & Representation & Basis Functions\\
  \hline
  $D_4$ & $\Gamma_1$ ($A_1$)  & 1, $k_x^2+k_y^2$,$k_z^2$\\
  &$\Gamma_2$ ($A_2$)&$k_xk_y(k_x^2-k_y^2)$\\
  &$\Gamma_3$ ($B_1$) &$k_x^2-k_y^2$\\
  &$\Gamma_4$ ($B_2$) &$k_xk_y$\\
  &$\Gamma_5$ ($E$) &$\{k_xk_z,k_yk_z\}$\\ \hline
  $C_{4v}$ & $\Gamma_1$ ($A_1$)  & 1, $k_x^2+k_y^2$,$k_z^2$\\
  &$\Gamma_2$ ($A_2$)&$k_xk_y(k_x^2-k_y^2)$\\
  &$\Gamma_3$ ($B_1$) &$k_x^2-k_y^2$\\
  &$\Gamma_4$ ($B_2$) &$k_xk_y$\\
  &$\Gamma_5$ ($E$) &$\{k_xk_z,k_yk_z\}$\\ \hline
  $D_{2d}$ & $\Gamma_1$ ($A_1$)  & 1, $k_x^2+k_y^2$,$k_z^2$\\
  &$\Gamma_2$ ($A_2$)&$k_xk_y(k_x^2-k_y^2)$\\
  &$\Gamma_3$ ($B_1$) &$k_x^2-k_y^2$\\
  &$\Gamma_4$ ($B_2$) &$k_xk_y$\\
  &$\Gamma_5$ ($E$) &$\{k_xk_z,k_yk_z\}$\\ \hline
  $C_{4}$ & $\Gamma_1$ ($A_1$)  & 1, $k_x^2+k_y^2$,$k_z^2$\\
  &$\Gamma_2$ ($A_2$)&$(k_x^2-k_y^2)$,$k_xk_y$\\
  &$\Gamma_3$ ($B_1$) &$-k_z(k_x+ik_y)$\\
  &$\Gamma_4$ ($B_2$) &$k_z(k_x-ik_y)$\\\hline
  $S_{4}$ & $\Gamma_1$ ($A_1$)  & 1, $k_x^2+k_y^2$,$k_z^2$\\
  &$\Gamma_2$ ($A_2$)&$(k_x^2-k_y^2)$,$k_xk_y$\\
  &$\Gamma_3$ ($B_1$) &$k_z(k_x-ik_y)$\\
  &$\Gamma_4$ ($B_2$) &$-k_z(k_x+ik_y)$\\\hline
  $D_{2}$ & $\Gamma_1$ ($A_1$)  & 1, $k_x^2$,$k_y^2$,$k_z^2$\\
  &$\Gamma_2$ ($A_2$)&$k_zk_x$\\
  &$\Gamma_3$ ($B_1$) &$k_xk_y$\\
  &$\Gamma_4$ ($B_2$) &$k_yk_z$\\\hline
  $C_{2v}$ & $\Gamma_1$ ($A_1$)  & 1, $k_x^2$,$k_y^2$,$k_z^2$\\
  &$\Gamma_2$ ($A_2$)&$k_zk_x$\\
  &$\Gamma_3$ ($B_1$) &$k_xk_y$\\
  &$\Gamma_4$ ($B_2$) &$k_yk_z$\\\hline
  $C_{2}$ & $\Gamma_1$ ($A_1$)  & 1, $k_x^2$,$k_y^2$,$k_z^2$,$k_xk_y$\\
  &$\Gamma_2$ ($A_2$)&$k_zk_x$,$k_zk_y$\\ \hline
  $C_{s}$ & $\Gamma_1$ ($A_1$)  & 1, $k_x^2$,$k_y^2$,$k_z^2$,$k_xk_y$\\
  &$\Gamma_2$ ($A_2$)&$k_zk_x$,$k_zk_y$\\ \hline
  $C_1$ & $\Gamma_1$ ($A_1$) &all components allowed \\
  \hline
\end{tabular}
\caption{Gap basis functions for different tetragonal, orthorhombic, and monoclinic non-centrosymmetric point groups. The $\Gamma_i$ representation labels are from Ref.~\onlinecite{kos}, the labels in brackets are also commonly used. The functions in brackets $\{f_1,f_2,...\}$ are the degenerate basis functions for irreducible representations of more than 1 dimension.}
\end{table}

\section{Magnetoelectric effects, Fulde-Ferrell-Larkin-Ovchinnikov states, and helical states} \label{helical}

A key new feature of non-centrosymmetric superconductors appears when magnetic fields are applied. In particular, symmetry allows new terms, Lifshitz invariants, in the Ginzburg-Landau free energy. From a symmetry point of view, the origin of these Lifshitz invariants is similar to the origin of the well known Dzyaloshinskii-Moriya interaction in spin systems without inversion symmetry. These Lifshitz invariants lead to a variety of novel effects, including a new magnetization term, and a helical superconducting state in the presence of a spatially uniform magnetic field (in which the superconducting order parameter develops a spatial variation, this state exhibits a non-trivial interplay with Fulde-Ferrell-Larkin-Ovchinnikov \cite{ff,lo} states), and supercurrents induced by a Zeeman field. In the following, we provide a detailed examination of the GL theory, which is followed by an overview of the corresponding microscopic theory.

\subsection{Ginzburg Landau free energy}

 Non-centrosymmetric superconductors allow the existence of Lifshitz invariants in the Ginzburg Landau (GL) free energy \cite{MS94,Edel96,Y02,agt03,KAS05,MinSam08}. These give rise to magnetoelectric effects \cite{LNE85,Edel95,Y02,Fuj05,lu08}, helical phases \cite{agt03,DF03,KAS05,dim07,agt07}, and a new anomalous magnetization \cite{LNE85,KAS05,OIM06,yip07,lu08}. Here we initially consider the GL theory for a single component order parameter and include a generalized Lifshitz invariant allowed by non-centrosymmetric structures. The form of this Lifshitz invariant depends on the specific point group symmetry and these are tabulated in Table V. The emphasis here is to highlight the physics arising from the Lifshitz invariants, so for simplicity, we ignore any anisotropy that appears in the usual GL free energy. Consequently, the GL free energy is (we work in units such that $\hbar=c=1$):
\begin{equation}
F=\int d^3r\left\{\alpha(T-T_{c0})|\eta|^2
+K\eta^*(D_x^2+D_y^2+D_z^2)\eta+\frac{\beta}{2}|\eta|^4+\frac{B^2}{8\pi}+f_{lif}\right\}\label{glfree}
\end{equation}

where $D_i=-i\nabla_i-2eA_i$ and ${\bf B}=\nabla\times {\bf A}$ and, the Lifshitz invariant  $f_{lif}$ is given by
\begin{equation}
f_{lif}= K_{ij}B_i[\eta^*(D_j\eta)+\eta(D_j\eta)^*].
\end{equation}

The GL equations are found by varying Eq.~\ref{glfree} with respect to ${\bf A}$ and $\eta$. This yields
\begin{equation}
\alpha(T-T_{c0})\eta+\beta|\eta|^2\eta+K{\bf D}^2\eta+K_{ij}[2h_i(D_j\eta)+i\eta\nabla_jB_i]=0
\end{equation}
and
\begin{equation}
{\bf J}_i=\frac{1}{4\pi}[\nabla\times({\bf B}-4\pi {\bf M})]_i=2eK[\eta^*(D_i\eta)+\eta(D_i\eta)^*]+4eK_{ji}B_j|\eta|^2
\label{GLcurrent}
\end{equation}
where
\begin{equation}
{\bf M}_i=-K_{ij}[\eta^*(D_j\eta)+\eta(D_j\eta)^*] \label{mag1}.
\end{equation}
In addition, the following boundary conditions are also needed (which originate from surface terms stemming from integration by parts in the variation of Eq.~\ref{glfree}):
\begin{equation}
[K\hat{n}_i(D_i\eta)+K_{ij}B_i\hat{n}_j\eta]_{boundary}=0 \label{boundary}
\end{equation}
where $\hat{n}_j$ is the component of the surface normal along $\hat{j}$. This is joined with the Maxwell boundary conditions: continuity of the normal component of ${\bf B}$ and continuity of the transverse components of  ${\bf H}={\bf B}-4\pi {\bf M}$ across the boundary. The appearance of ${\bf M}$ in Eq.~\ref{mag1} makes these boundary conditions non-trivial. The condition that no current flows through the boundary follows by taking the addition of the complex conjugate of Eq.~\ref{boundary} multiplied by $\eta^*$ and Eq.~\ref{boundary} multiplied by $\eta$, resulting in ${\bf J}\cdot {\hat n}|_{boundary}=0$. Note that the boundary conditions are valid only on a length scale greater than $\xi_0$, the zero-temperature coherence length. In the following,
the solutions to some problems are given to gain insight into the physics stemming from these Lifshitz invariants.

\begin{table}
\begin{tabular}{|c|c|}
\hline
Point Group & Lifshitz Invariants\\
  \hline
  $O$ & $K(B_xj_x+B_yj_y+B_zj_z)$ \\
  $T$ &  $K(B_xj_x+B_yj_y+B_zj_z)$\\
  $D_6$ & $K_1(B_xj_x+B_yj_y+B_zj_z)+K_2B_zj_z$ \\
  $C_{6v}$ & $K_{xy}(B_xj_y-B_yj_x)$ \\
  $C_6$ & $K_1(B_xj_x+B_yj_y+B_zj_z)+K_2B_zj_z+K_{xy}(B_xj_y-B_yj_x)$ \\
  $D_4$ & $K_1(B_xj_x+B_yj_y+B_zj_z)+K_2B_zj_z$ \\
  $C_{4v}$ & $K_{xy}(B_xj_y-B_yj_x)$ \\
  $D_{2d}$ & $K_{xy}(B_xj_y-B_yj_x)$ \\
  $C_4$ & $K_1(B_xj_x+B_yj_y+B_zj_z)+K_2B_zj_z+K_{xy}(B_xj_y-B_yj_x)$  \\
  $S_4$ & $K_1(B_xj_x-B_yj_y)+K_{xy}(B_yj_x+B_xj_y)$  \\
  $D_3$ & $K_1(B_xj_x+B_yj_y+B_zj_z)+K_2B_zj_z$ \\
  $C_{3v}$ & $K_{xy}(B_xj_y-B_yj_x)$ \\
  $C_3$ & $K_1(B_xj_x+B_yj_y+B_zj_z)+K_2B_zj_z+K_{xy}(B_xj_y-B_yj_x)$ \\
  $D_2$ & $K_1B_xj_x+K_2B_yj_y+K_3B_zj_z$ \\
  $C_{2v}$ & $K_{xy}B_xj_y+K_{yx}B_yj_x$ \\
  $C_2$ & $K_1B_xj_x+K_2B_yk_y+K_3B_zj_z+K_{yx}B_yj_x+K_{xy}B_xj_y$ \\
  $C_s$ & $K_{zx}B_zk_x+K_{zy}B_zj_y+K_{xz}B_xj_z+K_{yz}B_yj_z$\\
  $C_1$ & all components allowed \\
  \hline
\end{tabular}
\caption{Form of Lifshitz invariants for different non-centrosymmetric point groups. Here $j_i=\eta^*(D_i\eta)+\eta(D_i\eta)^*$. The point groups $T_d$, $D_{3h}$, and $C_{3h}$ are not listed since these point groups do not have Lifshitz invariants that are linear in $j_i$ and $B_i$.}
\end{table}

\subsection{Spatially uniform magnetic fields: Helical state}

The GL equations simplify when the magnetic field is spatially uniform. In particular, introducing the following redefined order parameter:
\begin{equation}
\tilde{\eta}=\eta \exp\big(i{\bf q}{\cdot {\bf x}}\big)=\eta \exp\big(i\frac{iB_jK_{jk}x_k}{K}\big) \label{hel}
\end{equation}
yields the following GL free energy for $\tilde{\eta}$ that no longer has any Lifshitz invariants
\begin{equation}
F=\int d^3r\left\{\Big[\alpha(T-T_{c0})-B_lK_{lm}B_jK_{jm}B_k\Big]|\tilde{\eta}|^2
+K_1\tilde{\eta}^*(D_x^2+D_y^2+D_z^2)\tilde{\eta}
 +\frac{\beta}{2}|\tilde{\eta}|^4+\frac{B^2}{8\pi}\right\}. \label{newGL}
\end{equation}
The resulting GL equations are those of a single component superconductor with an increase in $T_c$ due to the magnetic fields (in general, there is also a paramagnetic suppression to counter this increase in $T_c$).  The new GL equations follow from a minimization of Eq.~\ref{newGL} with respect to ${\bf A}$ and $\tilde{\eta}$. Importantly, the phase factor that appears in $\tilde{\eta}$  cancels the Lifshitz invariant contribution to the boundary condition. Additionally, the magnetization that follows from Eq.~\ref{newGL} is the same as that in Eq.~\ref{mag1} found earlier. Eq.~\ref{newGL} implies that some results from the usual GL theory apply (when magnetic fields are homogeneous). In particular:\\

\noindent i) Near the upper critical field, the vortex lattice solution coincides with that of Abrikosov.\\
ii) The critical current in thin wires will show no asymmetry under change of direction.\\
iii) $H_{c3}$, the surface critical field,  is unchanged with respect to usual superconductors (as found by DeGennes). Notably, the magnetic field induced change to $T_c$ does not change $H_{c3}$ to leading order in $(T_c-T)/T_c$).\\

\noindent {\it Helical State and Josephson interference experiments}\\

The key difference with respect to usual superconductors that arises from the combination of broken inversion symmetry and a magnetic field is that the order parameter develops a spatial modulation, as illustrated through Eq.~\ref{hel}. In particular,  a helical spatial dependence in the complex plane develops in $\tilde{\eta}$, leading to the name helical state to describe this thermodynamic phase.  When considering the bulk physics, this phase factor can be removed by a gauge transformation and consequently has no pronounced qualitative effect on the macroscopic bulk physics (note that as discussed later, it can have an important effect on the microscopic excitation spectrum). However, since the spatial variation is related to the phase of the order parameter, it is in principle possible to observe this through a Josephson interference experiment \cite{KAS05}. The details of such an experiment depend on the point group symmetry of the broken inversion material. Here we explicitly consider two-dimensional materials with point group symmetry $C_{6v}$, $C_{4v}$, $D_{2d}$, or $C_{3v}$, which all allow a Lifshitz invariant of the form $\kappa_{xy}(B_xj_y-B_yj_x)$ (which can stem microscopically from a Rashba spin-orbit interaction). For simplicity, we restrict ourselves to $C_{4v}$ symmetry, the results for $C_{6v}$, $D_{2d}$, and $C_{3v}$ will be the same. In this case a magnetic field applied perpendicular to the 4-fold symmetry axis (we label the 4-fold symmetry axis $\hat{z}$) and the magnetic field direction as $\hat{x}$)  will induce a spatial modulation perpendicular to both the field and 4-fold symmetry axis, that is ${\bf q}$ is along $\hat{y}$. If there is a centrosymmetric superconductor with a phase that is weakly linked to the non-centrosymmetric superconductor with an interface that has its surface normal along $\hat{x}$, then the development of the spatial modulation will lead to an Franuhofer interference pattern in the Josephson current, even though there in no magnetic flux through the junction.

\begin{figure}[t]
\begin{center}
  \includegraphics[width=0.7\columnwidth]{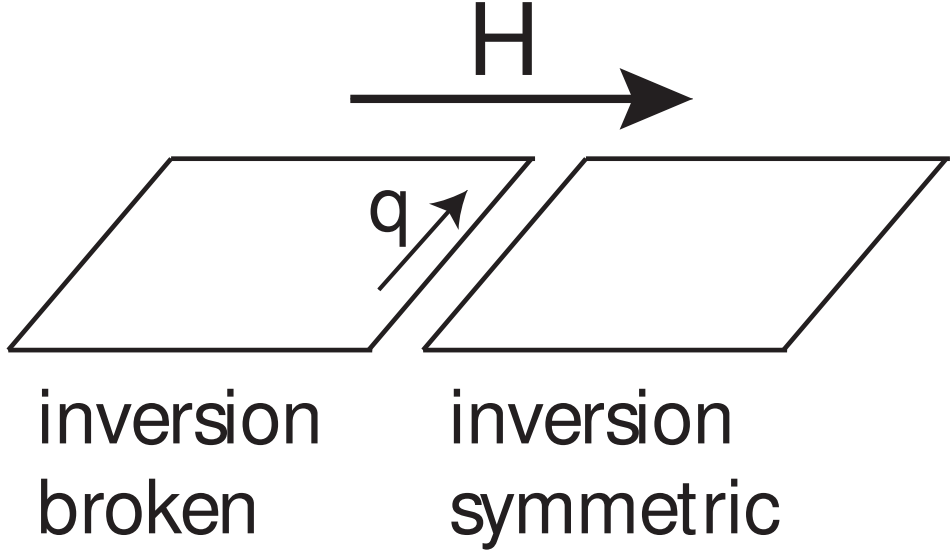}	
\end{center}
	\caption{Configuration for a Josephson interference pattern to observe the helical state in non-centrosymmetric supercondonductors.}
  \label{fig2}	
\end{figure}

This can be shown by considering the following free energy of the
junction
\begin{equation}
H_J=-t\int dx[\Psi_1(\vR)\Psi_2^*(\vR)+c.c.]
\end{equation}
where the integral is along the junction. The Josephson
current is
\begin{equation}
I_J=Im\Big [ t\int dx\Psi_1(\vR)\Psi_2^*(\vR)\Big ]
\end{equation}
If the junction length is $2L$, the integrating leads to
a maximum Josephson current of
\begin{equation}
I_J=2t|\Psi_{1}^0||\Psi_{2}^0|\frac{|\sin(qL)|}{|q L|}.
\end{equation}
Consequently, the Josephson current will
display an interference pattern for a field {\it perpendicular} to
the junction. In the usual case the Fraunhofer pattern
is observed for a magnetic field perpendicular to the
film (a magnetic flux then passes through the junction).\\

\noindent {\it Magnetoelectric Effect}\\

One of the first predictions of non-centrosymmetric superconductivity was that a spin-polarization is induced by a supercurrent \cite{Edel95}. Within the phenomenological theory, this spin-polarization is given by the magnetization in Eq.~\ref{mag1}, which is re-written here for convenience
\begin{equation}
{\bf M}_i=-K_{ij}[\eta^*(D_j\eta)+\eta(D_j\eta)^*].
\end{equation}
 Eq.~\ref{mag1} reveals that this magnetization appears when a phase gradient, that is a supercurrent, appears on the right hand side of this equation. It was later suggested that the converse effect exists, that is, a Zeeman field would induce a supercurrent \cite{Y02}. This follows from the expression for the current of Eq.~\ref{GLcurrent} in the limit that the usual GL current ($2eK[\eta^*(D_i\eta)+\eta(D_i\eta)^*]$) vanishes. This does not take into account the possibility discussed above, that the order parameter develops a spatial modulation in the presence of a Zeeman field (this modulation leads to a  nonvanishing $2eK[\eta^*(D_i\eta)+\eta(D_i\eta)^*]$). This phase modulated state ensures that the resultant total supercurrent is vanishing. However, in spite of the appearance of the helical state, it is possible to create this Zeeman field induced supercurrent through a geometry like that used to observe Little-Parks oscillations \cite{DF03}. This can be understood as follows, the supercurrent has two contributions, one is the usual GL current $2eK[\eta^*(D_i\eta)+\eta(D_i\eta)^*]$ and the other is the current that arises from the Lifshitz invariants. In the helical state, these two contributions cancel each other. However, by wrapping the superconductor in a cylinder, single valuedness of the order parameter does not allow the helical phase to develop and a non-zero current can develop since the two contributions non longer cancel. The resulting current will be periodic in the applied magnetic field \cite{DF03}.

\subsection{London Theory and Meissner State}

The Lifshitz invariants also lead to qualitatively new effects when the magnetic field is not spatially uniform. To see this, let us examine the usual penetration depth problem in which the magnetic induction decays exponentially from the surface where there is an applied field.  Let us work in the London limit and set $\eta=|\eta|e^{i\theta}$ with the assumption that the magnitude $|\eta|$ is spatially constant. We then minimize the GL free energy with respect to $\theta$ and ${\bf A}$ with arbitrary Lifshitz invariants, leaving the discussion of specific point group symmetries (which restrict the form of the Lifshitz invariants) to end of this Section. Minimization with respect to $\theta$ gives
\begin{equation}
K\nabla\cdot(\nabla\theta-2e{\bf A})+K_{ij}\nabla_iB_j=0
\end{equation}
which yields the continuity equation for the current ($\nabla \cdot {\bf J}=0$).  Minimization with respect to ${\bf A}$ gives
\begin{equation}
{\bf J}_i= \frac{1}{4\pi}[\nabla\times ({\bf B}-4\pi {\bf M})]_i=-\frac{1}{4\pi \lambda^2}[{\bf A}_i-\frac{1}{2e}\nabla_i\theta-
\sum_j \sigma_{ji}{\bf B}_j]
\label{London}
\end{equation}
with \begin{equation}
4\pi {\bf M}_i=\frac{1}{\lambda^2}\sum_j\sigma_{ij}({\bf A}_j-\frac{1}{2e}\nabla_j\theta),
\label{mag}
\end{equation}
 $1/\lambda^2=8\pi (2e)^2K|\eta|^2$ and $\sigma_{ij}=16\pi e \lambda^2 K_{ij}$.
 We consider the orientation shown in Fig.~3, with the  applied field along the $\hat{y}$ direction and the interface normal to be along the $\hat{z}$ direction. A rotation of the fields in the GL free energy allows this orientation to be generalized to all other orientations.
  It is assumed that the only spatial variations are along the interface normal ($z$). From $\nabla\cdot {\bf B}=0$ this assumption implies $B_z=0$. In addition, we set ${\bf A}=[A_x(z),A_y(z),0]$ , yielding ${\bf B}=(-\partial A_y/\partial z,\partial A_x/\partial z,0)$ and we choose a gauge where $\nabla\theta=0$. Eq.~\ref{London} yields
\begin{eqnarray}
\frac{\partial B_y}{\partial z}=&&\frac{1}{\lambda^2}\frac{\partial}{\partial z}[\sigma_{yy}A_y+\sigma_{zy}A_z]+\frac{1}{\lambda^2}A_x-\frac{1}{\lambda^2}\sigma_{xx}B_x \label{lon1}\\
\frac{\partial B_x}{\partial z}=&&\frac{1}{\lambda^2}\frac{\partial}{\partial z}[\sigma_{xx}A_x+\sigma_{zx}A_z]-\frac{1}{\lambda^2}A_y-\frac{1}{\lambda^2}\sigma_{yy}B_x \label{lon2}\\
4\pi J_z=&&0=A_z-\sigma_{zx}B_x-\sigma_{zy}B_y\label{lon3}.
\end{eqnarray}
Note that contributions from $\sigma_{xy}$ and $\sigma_{yx}$ do not appear. Taking derivatives of Eq.~\ref{lon1} and \ref{lon2} with respect to $z$, using Eq.~\ref{lon3} to eliminate $A_z$, we have
\begin{eqnarray}
(1-\frac{\sigma_{zy}^2}{\lambda^2})\frac{\partial^2 B_y}{\partial z^2}=&&\frac{1}{\lambda^2}B_y-\frac{\sigma_{xx}+\sigma_{yy}}{\lambda^2}\frac{\partial B_x}{\partial z}+\frac{\sigma_{zy}\sigma_{zx}}{\lambda^2}\frac{\partial^2 B_x}{\partial z^2}\\
(1-\frac{\sigma_{zx}^2}{\lambda^2})\frac{\partial^2 B_x}{\partial z^2}=&&\frac{1}{\lambda^2}B_x+\frac{\sigma_{xx}+\sigma_{yy}}{\lambda^2}\frac{\partial B_y}{\partial z}+\frac{\sigma_{zy}\sigma_{zx}}{\lambda^2}\frac{\partial^2 B_y}{\partial z^2}.
\end{eqnarray}
The corresponding boundary conditions are $B_i(z\rightarrow \infty)=0$ and
\begin{eqnarray}
H_y=&&B_y(z=0)-4\pi M_y(z=0)\\
0=&&B_x(z=0)-4\pi M_x(z=0)
\end{eqnarray}
where $H_y$ is the applied field. $M_x,M_y$ is given through Eq.~\ref{mag} and Eqs.~\ref{lon1}, \ref{lon2}, and \ref{lon3} can be used to eliminate ${\bf A}$ in favor of ${\bf B}$ and its derivatives. Setting $B_i=B_{i0}\exp(-\delta z/\lambda)$ yields a solution that can be found analytically. This solution is quite involved and it is more informative to consider the point groups $O$ and $C_{4v}$ as is done below. \\

\begin{figure}[t]
\begin{center}
  \includegraphics[width=0.7\columnwidth]{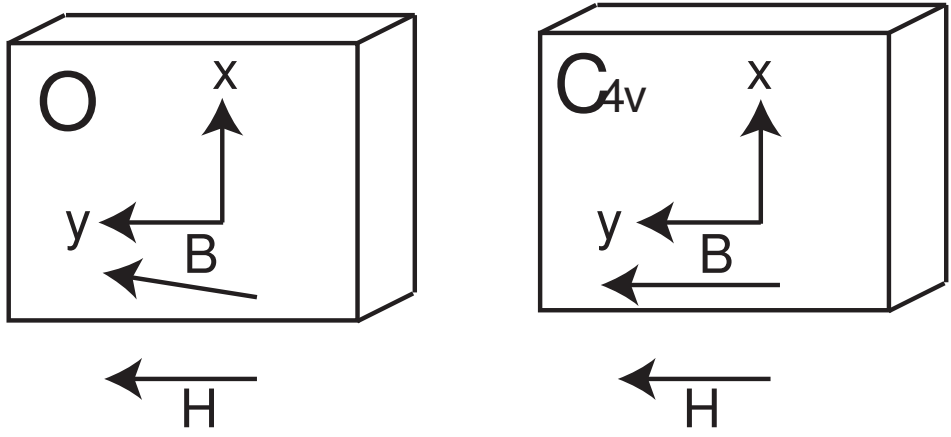}	
\end{center}
	\caption{London penetration for point groups $O$ (left) and $C_{4v}$ (right). The applied field is ${\bf H}$ and the magnetic induction at the surface is ${\bf B}$. In both cases the four-fold symmetry axis is along $\hat{z}$, which is along the surface normal. The magnetic induction decays with $z$ inside the superconductor. In both cases ${\bf B}$ is discontinuous at the interface due to magnetization in the superconductor that appears from the Lifshitz invariants. For point group $O$ this discontinuity is perpendicular to the applied field, while for point group $C_{4v}$ this discontinuity is along the applied field. }
  \label{fig3}	
\end{figure}

\noindent {\it $O$ point group}\\

 Materials with this point group are  Li$_2$Pt$_3$B \cite{Li2Pd3BRep,Li2Pt3BRep,Li2Pt3BNode} and Mo$_3$Al$_2$C \cite{Mo3Al2CRep,Mo3Al2CRep2}. Solutions to this problem can also be found in Refs.~\onlinecite{LNE85,lu08}. For point group $O$, the relevant Lifshitz invariant is  $K_1{\bf B}\cdot{\bf j}$. This is a scalar under rotations, so the solution does not depend on the surface normal orientation. In this case, the equations for ${\bf B}$ are
\begin{eqnarray}
\frac{\partial^2 B_y}{\partial z^2}=&&\frac{1}{\lambda^2} B_y+ \frac{\delta}{\lambda^2}\frac{\partial B_x}{\partial z}\\
\frac{\partial^2 B_x}{\partial z^2}=&&\frac{1}{\lambda^2} B_x- \frac{\delta}{\lambda^2}\frac{\partial B_y}{\partial z}.
\end{eqnarray}
where $\delta=-2\sigma_{xx}$ (note $\sigma_{xx}=\sigma_{yy}$).
The solution to first order in $\delta/\lambda$ is
\begin{eqnarray}
B_y=&&H_y\big[\cos\frac{\delta z}{\lambda^2}+\frac{\delta}{\lambda}\sin\frac{\delta z}{\lambda^2}\big]e^{-z/\lambda}\\
B_x=&&H_y\big[\frac{\delta}{\lambda}\cos\frac{\delta z}{\lambda^2}-\sin\frac{\delta z}{\lambda^2}\big]e^{-z/\lambda}.
\end{eqnarray}
This implies that the magnitude of the $B_x$ is discontinuous as it crosses the surface (though not that of $B_y$) and that ${\bf B}$ rotates inside the superconductor. In a slab, $B_x$ is of opposite sign on the opposite sides of the slab.\\

\noindent {\it $C_{4v}$ point group}\\

CePt$_3$Si \cite{CePt3Si2004} has this point group symmetry. For the point group  $C_{4v}$, the relevant Lifshitz invariant is by $K_{xy}\hat{z}\cdot {\bf B}\times {\bf j}$, leading to $\sigma=\sigma_{xy}=-\sigma_{yx}\ne0$. The solution now depends upon the relative orientation of the interface normal and the applied magnetic field and has been considered in Ref.~\onlinecite{yip07}. We explicitly consider the applied field oriented along the $\hat{y}$ direction and the interface normal either along or perpendicular to $\hat{z}$ (the four-fold symmetry axis). Initially consider the interface normal along the $\hat{z}$ direction. In this case $B_x=0$, and the usual London equation arises
\begin{equation}
\frac{\partial^2 B_y}{\partial z^2}=\frac{1}{\lambda^2}B_y\\
\end{equation}
with the unusual boundary condition $H_y|_{z=0}=(B_y+\frac{\sigma}{\lambda}B_y)|_{z=0}$ (due to existence of the magnetization in Eq.~\ref{mag}). This yields
\begin{equation}
B_y(z)=\frac{H_y}{1+\frac{\sigma}{\lambda}}e^{-z/\lambda}.
\end{equation}
 In this case, unlike the case with $O$ symmetry, there is no rotation of ${\bf B}$ across the sample surface. However, the ${\bf B}$ is discontinuous as the surface is crossed. Again, in a slab, the discontinuity in $B_y$ is opposite for opposite sides of the slab.

 Now consider the case that the interface normal is perpendicular to the $\hat{z}$ direction, say along the $\hat{x}$ direction (so that $B_x=0$).  In this case, $B_z=0$ and the solution for $B_y$ gives
\begin{equation}
B_y=\frac{H_y}{1-\frac{\sigma^2}{\lambda^2}}e^{-z/\tilde{\lambda}}
\end{equation}
where $\sigma=\sigma_{xy}$ and $\tilde{\lambda}=\lambda(1-\frac{\sigma^2}{\lambda^2})$. Again $B_y$ is discontinuous across the interface, however, since $\sigma/\lambda$ is presumably less than 1, the discontinuity is smaller than in the previous case. In addition, for a slab, the discontinuity is the same sign on opposite sides of the slab.

\subsection{Spatial structure of a single vortex}

For strongly type II superconductors, London theory can also give the field distribution of a vortex. The lack of inversion symmetry introduces some new physics. Again, the emphasis will be on point groups $O$ and $C_{4v}$ \cite{yip07,lu08,kas13}. Here we consider the parameter $\sigma_{ij}/\lambda$ to be small so that the Lifshitz invariants perturb the usual London solution. The usual London solution is given by $\theta=-\phi$ ($\phi$ is the polar angle) and, for a field applies along the ${\hat n}$, direction
\begin{equation}
{\bf B}=\frac{1}{2e\lambda^2}K_0(r/\lambda) \hat{n}
\end{equation}
where $K_0(x)$ is a modified Bessel function. The field distribution due to the Lifshitz invariants depends upon the point group symmetry. \\

\noindent{\it O point group}\\

The solution in this case was found originally in Ref.~\onlinecite{lu08}. The solution does not depend upon the field orientation and the modified London equation is
\begin{equation}
\nabla\times \nabla \times {\bf B}=-\frac{1}{\lambda^2}+2\frac{\delta}{\lambda^2}\nabla\times {\bf B}.
\end{equation}
The new term on the right hand side implies that, in addition to the usual field component along $\hat{z}$, there is a new component along $\hat{\phi}$. To first order in $\delta/\lambda$ the additional field is
\begin{equation}
B_{\phi}^{(1)}(x=r/\lambda)=\frac{\delta}{e \lambda^3}\Big\{K_1(x)\int_0^xx'dx'I_1(x')K_1(x')+I_1(x)\int_x^{\infty}x'dx'[K_1(x')]^2\Big\}
\end{equation}
where $I_1$ and $K_1$ are modified Bessel functions of the first kind. The asymptotic behavior of this transverse flux density is
\begin{equation}
 B_\phi \sim  \begin{cases}
               -\frac{\delta}{2e \lambda^3}[\frac{1}{x} + x \ln x - \frac{x}{2}], & x \rightarrow 0 \\
               \frac{\delta}{e \lambda^3} \sqrt{\frac{\mathstrut{\pi x}}{8}} e^{-x}, & x \rightarrow \infty.		\label{eq: asymp-Bphi}
              \end{cases}
\end{equation}
The corresponding current $J_z(x)$ (which is along the applied field direction) has the asymptotic behavior
 \begin{equation}
 J_z(x) \sim  \begin{cases}
               \frac{\mu}{4\pi \kappa^2} \ln \frac{1}{x}, & x \rightarrow 0 \\
               -  \frac{\mu}{4\pi \kappa^2}\sqrt{\frac{\mathstrut{\pi x}}{8}} e^{-x}, & x \rightarrow \infty.
              \end{cases}
\end{equation}
The divergence in $J_z$ for small $x$ is an artifact of the London theory and an examination of the full GL equations removes this \cite{kas13}. \\

\noindent{\it $C_{4v}$ point group}\\

 The fields that appear due to the Lifshitz invariants depend in this case upon the orientation of the applied field. For a field applied along the $\hat{y}$ direction, the solution for ${\bf B}$ is (correct  to order  $\sigma/\lambda$) \cite{yip07}
\begin{equation}
{\bf B}=\frac{1}{2e\lambda^2}K_0(|{\bf r}+\frac{\sigma}{\lambda}\hat{z}|/\lambda) \hat{y}.
\end{equation}
This implies that the maximum value of $B_y$ is shifted away from the vortex center. This shift was also found in a numerical solution of the GL equations \cite{OIM06}. For the applied field along the four-fold symmetry axis, the ${\bf B}$ field is unchanged from the usual solution. However, there is a magnetization along the radial direction \cite{yip07} (the radial magnetization was also found in the vortex lattice solution near $H_{c2}$ \cite{KAS05}).

\subsection{Vortex Lattice Solutions} \label{vortex}

Near the upper critical field, there have been a variety of studies of the Abrikosov vortex lattice \cite{KAS05,hia08,mat08,hia09}. Some of these studies predict novel lattice structures which physically stem from the interplay of FFLO physics and the helical phase discussed above \cite{hia08,mat08,hia09,agt14}. Within the phenomenological theory described here, this interplay can be captured by allowing $K$ and or $\beta$ to become negative in the GL free energy of Eq.~\ref{glfree}. In this section we will not consider these solutions, but focus on the role of the Lifshitz invariants. Near the upper critical field, the magnetic field is approximately spatially uniform and, as described earlier,  the vortex lattice is hexagonal (perhaps with a uniaxial anisotropy). The order parameter solution near the upper critical field takes the form
$\eta({\bf r})\propto \exp(i{\bf q}\cdot {\bf r}) \phi_0(x,y)$ where $\phi_0(x,y)$ is a lowest Landau level (LLL) solution and the field has been taken to be along the $\hat{z}$ axis. This solution, combining the helical phase factor with a LLL solution, has been named the helical vortex phase. The primary consequence is that the upper critical field is enhanced due to the presence of the Lifshitz invariants \cite{KAS05}. Note that due to the degeneracy of the LLL solution, there is the possibility of absorbing the helical phase factor. In particular, the LLL solution ${\tilde \phi}_0(x,y)=e^{i\tau_yx/l_H^2}\phi_0(x,y-\tau_y)$ (where $l_H$ is the magnetic length) is degenerate with $\phi_0(x,y)$, implying that the wavevector ${\bf q}$ can be removed in favor of a shift of origin, provided ${\bf q}$ is perpendicular to the applied magnetic field (which is true for point group $C_{4v}$ but not for point group $O$). In two dimensions for point group $C_{4v}$, the applied field is not screened, so the helical phase factor cannot be removed.  A second consequence of the Lifshitz invariants is on the spatial distribution of the ${\bf B}$ field. Just as in the single vortex solution, the magnetic induction can develop components that are not parallel to the magnetic field \cite{KAS05,lu08,kas13}. Finally, we note that in carrying out numerical solutions to the time-dependent GL equations, it is important to use the correct boundary conditions (those listed earlier), otherwise it is possible to find solutions in which vortices flow spontaneously \cite{OIM06}.

\subsection{Multi-component order parameters}

For non-centrosymmetric superconductors for which the superconducting order parameter contains multiple complex degrees of freedom, there will also exist Lifshitz invariants with physics similar to that described above. However, in this case there is a new possibility, that is the appearance of a helical phase when no magnetic fields are applied \cite{Li2Pt3BNode,MinSam08}.  This can occur when the ground state without magnetic field breaks time reversal symmetry \cite{sig91,UnconBook}. An example of this appears in the point group $O$ when the superconducting order parameter belongs either to one of the two three dimensional representations, for example for a spin singlet gap function $\psi({\bf k},{\bf r})=\eta_1({\bf r})k_yk_z+\eta_2({\bf r})k_xk_z+\eta_3({\bf r})k_xk_y$. Ignoring Lifshitz invariants, this state can have four possible ground states  \cite{sig91}
\begin{eqnarray}
\mbox{\boldmath$\eta$}_A=(\eta_1,\eta_2,\eta_3)\propto (1,0,0)\nonumber\\
\mbox{\boldmath$\eta$}_B=(\eta_1,\eta_2,\eta_3)\propto (1,1,1)\nonumber\\
\mbox{\boldmath$\eta$}_C=(\eta_1,\eta_2,\eta_3)\propto (1,i,0)\nonumber\\
\mbox{\boldmath$\eta$}_D=(\eta_1,\eta_2,\eta_3)\propto (1,\omega,\omega^2)\nonumber
\end{eqnarray}
where $\omega=e^{2i\pi/3}$. Symmetry allows the following Lifshitz invariant
\begin{equation}
iK(\eta_1^*D_y\eta_3+\eta_2^*D_z\eta_1+\eta_3^*D_x\eta_2 -c.c.).
\end{equation}
Writing $(\eta_1,\eta_2,\eta_3)=e^{i{\bf q}\cdot {\bf r}}(\eta_{0,1},\eta_{0,2},\eta_{0,3})$ with $\eta_{0,i}$ spatially independent yields
\begin{equation}
iK[q_y(\eta_{0,1}^*\eta_{0,3}-\eta_{0,1}\eta_{0,3}^*)+q_z(\eta_{0,2}^*\eta_{0,1}-\eta_{0,2}\eta_{0,1}^*)+q_x(\eta_{0,2}^*\eta_{0,3}-\eta_{0,2}\eta_{0,3}^*)],
\end{equation}
this reveals that the states with order $\mbox{\boldmath$\eta$}_C$ and $\mbox{\boldmath$\eta$}_D$ are unstable to developing a helical phase. For example, for $\mbox{\boldmath$\eta$}_C$, the ground state becomes $\mbox{\boldmath$\eta$} \propto e^{iqz}(1,i,0)$. The helical order develops in these two cases because the ground state breaks time-reversal symmetry.

\subsection{Microscopic derivation of Lifshitz invariants}

There have been a variety of  microscopic calculations of the Lifshitz invariants in Eq.~\ref{glfree} \cite{KAS05,Edel96,Y02,MinSam08}. These have been evaluated for temperatures close to $T_c$. Here the results are summarized. When there is only one helicity band crossing the Fermi energy, as would be the case for the surface state of a topological insulator \cite{san10}, the weak coupling limit gives
\begin{equation}
K_{ij}=-\frac{\mu_B N_0S_3}{2}\langle\phi^2({\bf k})
\hat{\bf{\gamma}}_i({\bf k})v_j({\bf k})\rangle \label{lif-1}
\end{equation}
where $N_0$ is the density of states at the Fermi energy, $\phi({\bf k})$ is a momentum dependent basis function for a one-dimensional representation of the superconducting point group (note that $\phi({\bf k})$ must be even parity), $\mu_B$ is the Bohr magneton,  and

\begin{equation}
S_3(T)=\pi  T\sum_n\frac{1}{|\omega_n|^3}=\frac{7\zeta (3) }{4\pi^2T^2}.
\label{S_3}
\end{equation}
When both helicity bands cross the Fermi energy (as is common), under the assumption that $\phi({\bf k})$ is the same for both helicity bands, then Eq.~\ref{lif-1} is multiplied by
\begin{equation}
\delta N = (N_+-N_-)/(N_++N_-). \label{relN}
\end{equation}
where $N_{\pm}$ are the partial density of states of the helicity bands at the Fermi energy (note that $N_0=N_++N_-$).\\

\subsection{Microscopic derivation of anomalous current and magnetization}

Expressions for the  anomalous magnetization Eq.~\ref{mag1} and related current Eq.~\ref{GLcurrent} have been found for all temperatures below $T_c$ \cite{Edel95,Y02}. In particular, in the limit of small magnetic fields and a small superconducting phase gradient ($\nabla \theta$), for circular 2D helicity bands in the clean limit with a Rashba spin-orbit interaction (that is $\mbox{\boldmath$\gamma$}({\bf k})=\alpha \hat{n}\times {\bf p}({\bf k})$), the Ginzburg Landau expressions Eq.~\ref{GLcurrent} and Eq.~\ref{mag1} for the supercurrent $J_x$ and the magnetic moment $M_y$  can be generalized to (here it is assumed that the field is applied along $\hat{y}$):
\begin{eqnarray}
J_x=&\rho_s\frac{\hbar \nabla_x \theta}{2m}-\kappa B_y \nonumber\\
M_y=&\frac{\kappa}{2} \hbar \nabla_x \theta
\end{eqnarray}
where $\rho_s$ is the superfluid density,
\begin{equation}
\kappa(T)=\frac{\mu}{4\pi \hbar^2}[p_{F+}\{1-Y(T,\Delta_+\}-p_{F-}\{1-Y(T,\Delta_-)\}] \label{kap}
\end{equation}
$p_{F,\pm}$ are the Fermi momenta for the two helicity bands, $\Delta_{\pm}$ are the corresponding gaps, $\mu$ is the Fermi energy, and $Y(x)$ is the Yoshida function. A key feature that emerges from this result is that, as found analogously in the above section for the Lifshitz invariants, $\kappa(T)\propto \delta N$ for $\delta N <<1$.

It has also been found that Fermi liquid corrections will alter the expression for the anomalous current \cite{Fuj05}. In particular, increased ferromagnetic correlations enhance the anomalous current. This can play an important role in heavy Fermion materials, for which the anomalous current as given through Eq.~\ref{kap} is unchanged but for which the effective mass enhancement suppresses the usual supercurrent \cite{Fuj05}. Consequently, the relative effects of the Lifshitz invariants may be significantly enhanced in these materials.

\subsection{Macroscopic theory of FFLO and helical phases}

Given that the helical phase occurs as a consequence of a Zeeman field, microscopic descriptions of this state often involve an interplay with the closely related Fulde-Ferrell-Larkin-Ovchinnikov (FFLO) state \cite{ff,lo}. It is useful to understand this interplay within a macroscopic description prior to discussing the microscopic results. The starting point for this is the GL free energy given in Eq.~\ref{glfree}. As was pointed out in Ref.~\onlinecite{buz97}, a Zeeman field can cause the phenomenological parameter $K$ to become negative for weak-coupling theories of singlet superconductors, signaling a transition into a spatially modulated state. When this occurs, terms with fourth order derivatives need to be added in the GL free energy to ensure that the energy is bounded below. Intriguingly, it is also found that the phenomenological parameter $\beta$ also becomes negative at a Zeeman field comparable to (or equal to) that at which $K$ becomes negative \cite{buz97}. This provides the macroscopic explanation for why the so called Larkin-Ovchinnikov (LO) phase is often stable. In this phase, in the simplest case, the order parameter takes the form $\eta_{LO}=\eta_0\cos ({\bf q}{\cdot {\bf x}})$. If $\beta>0$, then the so called Fulde-Ferrell (FF) phase is stable, where $\eta_{FF}=\eta_0\exp^{i{\bf q}{\cdot {\bf x}}}$. The near simultaneous sign change of both $K$ and $\beta$ with Zeeman field (in weak-coupling theories) imply that the detailed FFLO phase diagram depends sensitively on the nature of impurity scattering and anisotropy in the singlet gap function \cite{buz97,agt01}. Indeed, a key difference between the helical phase in non-centrosymmetric materials and the FFLO phase, is that since the helical phase is determined by symmetry arguments, it will be robust in the presence of impurities; this is not the case for the FFLO phase.

More recently, it has been understood that important insight into FFLO and related pair density wave (PDW) states can be gained by expressing the GL free energy in terms of a plane-wave expansion of the order parameter, for example using $\{\psi_q({\bf x}),\psi_{-q}({\bf x})\}$ where $\eta({\bf x})=\psi_q({\bf x})e^{i{\bf q}{\cdot {\bf x}}}+\psi_{-q}({\bf x})e^{-i{\bf q}{\cdot {\bf x}}}$ \cite{agt08,DF03,dim07,ber09,rad11}. In such a basis, the GL free energy becomes (here we have not included the effect of broken parity symmetry which we discuss below)
\begin{equation}
F=\int d^3r\left\{\alpha(|\psi_q|^2+|\psi_{-q}|^2)+\frac{\beta_1}{2}(|\psi_q|^2+|\psi_{-q}|^2)^2+\beta_2|\psi_q|^2|\psi_{-q}|^2
+K(|{\bf D}\psi_q|^2+|{\bf D}\psi_{-q}|^2)  \right\}\label{glfree-2}.
\end{equation}
A feature that immediately emerges from Eq.~\ref{glfree-2}, is that the symmetry of the free energy is higher than the usual $U(1)$ gauge symmetry. In particular, it can be seen that the free energy is independent of the values of the phases of both $\psi_q$ and $\psi_{-q}$, implying a $U(1)\times U(1)$ symmetry. The emergence of this second $U(1)$ degree of freedom physically describes the breaking of translational symmetry breaking by the FFLO order parameter and is a general feature of such theories. It is important in understanding the topological defects of the theory, which no longer include just vortices, but also dislocations due to the translational symmetry breaking. Intriguingly, it has been found that defects combining half a vortex with half a dislocation also appear \cite{agt08,ber09-2,rad09}. Such defects play an important role in the vortex physics, where a conventional vortex can decay into a pair of half-flux vortices \cite{agt08}, and in the defect mediated melting of these phases, which can give rise to new phases, such as a spatial homogeneous charge 4e superconductor \cite{ber09-2,rad09}.

The approach of the previous paragraph is perhaps the clearest to understand the interplay between the helical and FFLO phases when inversion symmetry is broken. In particular, the simplest means to capture this interplay is to introduce a term
\begin{equation}
\epsilon(|\psi_q|^2-|\psi_{-q}|^2)
\end{equation}
to Eq.~\ref{glfree-2}. This term explicitly breaks both parity and time-reversal symmetries and therefore appears in broken parity materials only when a magnetic field is applied. The consequences of this term are that either $\psi_q$ or $\psi_{-q}$ will appear at a phase transition and that a second phase transition can occur into a LO-like phase, for which both components $\psi_q$ or $\psi_{-q}$ are non-zero. An additional consequence of this term in the free energy is that the one-half flux vortices of the inversion symmetric FFLO phases now become fractional flux quanta vortices with $\Phi_1+\Phi_2=\Phi_0$ and $\Phi_1\ne \Phi_0/2$. This can lead to the development of Skyrmion vortex phases, as is found within the microscopic theory discussed in Ref.~\onlinecite{agt14}.

\subsection{Microscopic Theory of the Helical and FFLO Phases}

The origin of the helical phase within a microscopic description follows from the observation that once both parity and time-reversal symmetries are both broken, there is no guarantee that single-particle states at momentum ${\bf k}$ and $-{\bf k}$ are degenerate. Consequently, the possibility that Cooper pairs with a finite center of mass momentum (that is pairing between single particle states with momenta ${\bf k}+{\bf q}/2$ and $-{\bf k}+{\bf q}/2$, leading to a gap function $\Delta({\bf x})\propto e^{i{\bf q}\cdot {\bf x}}$) should be considered. Within a weak-coupling theory, finite momentum pairing states can appear once a Zeeman magnetic field is turned on.  In particular, including the Zeeman field through
\begin{equation} H_Z=-\sum_{{\bf k},\alpha,\beta}\mu_B{\bf H}\cdot \mbox{\boldmath$\sigma$} _{\alpha\beta}
    a^\dagger_{{\bf k}\alpha}a_{{\bf k}\beta}\end{equation} leads to the following single particle
excitation spectrum
\begin{equation}
    \xi_{\pm}({\bf k},{\bf H})=\xi({\bf k})\pm \sqrt{\mbox{\boldmath$\gamma$}^2({\bf k})-2\mu_B\mbox{\boldmath$\gamma$}({\bf k})\cdot {\bf H}+
   \mu_B^2{\bf H}^2}.
\label{e3}
\end{equation}
Taking $|\mbox{\boldmath$\gamma$}|>>|H|$ (and ignoring small
regions of momentum space where $\mbox{\boldmath$\gamma$}=0$ may occur) yields
\begin{equation}
\xi_{\pm}({\bf k},{\bf H}) \approx \xi({\bf k})
\pm\mu_B\hat{\mbox{\boldmath$\gamma$}}({\bf k})\cdot{\bf H}.
\end{equation}
As discussed in the introduction, the origin of finite momentum pairing states are a consequence of this dispersion.  For a Rashba spin-orbit interaction,
with $\mbox{\boldmath$\gamma$}=\gamma_{\perp}(k_y\hat{x}-k_x\hat{y})$, an otherwise parabolic dispersion, and
a magnetic field along $\hat{x}$, the Fermi surfaces remain circular with centers
shifted along the $\hat{y}$ direction (the two Fermi surfaces are shifted in opposite directions).  Finite momentum
Cooper pairs are stable because pairing now occurs through the new center of one of the Fermi surface. In this case, the same momentum vector
$\vq$ can be used to pair {\it every} state on one of the two
Fermi surfaces, illustrating that the finite momentum pairing state can be very stable. For an arbitrary ASOC, a finite center of mass pairing can occurs if
the paired states are degenerate, that is if
$\xi_{\pm}({\bf k}+{\bf q},{\bf H})=\xi_{\pm}(-{\bf k}+{\bf q},{\bf H})$, which yields $\hbar
\vq\cdot \vv_F=\mu_B\vH\cdot\hat{\mbox{\boldmath$\gamma$}}({\bf k})$. For the usual FFLO superconductor, the equivalent condition is
$\hbar \vq\cdot \vv_F=\mu_B|\vH|$ which typically is satisfied over a smaller phase space than $\hbar
\vq\cdot \vv_F=\mu_B\vH\cdot\hat{\mbox{\boldmath$\gamma$}}({\bf k})$, indicating an increased relative stability of finite momentum pairing states when inversion symmetry is broken.

The bulk of microscopic studies of finite momentum pairing states have focussed on a Rashba spin-orbit interaction in two-dimensions \cite{BG02,DF03,dim07,agt07,mic12,lod13,agt14,hou15}. There have also been few studies in three dimensions \cite{KAS05,Sam08,mat08,hia09,agt07,yan07,san10}, including point groups $C_{4v}$ and the point group $O$, which develops a helical modulation along the direction of the applied magnetic field and therefore easily co-exists with a vortex lattice phase. These studies have resulted in a clear physical picture of the interplay between the helical phase and the  FFLO phase, the role of disorder on these phases, and how to probe these phases. Without going into more detailed microscopic arguments, some of the main results for a Rashba interaction are presented in the remainder of this section.

When no disorder is present, then the interplay between the FFLO and the helical phases plays an important role in the underlying physics. This interplay manifests itself with the choice of ${\bf q}$ in forming finite center of mass momentum states. In particular, as discussed above, with a Rashba spin-orbit coupling and a magnetic field along the $\hat{x}$ direction, the two Fermi surfaces prefer opposite ${\bf q}$ vectors. This gives rise to competition between the single-q (helical) and multiple-q (FFLO-like) phases. A central parameter in determining the stable ground state is $\delta N$ introduced in Eq.~\ref{relN} \cite{dim07,agt07,Sam08,hia09,mic12,lod13,hou15}. When $\delta N$ is small, both the helical and FFLO-like phases appear. In the low field limit, the helical phase appears with
$|{\bf q}|=q\approx\delta N H \mu_B/v_F$. As the field is increased above the typical Pauli field, then the FFLO-like phase appears through a phase transition while $q$ increases in magnitude to a value $q\approx H\mu_B/v_F$. As $\delta N$ increases, the FFLO-like phase becomes less stable and eventually disappears. For a Rashba interaction in 2D with a parabolic dispersion, the FFLO-like phase no longer exists when $\delta N>0.25$ \cite{agt14}.  In the clean limit, the helical phase can be observed through density of states measurements. At high fields, the density of states
in the helical phase reveals a gap for the Fermi surface for which pairing is favored by ${\bf q}$ and a residual normal state (temperature independent) density of states for the Fermi surface that would prefer to pair with a momentum $-{\bf q}$. The appearance of this residual density of states as field is increased would provide a signature of the helical phase.

Disorder also tends to remove the FFLO-like phases, leaving only the helical state as stable. In the moderate disorder limit, for which $1/\tau<<\Delta_{so}$ where $\tau$ is the relaxation time and $\Delta_{so}=\alpha k_F$, is the strength of the Rashba spin-orbit coupling, significant theoretical progress can be made \cite{dim07,mic12,hou15}. In particular, it has been shown that
\begin{equation}
{\bf q}=-\frac{4\delta N}{v_F}\mu_B{\bf H}\times {\hat{z}}
\end{equation}
throughout the phase diagram (this assumes that $\Delta_{so}<<\epsilon_F$) and the upper critical field for in-plane fields is given by
\begin{equation}
\mu_B H_{c2}=\frac{1}{2}\sqrt{\frac{\hbar \Delta_0}{\tau}}
\end{equation}
where $\Delta_0$ is the magnitude of the superconducting gap at zero field and at zero temperature. The increase of the critical field with disorder has been noted in a series of publications \cite{dim07,mic12,hou15} and a related increase has also been found for superconductors with ASOC consistent with cubic $O$ point group symmetry \cite{Sam08}.
We note that strong spin-orbit coupling and strong disorder appear to be generic ingredients for a recent observation that $T_c$ is increased in 2D materials when an in-plane magnetic field is applied \cite{gar11}. However, the results of the weak-coupling theories presented above cannot account for this observed increase in $T_c$, though they naturally give rise to a $T_c$ that is not as strongly suppressed by in-plane fields as in usual centrosymmetric singlet superconductors. Finally, we note that there is experimental support for the helical phase in monolayer Pb films \cite{sek13}.

\section{Topological superconductivity and edge states} \label{toposec}

With the realization that spin-orbit coupling can drive topological insulator states, with edge states protected by the topology of the bulk wavefunction \cite{kan05a,kan05b}, a deeper understanding of bound states inside vortex cores or at the edge of superconductors has emerged. Indeed topological materials and superconductivity has generated a large literature and is a rapidly evolving field which is covered by a series of recent review articles \cite{has10,qi11,miz16,chi15,yip14,sam15,sch15}. Here we focus on the implications of these recent results on non-centrosymmetric superconductors. In this context, we emphasize three aspects:\\
1- Topological edge states in fully gapped superconductors \cite{vol88,vor08,lu08,lu10,tan09,qi09,qi10,sch08,sam15,sam15-2,sch15}.\\
2- Topological edge states in nodal broken inversion superconductors \cite{ini07,vor08,lu10,yip14,sch11,sch12,sch13,sch15,tim15}.\\
3- The use of noncentrosymmetric superconductors to create single Majorana modes \cite{qi11,Majoran1,sat10,fu08}.\\

A key result is that the degree of singlet-triplet mixing plays a central role in determining if the superconductor is topological. This is particularly relevant when the pairing is in the isotropic channel, that is, the symmetry of the superconducting state is the same as that of a usual $s$-wave superconductor. In this case, the predominantly spin-singlet pairing states are not topological, while the predominantly spin-triplet states are. This applies to fully gapped superconducting states and to mixed singlet-triplet superconducting states that have line nodes. If the pairing is not in the isotropic channel, then topological arguments used in the context of understanding Andreev bound states in unconventional superconductors (for example the well known bound states of a $d_{x^2-y^2}$ superconductor along a $(1,1,0)$ surface \cite{hu94,cov97}) also apply. In this Section, the gap functions are defined through the pairing Hamiltonian given in Eq.~\ref{pairing}. When just two helicity bands are included, these gap functions are given by Eq.~\ref{mix} (note that the discussion below allows for the existence of more than two bands).

\subsection{Fully gapped noncentrosymmetric superconductors}

For fully gapped superconductors, the "ten-fold way" topological classification applies \cite{has10,qi11,chi15,sch08}. Within this classification, non-centrosymmetric superconductors with time-reversal symmetry belong to class D-III. For this class, it is known that in 1D and 2D, there is a $Z_2$ topological index that characterizes topological superconductivity, while in 3D there is an integer $Z$ index. Focussing first on the 3D case, the $Z$ index can be expressed as a winding number over all of momentum space \cite{sch08}. This is not always useful for superconductivity, for which the pairing often only occurs for electrons within a small energy window near the Fermi surface. In this limit, the corresponding winding number has been found to be \cite{qi10,sam15-2}
\begin{equation}
N_W=\frac{1}{2}\sum_s \textrm{sgn}(\Delta_s) C_{1s}
\label{Z3D}
\end{equation}
where the $s$ sums over disconnected Fermi surfaces, $\textrm{sgn}(\Delta_s)$ gives the sign of the gap function on the Fermi surface labelled by $s$, and $C_{1s}$ is the first Chern number of the Fermi surface $s$ and is computed in the normal state without superconductivity. In the context of the pseudo-spin basis discussed earlier the sum over $s$ involves two Fermi surfaces and for each of these, $C_{1s}$ can be interpreted as the monopole charge due to the spin-orbit induced spin texture. An illustrative example is a superconductor with point group $O$, which allows $\mbox{\boldmath$\gamma$}({\bf k})\propto k_x\hat{x}+k_y\hat{y}+k_z\hat{z}$. This yields a pair of spherical Fermi surfaces in the small $k$ limit. In this case the spin texture on each Fermi surface has the same form as the electric field due to a point charge at the origin, with the two Fermi surfaces having opposite monopole sign, $C_{1s}=(-1)^s$. Consequently, only when the sign of the gap is opposite on both these Fermi surfaces do we have a non-zero topological index, implying a topological superconductor. As discussed earlier, this corresponds to the situation that the superconducting state is predominantly spin-triplet (the predominantly spin-singlet case is not topological). From the bulk-boundary correspondence, a non-zero $N_W$ implies that the there exist edge states. In the context of the point group $O$ discussed in the above paragraph, the corresponding edge states are two-dimensional Majorana cone states \cite{qi11,sch15,chi15} which are Majorana edge states with a two-dimensional conical dispersion in the surface Brillouin zone momenta.

In 2D, the associated topological index for time-reversal invariant non-centrosymmetric superconductors is \cite{qi10,Majoran1,sam15}
\begin{equation}
N_{2D}=\Pi_s [\textrm{sgn}(\Delta_s)]^{m_s}
\end{equation}
where the product is over all disconnected Fermi surfaces and $m_s$ is the number of time-reversal invariant points contained by the Fermi surface with label $s$ (time-reversal invariant points satisfy ${\bf k}={\bf G}/2$ where ${\bf G}$ is a reciprocal lattice point). A relevant example in this case is a Rashba spin-orbit interaction in 2D for which $\mbox{\boldmath$\gamma$}({\bf k})\propto k_x\hat{y}-k_y\hat{x}$ in the small momentum limit. In this case there is a pair of circular Fermi surfaces and the only time-reversal invariant point contained by either of these Fermi surfaces is ${\bf k}=0$, leading to $m_s=1$. Consequently, when the two gaps have opposite sign, the superconductor is topological, supporting a pair of helical Majorana edge states. These two states are partners under time-reversal symmetry and exhibit linear dispersion in the momentum at the edge \cite{qi10,Majoran1,tan09,sam15,sch15}. These states support spin-currents which have lead to a variety of investigations \cite{vol88,vor08,tan09,chi15}. Again, topological superconductivity corresponds to the situation that the superconducting state is predominantly spin-triplet (the predominantly spin-singlet case is not topological). In addition to the helical pair of edge modes, usual single-flux quantum vortices also contain a pair of Majorana modes \cite{lu08,sat09,qi10}. Since these vortices do not contain isolated Majorana modes, they will not obey non-Abelian statistics that are useful in the context of quantum computing \cite{sat09,qi10}.

We note that some related results exist for odd-parity superconductors with parity symmetry. In this case, a fully gapped odd-parity superconductor is topological if the Fermi surface encloses an odd number of time-reversal invariant points \cite{sat09,fu10}.

\subsection{Nodal noncentrosymmetric superconductors}

It was realized that requiring a full superconducting gap was not required to generate edge states \cite{sato06,ini07,vor08,lu10,yip14,sch11,sch12,sch13,sch15,tim15}. For pairing states with the same symmetry as the crystal ($s$-wave-like), these states appeared in the predominantly spin-triplet case.  When the gap has nodes, then the topological arguments given in the previous subsection no longer apply. However, it was found that other topological invariants can be used to describe the edge states that appear \cite{sato11,sch11,sch12}. In particular it was found that three types of edge states can be associated with non-zero topological invariants; namely flat bands, Fermi arcs, and Majorana helical states in 2D or a Majorana cone in 3D.

The Majorana flat bands states are related to the well known Andreev bound states of $d$-wave superconductors. These bound states were shown to exist on a surface with normal $\hat{{\bf n}}$ if $\Delta({\bf k}_\parallel,k_n)\Delta({\bf k}_\parallel,-k_n)<0$ where ${\bf k}_\parallel$ is a wavevector in the surface Brillouin zone )and is perpendicular to $\hat{{\bf n}}$. More recently, a related integer topological invariant \cite{sato11,sch12} has been developed to describe superconducting topological Majorana flat band states in the $\pm$ helicity bands
\begin{equation}
W_{lmn}({\bf k}_\parallel)=-\frac{1}{2}\sum_{\nu=\pm,\epsilon_{\nu}({\bf k})=0}\textrm{sgn}[\partial_{k_n}\epsilon_{\nu}({\bf k})]\textrm{sgn}[\Delta_{\nu}({\bf k})]
\end{equation}
where $(lmn)$ are the Miller indices describing the surface, $\epsilon_{\nu}({\bf k})$ is the electron dispersion on helicity band $\nu$, $\epsilon_{\nu}({\bf k})=0$ denotes that the sum is over all points for which the trajectory across the 3D Brillouin zone crosses the helicity Fermi surface $\nu$, this trajectory is given by the straight line $({\bf k}_\parallel,k_n)$ with $k_n$ varying from one side of the zone to the other and ${\bf k}_\parallel$ is fixed. When this index is non-zero, then flat bands states will exist on the corresponding edge.  Typically, such trajectories cross the Fermi surface at two points and the sign of $\partial_{k_n}\epsilon_{\nu}({\bf k})$ is opposite on these two points, in which case this invariant simplifies \cite{sato11,sch12}
\begin{equation}
W_{lmn}({\bf k}_\parallel)=-\frac{1}{2}\sum_{\nu=\pm}\Big \{\textrm{sgn}[\Delta_{\nu}({\bf k}_{F,\nu})]-\textrm{sgn}[\Delta_{\nu}(\tilde{{\bf k}}_{F,\nu})]\Big\}
\end{equation}
where ${\bf k}_{F,\nu}$ is one crossing point of the trajectory on the helicity Fermi surface $\nu$ and  $\tilde{{\bf k}}_{F,\nu}$ is the other crossing point. This immediately implies that if line nodes occur on one of the two helicity bands due to spin-singlet and spin-triplet mixing in the $s$-wave-like pairing channel, a surface has flat band states for the  surface BZ momentum region that is bounded by the projections of the line nodes of the bulk gap onto the surface BZ. An example of using this invariant is given below. These flat band states give rise to a zero bias conductance peak that can be seen through scanning tunneling spectroscopy \cite{sato06,ini07,vor08,lu10,yip14,sch11,sch12,sch13,sch15,tim15}. The stability of these flat bands at the edge have been examined for stability against disorder \cite{que14} and also for intrinsic instabilities \cite{tim15}. When time reversal symmetry is broken, these flat bands develop into a chiral dispersion, allowing for edge currents to flow \cite{stb13}.

Fermi arcs can appear if the ASOC satisfies a particular symmetry and if there is more than one nodal line on a Fermi surface \cite{sch12}. In particular, if the  ASOC satisfies
$\mbox{\boldmath$\gamma$}({\bf k}_i,k_0)=\mbox{\boldmath$\gamma$}({\bf k}_i,-k_0)$, along some direction $\hat{u}$ orthogonal to ${\bf k}_i$, then a Fermi arc will connect the projection of two nodal rings on the surface BZ for a surface normal is orthogonal to ${\bf u}$ \cite{sch12}. Each point of this Fermi arc has a pair of linearly dispersing Majorana helical modes.  An example of when these Fermi arcs appear is given below.

Finally, Kramers degenerate Majorana cone states can appear at time-reversal invariant points ${\bf K}_{\parallel}$ of the surface BZ. These states appear if the $Z_2$ invariant $N_{{\bf k}_\parallel}= \textrm{sgn}[\Delta_{+}({\bf k}_{F,+})]\textrm{sgn}[\Delta_{-}({\bf k}_{F,-})]$ is non-zero \cite{sch11,sch12}. Here ${\bf k}_{F,\pm}=(k_{n\pm},{\bf K}_{\parallel})$ are the two points on the helical Fermi surfaces that are crossed by varying $k_{n}$ with fixed ${\bf K}_{\parallel}$ and subject to $\partial_{k_n}\epsilon_{\nu}({\bf k})>0$ at these points.

An instructive example that illustrates the use of the above topological invariants to understand the resultant surface states is  a Rashba spin-orbit coupling $\mbox{\boldmath$\gamma$}({\bf k})\propto k_x\hat{y}-k_y\hat{x}$ on a spherical Fermi surface \cite{sch12,yip14}. In this case, the gap functions on the two helicity bands are given by $\Delta_{\pm}(\theta,\phi)=\Delta_s\pm\Delta_t|\sin(\theta)|$ where $\theta,\phi$ are the spherical angles that describe the position on the Fermi surface (we take $\Delta_s$ and $\Delta_t$ to be positive).  If $\Delta_t>\Delta_s$, then $\Delta_-$ has two line nodes at fixed $\theta=\theta_0,\pi-\theta_0$. For a surface normal along the $\hat{z}$ direction, the above topological invariants all show that are no topological edge states, this is because every trajectory through the BZ along the $\hat{z}$ direction crosses the FS at two points for which the gap has the same value. If the surface normal is rotated slightly away from the $\hat{z}$ axis, then there will be surface flat bands arising for the projections of the nodal planes on the surface BZ (there are no flat bands where the projections of each nodal ring overlap on the surface BZ). Now consider a surface normal along $\hat{x}$. In this case the symmetry of the Rashba interaction $\mbox{\boldmath$\gamma$}({\bf k}_\parallel, k_z)=\mbox{\boldmath$\gamma$}({\bf k}_\parallel, -k_z)$ applies and there is Fermi arc that runs along $k_y$ from the projection of one nodal line to the other on the surface BZ. Rotating slightly away from an $\hat{x}$ normal, towards $\hat{z}$ implies that the same symmetry can no longer be used. However, there will be a Majorana cone state at the origin since this is a time-reversal invariant point and the $Z_2$ index $N_{{\bf k}_\parallel}= \textrm{sgn}[\Delta_{+}({\bf k}_{F,+})]\textrm{sgn}[\Delta_{-}({\bf k}_{F,-})]$ is non-zero for ${\bf k}_{\parallel}=0$. These surface states are examined in detail in Ref.~\onlinecite{sch12}. It is also worthwhile noting that if the spin-singlet position of the gap has symmetry imposed nodes (for example having $d_{xy}$ symmetry), then the modes at the edge are also interesting \cite{tan10,sato11,sch12}.

\subsection{Isolated Majorana modes in vortex cores}

It was realized that vortices containing isolated Majorana bound states obey non-Abelian statistics \cite{vol99,qi11}. This has a number of dramatic consequences, perhaps the most important of which is in topological quantum computing. This subject is reviewed in detail in Ref.~\onlinecite{qi11}, so we briefly discuss the role of non-centrosymmetric superconductors in this context. In particular, it was realized that a two-dimensional material with a Rashba spin-orbit interaction in the low density limit has the potential to have a vortex with a single Majorana bound state in the vortex core \cite{sau10,sat09,ali10}. In particular if a Zeeman magnetic field is applied such that one of the two helicity bands becomes gapped, so that only one helicity band remains,  a usual superconducting vortex in the remaining helicity band will have a single Majorana mode.

\section{Summary and prospects}

We have reviewed the properties of both strongly and weakly correlated NCS and summarised the diverse observed properties. We also provided a theoretical understanding of these properties. Theory indicates that the ASOC plays an essential role in understanding much of the physics of NCS materials, though electronic interactions are also important to drive some of the new physics, like singlet-triplet mixing. This latter aspect has been a priority from the experimental point of view and we have discussed the effects of the ASOC on the superconducting properties and the extent to which the results give evidence for the presence or absence of singlet-triplet mixing.

A diverse range of properties have been observed in NCS and some systems showing evidence for either fully gapped or nodal multiband superconductivity are compatible with a singlet-triplet mixing.  In particular for the Li$_2$(Pd$_{1-x}$Pt$_x$)$_3$B system,  the evolution from nodeless to nodal two band superconductivity upon increasing the ASOC by substituting Pt for Pd, is strong evidence for mixed parity pairing. Other systems showing multiple gaps are compatible with singlet triplet mixing, but in some cases this may also arise from conventional $s$-wave multiband superconductivity, particularly when there is not a clear relationship between the ASOC and the pairing state. Superconducting interfaces and heterostructures, as well as other low dimensional systems also provide a further opportunity for examining the effects of inversion symmetry breaking. An advantage of studying these systems is that the strength of the ASOC can often be tuned controllably.

It is clear that despite the fact that mixed parity pairing is predicted to be a general feature of NCS in the presence of ASOC, many weakly correlated NCS show behaviour consistent with single band $s$~wave superconductivity. A simple interpretation of this is that if the triplet component is very weak then there will be two gaps of very similar magnitude dominated by spin singlet pairing and the behaviour will be indistinguishable from ordinary BCS superconductors. The parameter $E_{ASOC}/T_c$, which compares the relative energies of the  antisymmetric spin orbit coupling and superconducting condensate, may give an indication of promising systems to look for significant singlet-triplet mixing  \cite{BiPdYuan}. However, measurements show a large ASOC  appears to be a necessary but not sufficient condition for the presence of significant mixed parity pairing. This also requires pairing interactions in both the singlet and triplet channels, which can explain why it may be present in some Ce based NCS, but absent in their isostructural, weakly correlated analogues. In this context, through the observation of singlet-triplet mixing, NCS materials can provide new insight into the role of correlations in superconducting materials. While the presence of heavy atoms provides a guide for finding compounds with strong ASOC, it may also be necessary to examine detailed electronic structure calculations, particularly to look for systems where the ASOC causes significant splitting in the vicinity of the Fermi level.

Measurements of the spin susceptibility and $H_{c2}(T)$ also provide evidence for the effects of ASOC on superconductivity.  However an important consequence of the ASOC is that it is not possible to distinguish between the singlet and protected triplet state using these measurements and therefore they can not be used to probe for the presence of singlet-triplet mixing. Perhaps the best evidence for an anisotropy of the spin susceptibility compatible with the effects of ASOC comes from measurements of CeIrSi$_3$, where there is qualitative agreement between theory and experiment. However in CePt$_3$Si, the isotropicconstant Knight shift and small anisotropy of  $H_{c2}(T)$ are difficult to understand and measurements of both these systems may be complicated by the presence of strong electronic correlations. {Anisotropy in $H_{c2}$ shows a better agreement between theory of the ASOC and experiment, most strikingly in the 2D Ising superconductors MoS$_2$ and NbSe$_2$. Zero field $\mu$SR measurements  can give clear evidence for the TRS breaking expected for some spin-triplet states. However the symmetry analysis of LaNiC$_2$ demonstrates that this can in fact indicate an absence of singlet-triplet mixing, rather than its presence.

Despite the interesting results already reported, there is a considerable amount of further work which can be carried out and a number of open questions.

\begin{itemize}

\item The predicted magnetoelectric effects and related physics, which are required to exist solely due to symmetry arguments, have not yet been observed.

\item More detailed experimental and theoretical work is required to clearly demonstrate the presence of mixed singlet-triplet pairing. It is necessary to identify further NCS where there is evidence for such a mixed parity state. In particular, it is desirable to directly probe for the presence of a spin-triplet component, which may be measured by looking for Majorana edge states and Majorana modes.

\item The role of the different parameters in determining whether there is significant singlet-triplet mixing remains unclear. In particular it is important to clarify the relative importance of the ASOC strength and the presence of electronic correlations.

\item While low dimensional superconducting systems have provided a valuable opportunity to controllably tune the ASOC, the nature of the pairing symmetry in these systems has yet to be clarified, requiring further study.

\item Several NCS have been identified as promising candidates to display topological superconductivity, due to the presence of a topologically non-trivial band structure. Further experimental work is required to confirm whether topological superconducting states exist in these materials.

\end{itemize}

\section{Acknowledgements} We thank Philip Brydon, Kirill Samokhin, Manfred Sigrist, Sungkit Yip and Jorge Quintanilla for useful discussions. DFA was supported by the NSF via DMREF-1335215. The work at Zhejiang University was supported by the National Natural Science Foundation of China (No. 11474251), National Key R\&D Program of the MOST of China (No. 2016YFA0300202), the Science Challenge Program of China and the Max Planck Society under the auspices of the Max Planck Partner Group of the Max Planck Institute for Chemical Physics of Solids. MBS acknowledges a visiting research appointment in the National High Magnetic Field Laboratory at Los Alamos National Laboratory.

\bibliographystyle{unsrt}
\bibliography{ThesisBib}

\end{document}